\documentclass[reprint, superscriptaddress, nofootinbib,nobibnotes,amsmath,amssymb, aps,prd]{revtex4-2}

\RequirePackage[colorlinks=true
,urlcolor=magenta
,anchorcolor=blue
,citecolor=magenta
,filecolor=blue
,linkcolor=blue
,menucolor=blue
,linktocpage=true
,pdfproducer=medialab
,pdfa=true
]{hyperref}

\usepackage{amsmath,amssymb,amsthm,amsfonts}
\usepackage{cprotect}
\usepackage{graphicx,tabularx}
\usepackage{float}
\usepackage{multirow}  
\usepackage{cancel}
\usepackage{comment}
\usepackage{physics}
\usepackage{array}
\usepackage{enumitem}
\newcolumntype{P}[1]{>{\centering\arraybackslash}p{#1}}

\usepackage{xcolor}
\usepackage{dcolumn}
\usepackage{bm}

\usepackage{bm,epstopdf,natbib,hyperref,color,verbatim,multirow,bm,tikz,xcolor}
\hypersetup{colorlinks=true,urlcolor=blue,citecolor=blue,linkcolor=blue,menucolor=blue,anchorcolor=blue,filecolor=blue}
\everymath{\displaystyle}
\definecolor{lime}{HTML}{A6CE39}
\DeclareRobustCommand{\orcidicon}{
	\begin{tikzpicture}
	\draw[lime, fill=lime] (0,0) 
	circle [radius=0.16] 
	node[white] {{\fontfamily{qag}\selectfont \tiny ID}};
	\draw[white, fill=white] (-0.0625,0.095) 
	circle [radius=0.007];
	\end{tikzpicture}
	\hspace{-3mm}
}
\foreach \x in {A, ..., Z}{\expandafter\xdef\csname orcid\x\endcsname{\noexpand\href{https://orcid.org/\csname orcidauthor\x\endcsname}
			{\noexpand\orcidicon}}
}

\def\lsim{\:\raisebox{-0.9ex}{$\stackrel{\textstyle<}{\sim}$}\:}
\def\gsim{\:\raisebox{-0.9ex}{$\stackrel{\textstyle>}{\sim}$}\:}

\preprint{LA-UR-24-33252}

\begin{document}


\title{The Sound of Dark Sectors in Pulsar Timing Arrays}

\author{Amitayus Banik\hspace{-1mm}\orcidA{}}
\email{abanik@cbnu.ac.kr}
\affiliation{Department of Physics, Chungbuk National University, Cheongju, Chungbuk 28644, Korea}
\affiliation{Research Institute for Nanoscale Science and Technology, Chungbuk National University, Cheongju, Chungbuk 28644, Korea}

\author{Yanou Cui\hspace{-1mm}\orcidB{}}
\email{yanou.cui@ucr.edu}
\affiliation{Department of Physics and Astronomy, University of California, Riverside, CA 92521, USA}

\author{Yu-Dai Tsai\hspace{-1mm}\orcidC{}}
\email{y.tsai@sheffield.ac.uk,yudaitsai.academic@gmail.com}
\affiliation{School of Mathematical and Physical Sciences, University of Sheffield, Sheffield S3 7RH, UK}
\affiliation{The University of Manchester, Manchester M13 9PL, UK}
\affiliation{Los Alamos National Laboratory (LANL), Los Alamos, NM 87545, USA}

\author{Yuhsin Tsai\hspace{-1mm}\orcidD{}}
\email{ytsai3@nd.edu}
\affiliation{Department of Physics and Astronomy, University of Notre Dame, IN 46556, USA}

\begin{abstract}
A phase transition in the dark sector (DS) presents a promising explanation for the stochastic gravitational wave (GW) signals detected in recent observations by Pulsar Timing Arrays (PTAs). Instead of focusing solely on fitting data with phenomenological parameters, we systematically delineate simple, underlying DS models at the microscopic, Lagrangian level and uncover the conditions required to yield the GW spectrum observed by PTAs. Within this framework, the characteristic scale of the dark sector is ${\cal O}$(MeV), with complementary tests at future energy- and intensity-frontier experiments.
\end{abstract}

\maketitle


\section{Introduction}
\label{sec:intro}
Recent measurements from several Pulsar Timing Arrays (PTAs), including NANOGrav, EPTA, InPTA, PPTA, CPTA and MPTA~\cite{NANOGrav:2020bcs, NANOGrav:2023gor, EPTA:2021crs, EPTA:2023fyk, EPTA:2023xxk, Goncharov:2021oub, Reardon:2023gzh, Xu:2023wog,Miles:2024seg}, have provided strong evidence for quadrupolar Hellings-Downs correlations \cite{Hellings:1983fr}, potentially indicating the discovery of a stochastic gravitational wave background (SGWB) in the nanohertz frequency range. These findings signal a new chapter in gravitational wave (GW) cosmology and prompt the critical question of the origin of this observation. A potential origin of the signal is GW emission from a large population of coalescing supermassive black holes~\cite{Haehnelt:1994wt, Rajagopal:1994zj, Jaffe:2002rt, Wyithe:2002ep, Sesana:2004sp, Sesana:2008mz, Burke-Spolaor:2018bvk, Middleton:2020asl, NANOGrav:2023hfp}. However, the specifics of the astrophysical mechanisms capable of generating the observed signal remain an active area of investigation~\cite{1980Natur.287..307B, Milosavljevic:2002ht}.
Alternatively, beyond the Standard Model (SM) physics offers compelling cosmological sources of the signal~\cite{Ellis:2020ena, Ashoorioon:2022raz, NANOGrav:2023hvm, Ferreira:2022zzo, Bringmann:2023opz,  Ellis:2023tsl, Kume:2024adn, Kitajima:2023cek, Bai:2023cqj, Benetti:2021uea, Chowdhury:2023xvy, Fujikura:2023lkn, Winkler:2024olr}, and precise measurements of the signal may open up exciting opportunities to probe the microscopic details of the underlying new physics in the near future. In this work, we focus on the generally appealing possibility that the PTA signal originates from SGWB generated during a phase transition (PT) in a dark sector (DS), and explore the connections between the specifics of the GW spectrum and the characteristics of the DS.

GWs originating from PTs \cite{Caprini:2015zlo, Athron:2023xlk} have been widely studied in the context of the electroweak PT, whereby beyond the Standard Model (SM) physics is required to render it strong first-order for generating detectable GWs (see, e.g., Refs.~\cite{Kamionkowski:1993fg, Apreda:2001us, Grojean:2006bp, Ashoorioon:2009nf, Kakizaki:2015wua, Vaskonen:2016yiu, Dorsch:2016nrg, Beniwal:2017eik, Ellis:2018mja, Chatterjee:2022pxf}). In these cases, the predicted GW signal can be within the LISA frequency range~\cite{LISA:2017pwj, LISA:2022kgy}, offering potential complementarity between related collider searches and future GW observations. Recent years have seen increasing interest in PTs occurring solely within a DS~\cite{Schwaller:2015tja, Jaeckel:2016jlh, Breitbach:2018ddu, Dent:2022bcd, Morgante:2022zvc, Pasechnik:2023hwv, Koutroulis:2023wit, Feng:2024pab}, with its own scalar potentials and, potentially gauge symmetries, which allows for a strong first-order PT. GW signatures would serve as a promising means for probing such a DS, in particular when its non-gravitational interaction with the SM is absent or feeble.

Most of the existing related studies~\cite{Bringmann:2023opz, Addazi:2023jvg, Ghosh:2023aum,Winkler:2024olr} focused on parameterizing DS PTs with macroscopic quantities, such as the energy density of DS, the temperature and duration of the PT. Only a few studies~\cite{DiBari:2023upq,Han:2023olf,Li:2023bxy,Salvio:2023ynn, Salvio:2023blb} have investigated microscopic details of the DS that underlie the observed PTA spectra. To address this gap, we delineate minimal DS models with perturbative dynamics that can realize a strong first-order PT, involving gauge and fermionic fields, as inspired by the electroweak sector of the SM. We perform a systematic general study of the first-order PT in such models and the resultant GW signatures, focusing on fitting the PTA observations through the recently released NANOGrav 15-year dataset~\cite{NANOGrav:2023gor} as implemented in the publicly available \verb|PTArcade|~\cite{Mitridate:2023oar} framework.
Based on the PTA fit, we also identify a plausible cosmological history for the DS, and discuss possible complementarity with future laboratory experiments. 

The rest of this article is organized as follows. We first introduce the minimal DS models of interest and demonstrate the conditions for realizing a strong first-order PT. Then, we estimate the SGWB from the benchmark models and analyze how well they fit the PTA data. Following that, we discuss the implications for the DS cosmology based on the properties of the DS model as required by fitting PTA observations. Finally, we conclude.  

\section{Dark Sector Phase Transitions}
\label{sec:DSPT}
We start by systematically laying out the minimal model setup that enables us to perform a perturbative analysis and demonstrate the key characteristics of models that generate GWs consistent with the PTA observations. A PT typically originates from the spontaneous breaking of a symmetry, which can be triggered by scalar degrees of freedom acquiring vacuum expectation values (VEVs). To realize a strong first-order PT enabling detectable GW signals, additional bosonic degrees of freedom interacting with this scalar are typically required.  We therefore consider a DS with a gauged dark U$(1)$ symmetry, spontaneously broken when a complex scalar $\Phi$ acquires a VEV. To allow for a Yukawa coupling $y_D$, we introduce a dark U$(1)$ charged fermion $(\Psi)$ and a dark U$(1)$ singlet fermion $(\chi)$. All these fields are singlets under the SM gauge group. 

The Lagrangian relevant to the PT is then of the form 
\begin{align}
    \mathcal{L}_{\rm{DSPT}} \supset (D^{\mu}\Phi)^{\dagger}(D_{\mu}\Phi) + \mu^2\,\Phi^{\dagger}\Phi - \frac{\lambda}{2}(\Phi^{\dagger}\Phi)^2 - y_D\,\bar{\Psi} \Phi \chi \,,
    \label{eq:ds_lagrangian}
\end{align}
where the covariant derivative $D_{\mu}\Phi = \partial_{\mu}\Phi +i g_D A'_{\mu}\Phi$ couples the scalar to the dark gauge boson $A'_{\mu}$ with coupling strength $g_D$. We take $\mu^2 > 0$ and $\lambda > 0$ to ensure the stability of the potential and allow $\Phi$ to acquire a nonzero VEV $v_0$,
\begin{align}
    \langle\Phi\rangle = \frac{v_0 + \phi}{\sqrt{2}}\,.
    \label{eq:VEV}
\end{align}
Here, $\phi$ is the physical, dark Higgs, and $v_0 = \sqrt{\mu^2/\lambda}$ is the VEV at $T=0$, and we adopt unitary gauge to gauge away the Goldstone boson. 

For generality, we also consider larger gauge groups such as SU$(N)$, by promoting $\Phi$ and $\Psi$ to multiplets in representations of SU$(N)$. However, in order to eliminate relic massless degrees of freedom after the PT, which can be subject to strong constraints from Big Bang Nucleosynthesis (BBN) and the Cosmic Microwave Background (CMB) observations, in particular, in the form of the effective number of new relativistic neutrino degrees of freedom $\Delta N_{\rm{eff}}$ \cite{Kawasaki:2004qu, Hannestad:2004px,Hasegawa:2019jsa,Planck:2018vyg}, we assume maximal breaking of SU$(N)$. This ensures the $N^2-1$ gauge bosons acquire equal masses from the PT. For SU$(2)$, we take $\Phi$ in the fundamental representation, with the VEV configuration $\langle\Phi\rangle = [0,\, (v_0 + \phi)/\sqrt{2}]^{T}$,
which maximally breaks SU$(2)$. For larger $N$, the VEV configuration needs to be specifically tailored: e.g., for SU$(3)$, a possibility is that $\Phi$ is in the adjoint representation, with unequal VEVs along the diagonal. 

Because a fully secluded dark sector could conflict with the $\Delta N_{\rm eff}$ constraints discussed above, we assume that the DS was initially in thermal contact with the SM radiation bath, with portal interactions enabling efficient entropy transfer to the visible sector prior to or during BBN. The DS PT dynamics can be modified by SM fields coupled to $\Phi$, as discussed in \ref{app:rev_FOPT}; we address the cosmological implications of such portal couplings in Sec.~\ref{sec:DS_cosmo}. For the range of portal couplings considered here, however, these effects are parametrically suppressed relative to the intrinsic DS interactions. As a result, the leading features of the PT and the resulting GW signal are governed primarily by the DS dynamics, which is the core focus of this work.

Based on the DS models presented, we can then calculate the phenomenological parameters that characterize the PT: $T_*$, the temperature at which the PT occurs; $\beta/H_{\ast}$, the inverse time duration of the PT; and $\alpha_{\ast}$, the latent heat released in the PT, normalized to the total radiation density of the universe at the time of the PT. These quantities are derived from the bubble nucleation rate of the true vacuum, given by \cite{Coleman:1977py,Callan:1977pt, Linde:1980tt, Linde:1981zj}
\begin{align}
    \Gamma(T) &= A(T)\left(\frac{S_3}{2\pi T}\right)^{\frac{3}{2}}\exp\left(-\frac{S_3}{T}\right)
    \label{eq:Gamma}
\end{align}
where $S_3$ is the O(3)-symmetric Euclidean action for the ``bounce'' solution
\begin{align}
    &S_3 = 4\pi\int_0^\infty dr \, r^2\left[\frac{1}{2}\left(\frac{d\varphi_b}{dr}\right)^2+V_{\rm{1-L}}(\varphi_b,T)\right]  \\
    &\text{with}\quad  \frac{d^2 \varphi_b}{dr^2} + \frac{2}{r}\frac{d\varphi_b}{dr} = V'_{\rm{1-L}}(\varphi_b, T)\,,
\end{align}
subject to the boundary conditions $\varphi_b (\infty) = 0$ and $\partial_r\varphi_b (0) = 0$.  Here, in the computation of $S_3$, which we obtain through the \verb|Mathematica| package \verb|FindBounce|~\cite{Guada:2020xnz}, we require the one-loop effective potential at finite-temperature $V_{\rm 1-L}$, which encodes the model dependence. The precise details of our calculation of $V_{\rm 1-L}$ are provided in Appendix~\ref{app:rev_FOPT}. Finally, for the pre-factor in \eqref{eq:Gamma}, we approximate\footnote{We refer readers to Refs.~\cite{Croon:2020cgk,Gould:2021ccf,Gould:2021dzl} for detailed discussions on the calculation of this pre-factor.} $A(T) \approx T^4$, based on dimensional grounds~\cite{Linde:1980tt,Linde:1981zj}. 

\begin{figure}[t!]
\centering
\includegraphics[width=0.43\textwidth]{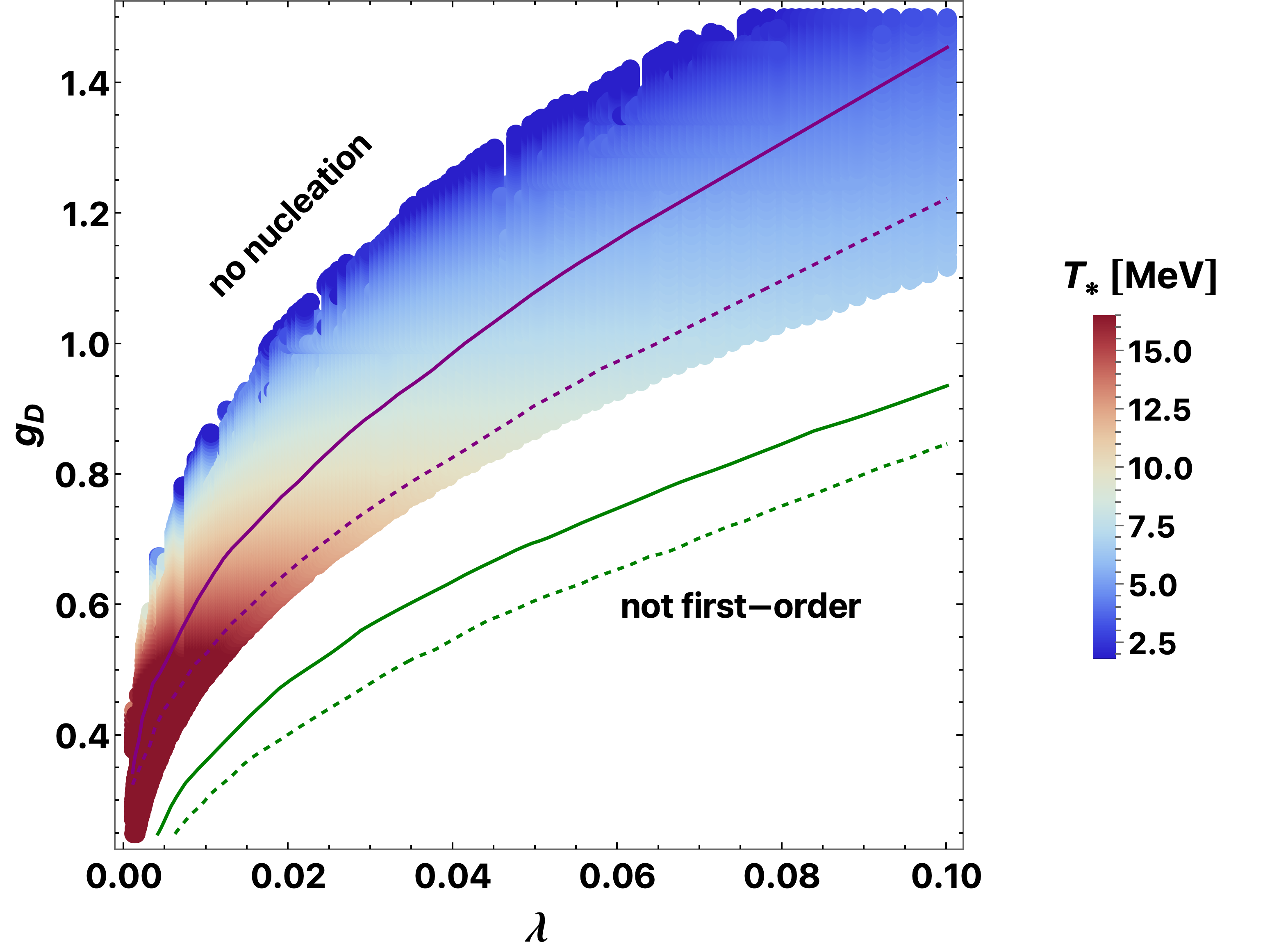}\\[0.5mm]
\includegraphics[width=0.42\textwidth]{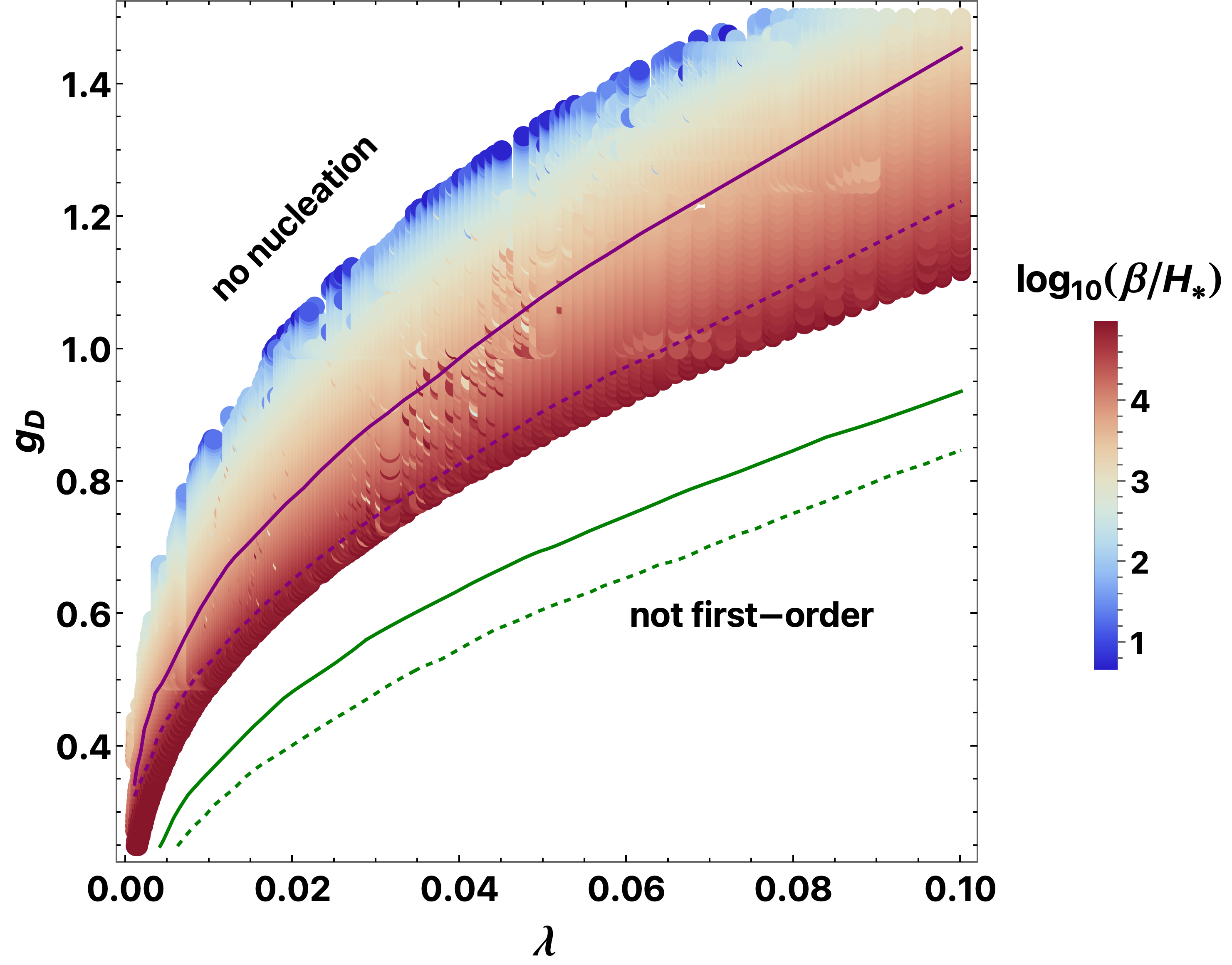}\\[0.5mm]
\includegraphics[width=0.43\textwidth]{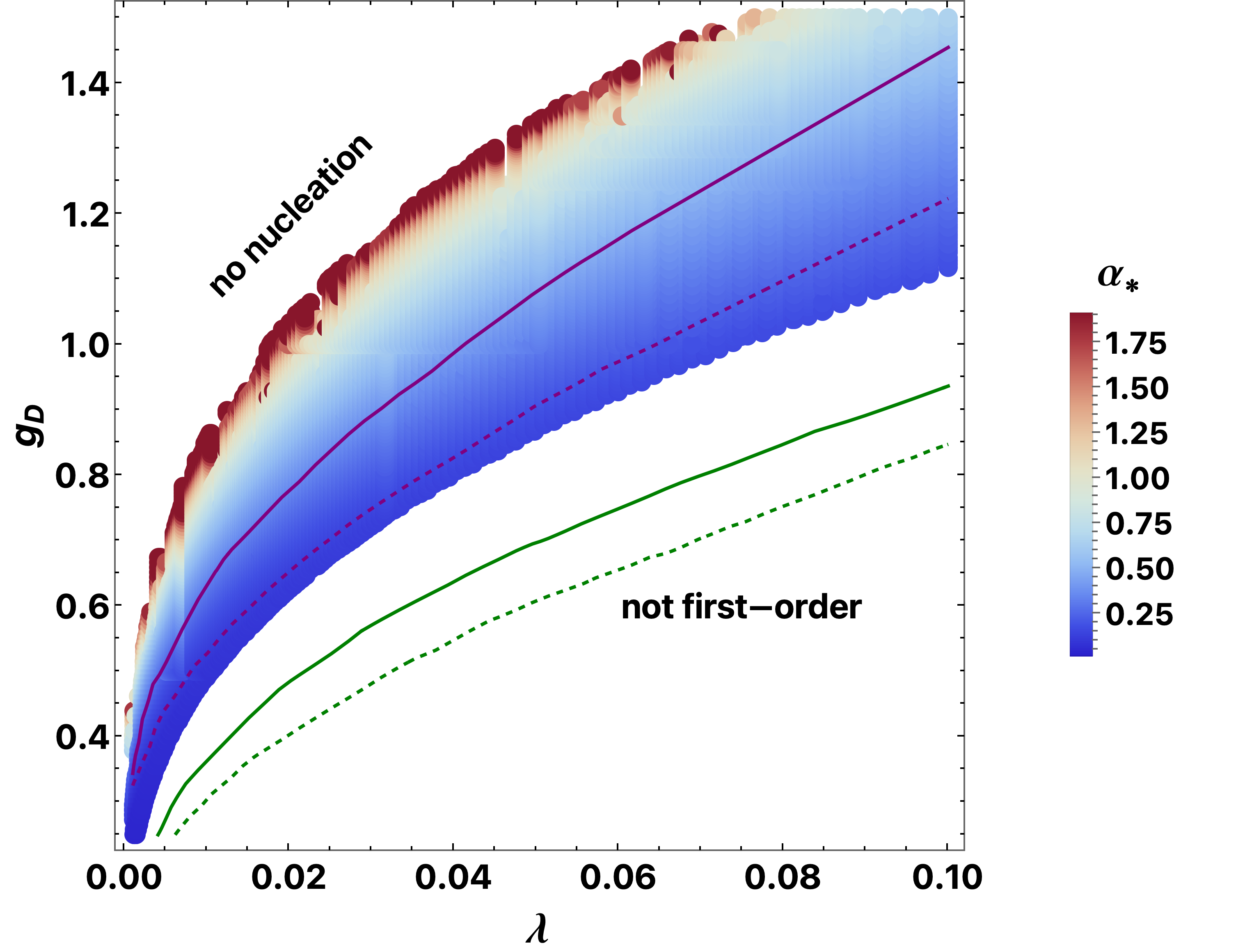}
\caption{Density plots of PT parameters $T_*$, $\beta/H_{\ast}$, $\alpha_{\ast}$ in the $(\lambda,\,g_D)$ plane for $y_D = 0$ and fixed $\mu = 2$~MeV. Colored regions illustrate the case with the U$(1)$ model, with boundaries of no bubble nucleation and non-first-order PTs marked by solid black and dashed-gray lines,  respectively. Similar features are obtained for SU$(2)$ and SU$(3)$ where we only show the boundaries (without colored shades) to avoid over-cluttering the figure: purple lines show the boundary for SU$(2)$ (solid) and SU$(3)$ (dashed) above which the nucleation criterion is not met, and dark green lines give the boundary above which the PT remains first-order for SU$(2)$ (solid) and SU$(3)$ (dashed).}
\label{fig:PT_gauge}
\end{figure}

The bubble nucleation temperature $T_n$ is then determined by comparing the rate in \eqref{eq:Gamma} to the Hubble time, 
\begin{equation}
    \Gamma(T_n) \sim H^4(T_n)\,,\quad \text{where} \quad H^2(T) = \frac{8\pi G_N}{3}\,\rho_R(T)\,,
    \label{eq:nucl_cond}
\end{equation} 
where the radiation density is given by
\begin{align}
    \rho_R(T) &= \frac{\pi^2}{30}(g_{*,\rm SM} + g_{*,\rm DS})\, T^4 
         \equiv \frac{\pi^2}{30}g_{*,\rm tot}\,T^4
\end{align}
and we have assumed a DS thermalized with the SM. Here, $g_*$ refers to the relativistic degrees of freedom. As we have considered phase transitions that occur during radiation domination and do not encounter significant supercooling~\cite{Ellis:2018mja}, in the parameter space scanned, we approximate the nucleation temperature as the temperature of the PT, that is, $T_* = T_n$. The PT inverse duration and strength are then given by
\begin{align}
    &\frac{\beta}{H_{\star}} \equiv T_* \,\left[\frac{d}{dT}\left(\frac{S_3}{T}\right)\right]\bigg|_{T_*}\,,\\
    &\alpha_{\ast} \equiv \frac{1}{\rho_R(T_*)}\left(\Delta V\bigg|_{T*} - \frac{T_*}{4}\,\frac{\partial \Delta V}{\partial T}\bigg|_{T_*}\right)\,,
\end{align}
where $\Delta V$ is the potential difference between the true and false vacua. 

The GW spectrum depends on these PT parameters, which in turn depend on the underlying model parameters such as $\lambda$ and $g_D$. Therefore, in order to facilitate our discussion in the context of the PTA data, we systematically analyze how the phenomenological PT parameters depend on the parameters in the model's Lagrangian.

To begin, we consider the U$(1)$ model with $y_D = 0$. The dimensionful parameter $\mu$, which sets the symmetry-breaking scale through the VEV in \eqref{eq:VEV}, in turn determines the scale of the PT temperature $T_*$ as demonstrated in Fig.~\ref{fig:model_PTparam1} of Appendix~\ref{app:PT_params}. Given that the peak frequency of the GW spectrum is proportional to $T_*$, and that a PT in the MeV range typically results in a peak frequency of GW in the PTA range \cite{Breitbach:2018ddu, Bai:2021ibt, Bringmann:2023opz}, we set $\mu = 2$ MeV as the benchmark for our discussion. The precise value of $T_*$ is subsequently determined by the couplings, as seen in the top plot of Fig.~\ref{fig:PT_gauge}. We then study the dependence of the remaining PT parameters, $\beta/H_*$ and $\alpha_*$, on the bosonic couplings $\lambda$ and $g_D$, which we show in the middle and bottom plots of Fig.~\ref{fig:PT_gauge}. We observe that these PT parameters vary monotonically across the parameter space. As the couplings approach the boundary above which bubble nucleation ceases (in the upper-left region), $\alpha_*$ increases and $\beta/H_*$ decreases. This corresponds to the strongest and longest PTs characterized by $\alpha_* \sim 2$ and $\beta/H_* \sim 10$. Conversely, moving towards the lower-right region where the transition is no longer first-order, PTs become weaker and shorter, indicated by smaller $\alpha_*$ and larger $\beta/H_*$. These trends are consistent with observations in similar models, e.g., Refs.~\cite{Baldes:2018emh,Madge:2018gfl}. 

\begin{figure}[t!]
\centering
\includegraphics[width=0.44\textwidth]{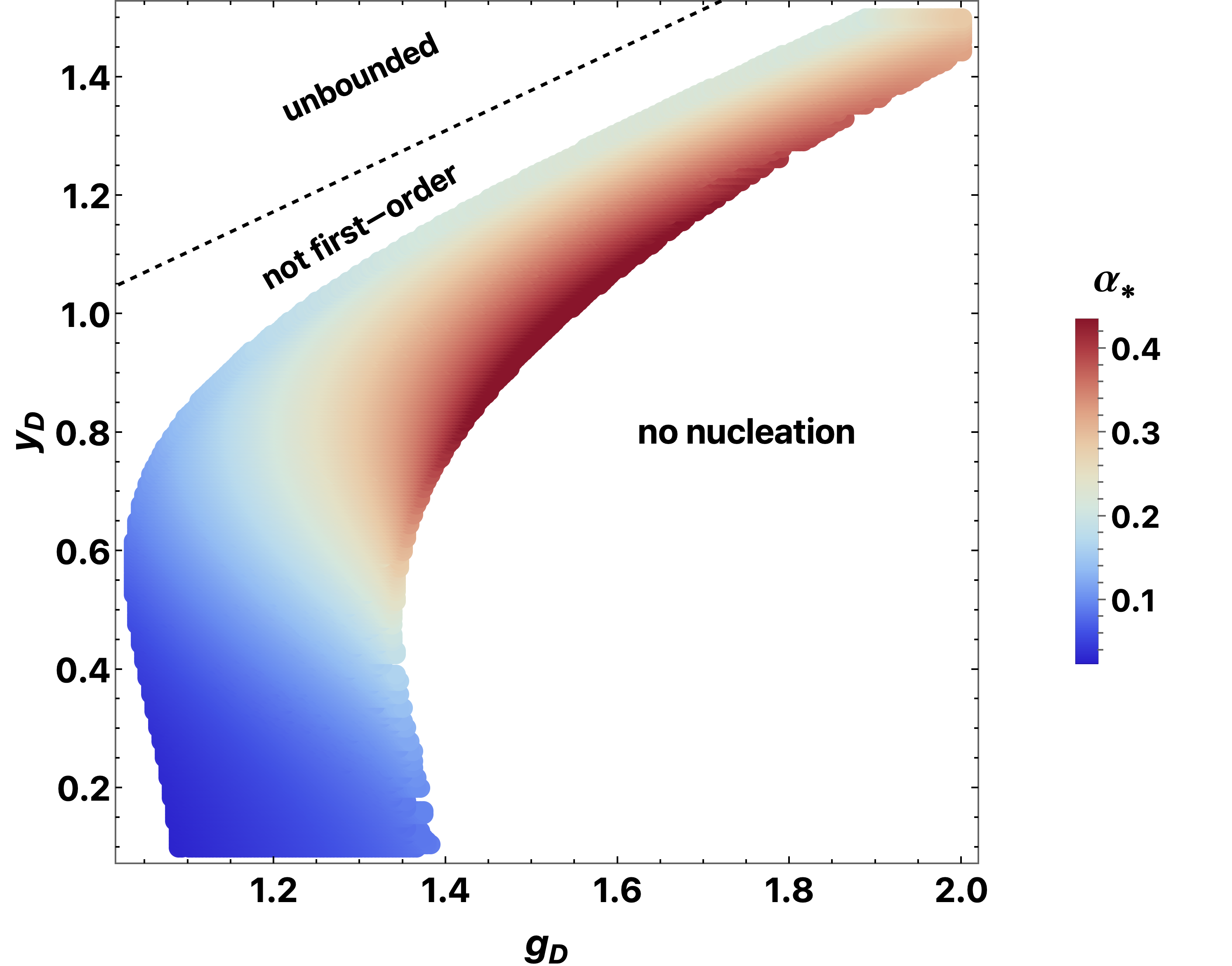}\\[1mm]
\includegraphics[width=0.44\textwidth]{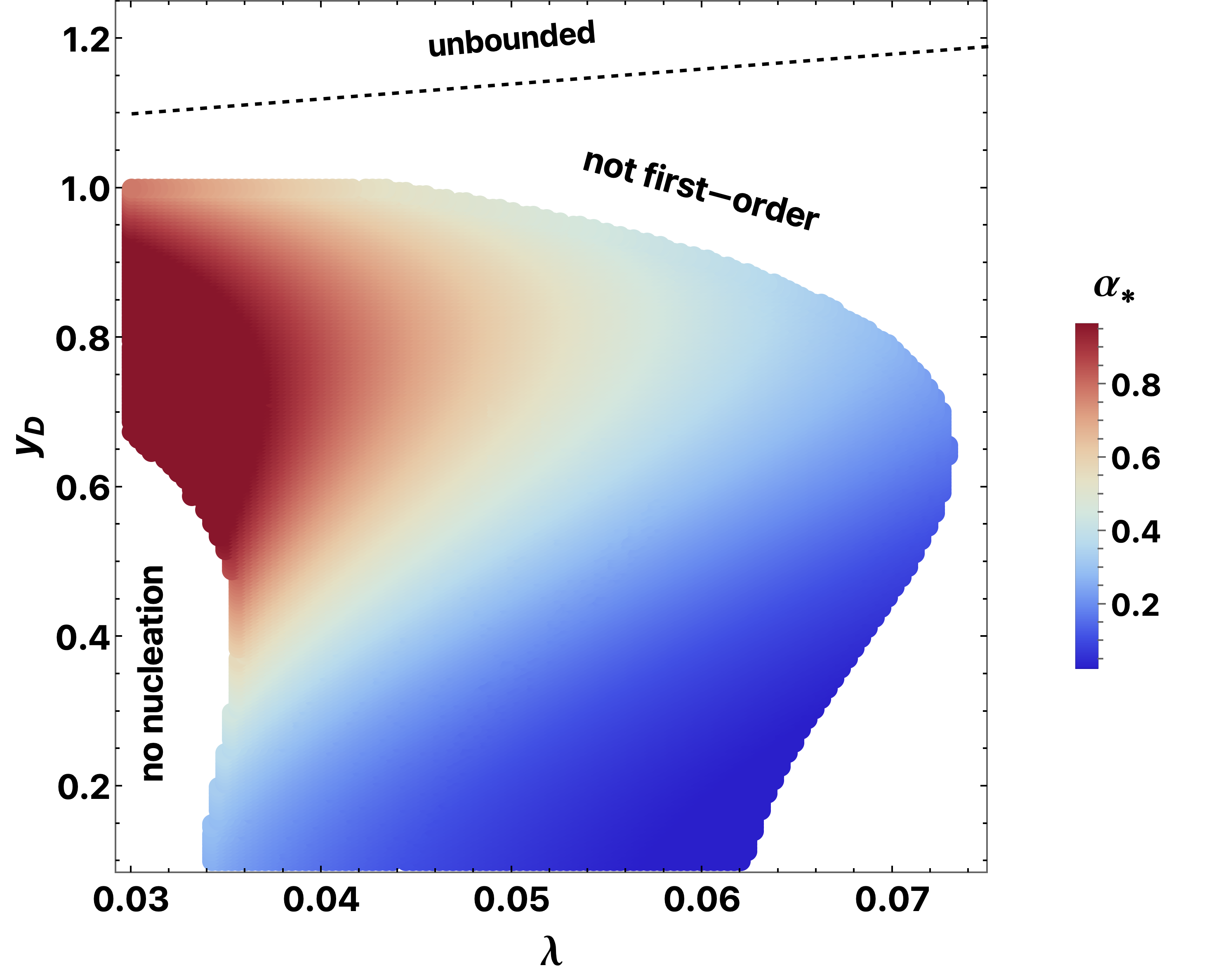}
\caption{Density plots of $\alpha_*$ for the U(1) model with fixed $\mu = 2$~MeV. \textit{Top:} The dependence of $\alpha_*$ in the $(g_D,y_D)$ plane with fixed $\lambda = 0.05$. \textit{Bottom:} In the $(\lambda_D,y_D)$ plane with fixed $g_D = 1.2$.}
\label{fig:PT_ferm}
\end{figure}

While further details on the exact nature of these trends are provided in Appendix~\ref{app:rev_FOPT} and Appendix~\ref{app:PT_params}, a qualitative understanding can be gained as follows by increasing the gauge coupling while fixing $\lambda$, a potential barrier forms which requires more energy to tunnel through to the true vacuum; and beyond a certain value of $g_D$ the scalar field can no longer tunnel to the true minimum due to the barrier. For larger symmetry groups such as SU$(2)$ and SU$(3)$, the aforementioned trends for $\alpha_*$ and $\beta/H_*$ observed in the U$(1)$ model persist qualitatively. However, the parameter space ``tilts": for the same value of $\lambda$, a smaller $g_D$ is required, as indicated by the colored boundaries in Fig. \ref{fig:PT_gauge}. This is due to the increase in the number of massive gauge bosons that appear at the loop level, thereby modifying the effective potential.

Finally, we discuss the effects of $y_D \neq 0$ on the existing parameter space. Note that the fermionic contribution to the effective potential carries a negative sign, in contrast to the positive contribution from bosons. Therefore, a certain target of the potential's form may be maintained by tuning the bosonic couplings to compensate for the fermionic contribution, thereby opening up new regions of parameter space that yield comparable values of the PT parameters. This effect is evident in Fig.~\ref{fig:PT_ferm}, which gives the dependence of $\alpha_*$ of $y_D$, $g_D$, and $\lambda$ in their respective planes. For instance, a shift in the ``no nucleation'' regions is observed:  in the top plot of Fig.~\ref{fig:PT_ferm}, for fixed $\lambda = 0.05$, the boundary is located at $(g_D, y_D) =(1.5,1.0)$, whereas in the bottom plot of Fig.~\ref{fig:PT_gauge} (where $y_D = 0$), it is at $g_D = 1.25$. Additionally, the bottom plot of Fig.~\ref{fig:PT_ferm} demonstrates that an allowed region for $\lambda=0.03$ emerges at $y_D = 0.7$ for fixed $g_D = 1.2$. For completeness, we show the other PT parameters in Fig.~\ref{fig:model_PTparam2} of Appendix~\ref{app:PT_params}. Nevertheless, for a fixed $g_D$ or $\lambda$, as $y_D$ increases further, the negative contribution from the fermions may result in the PT becoming a cross-over, and beyond a certain value, the effective potential becomes unbounded.

In summary, the PT parameters are mainly affected by bosonic couplings to the scalar driving the PT. We numerically scanned for $\sim 25000$ points in parameter space, within the perturbative ranges of $ \lambda \in [10^{-3},\, 10^{-1}]$, $g_D \in [0.2,\,2.0]$ and $y_D \in [0,\,1.5]$ for the gauge groups U(1), SU(2) and SU(3). As we shall see, combinations of the couplings that result in a large $\alpha_*$ and a small $\beta/H_*$ are preferred to achieve the sufficient amplitude of GWs as observed by PTAs.

\section{GW Signal for PTA}
\label{sec:GW+PTA}

A first-order PT proceeds through the nucleation of bubbles of the true phase, from which GWs can be generated from their collisions (bc) \cite{Kosowsky:1992rz, Kosowsky:1992vn, Huber:2008hg, Weir:2016tov}, as well as through sound waves (sw) \cite{Hindmarsh:2013xza, Hindmarsh:2015qta, Hindmarsh:2017gnf} and magneto-hydrodynamic turbulence (turb) \cite{Kosowsky:2001xp, Dolgov:2002ra, Caprini:2009yp} induced by the interaction of the bubbles with the surrounding plasma,
\begin{align}
    \Omega_{\rm{GW}}h^2(f) = \Omega^{\rm{bc}}_{\rm{GW}}h^2(f) + \Omega^{\rm{sw}}_{\rm{GW}}h^2(f) + \Omega^{\rm{turb}}_{\rm{GW}}h^2(f)\,,
    \label{eq:omega_gw}
\end{align}
where $\Omega_{\rm GW}$ is today's GW energy density (per logarithmic momentum interval) over the critical energy density.

The relative importance of the different contributions depends on the microphysics of the first-order PT. For example, if the scalar field has strong interactions with the surrounding plasma, the sound wave and turbulence contributions tend to be larger, which is our case for a strong first-order PT. In Appendix~\ref{app:gw_spec}, we review the analytical expressions used to obtain the GW spectra for our analysis.

The GW spectrum depends on the PT parameters in the following manner: $T_*$ determines the peak location, while $\alpha_*$ and $\beta/H_*$ determine the amplitude and shape of the spectrum. These, in turn, depend on the Lagrangian parameters, as discussed in the previous section, translating the dependence of the GW spectrum down to the model parameters: $\{\mu,\,\lambda,\,g_D,\,y_D,\,N\}$. Consequently, we study the model parameter space that is consistent with current PTA observations. For this purpose, we make use of the publicly available code \verb|PTArcade|~\cite{Mitridate:2023oar} to implement the PTA likelihood using the NANOGrav 15-year dataset~\cite{NANOGrav:2023gor}.

\begin{figure}[t!]
\centering
\includegraphics[width=0.45\textwidth]{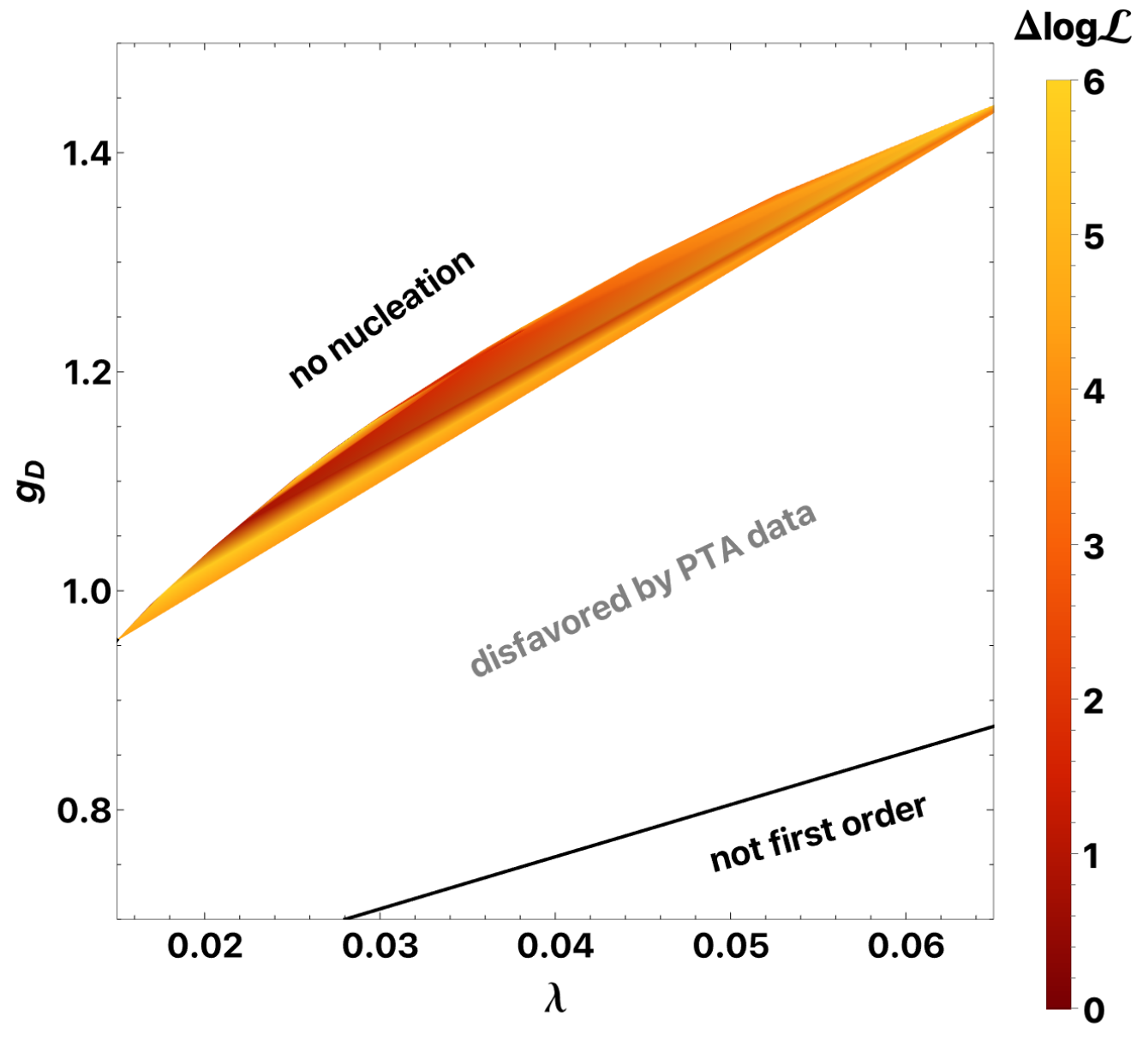}\\
\includegraphics[width=0.45\textwidth]{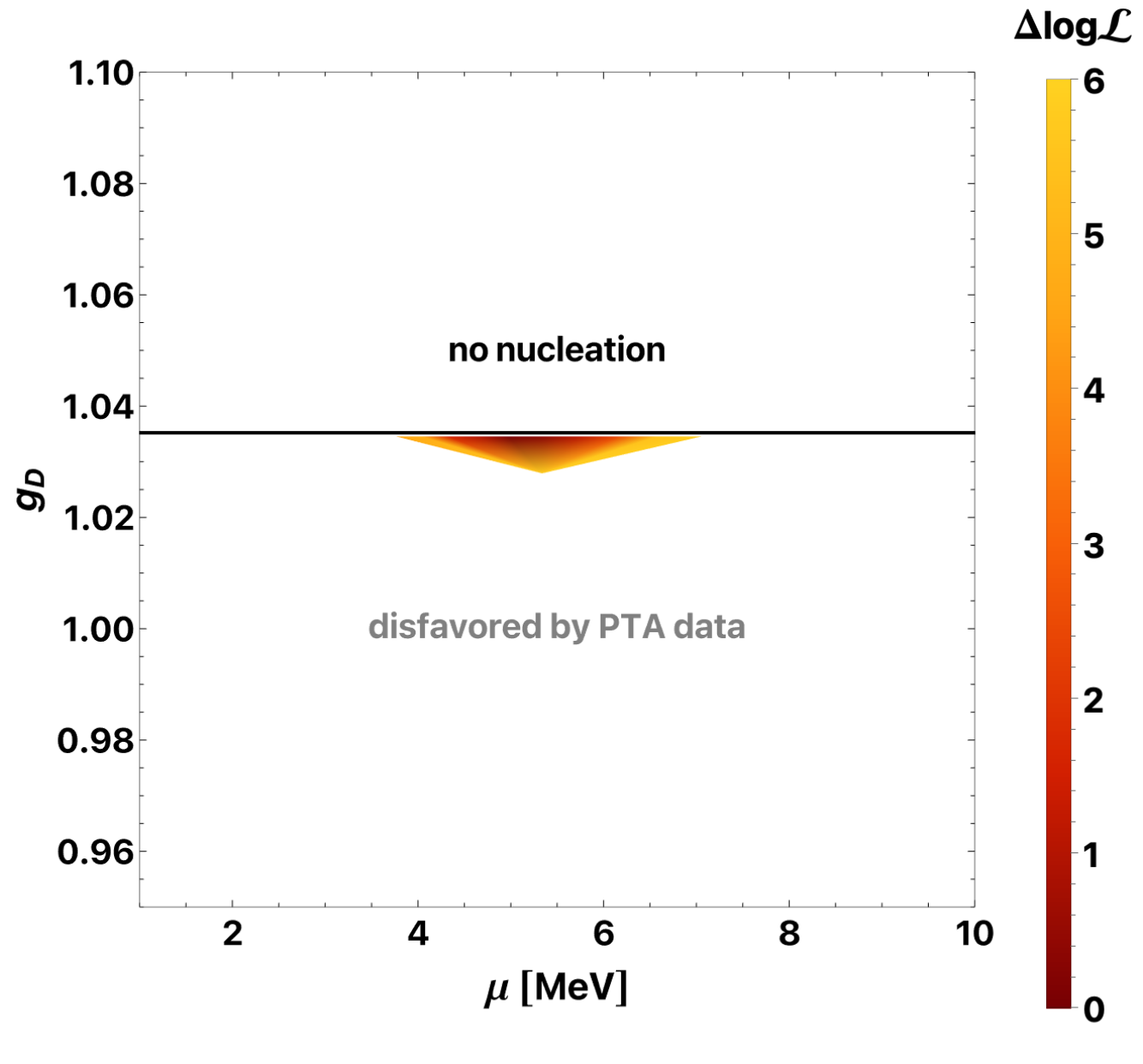}
\caption{Colored regions corresponding to $\Delta\log{\cal L}$ values for up to a $3\sigma$ fit, obtained from the PTArcade analysis of the NANOGrav 15-year dataset, for the U$(1)$ model with $y_D = 0$.
\textit{Top:} The coupling combination $(\lambda, g_D)$ preferred, with fixed $\mu = 4$~MeV as an example, to identify the optimal transition strength and duration, for the overall GW amplitude. 
\textit{Bottom:} The combination of $(\mu, g_D)$ with fixed $\lambda = 0.02$ as an example, to identify the optimal transition temperature for the GW peak frequency.
}
\label{fig:likelihood}
\end{figure}

The GW spectrum given by \eqref{eq:omega_gw} depends on the PT parameters $\{T_*, \alpha_*,\beta/H_*\}$, as described in Appendix~\ref{app:gw_spec}. We thus implement this spectrum into \verb|PTArcade|, with the following priors on the PT parameters:
\begin{equation}
    T_* \in [1, 100]\,{\rm MeV}\,,\quad 
    \alpha_* \in [10^{-3}, 2]\,,\quad
    \beta/H_* \in [10, 10^6]\,.
\end{equation}
These priors, in particular the ones on $\alpha_*$ and $\beta/H_*$, are motivated by the results of our parameter scans as discussed in Sec.~\ref{sec:DSPT}. Based on the sampled likelihood, we then determine the point that reports the maximum logarithmic likelihood $\log {\cal L}_{\rm max}$, which corresponds to the PT parameters that best-fit the observed PTA dataset. Now, using the results of our parameter scans, we compute:
\begin{align}
    \Delta \log {\cal L}_i = \log {\cal L}_{\rm max} - \log {\cal L}(x_i) \,,
    \label{eq:DeltaLogL}
\end{align}
where $\log {\cal L}(x_i)$ refers to the log-likelihood of the PT parameters corresponding to the $i^{\rm th}$ Lagrangian parameter set for the DSPT model under consideration. Physically, \eqref{eq:DeltaLogL} quantifies how disfavored a given point is relative to the best-fit point, such that smaller values of $\Delta \log {\cal L}$ indicate a better fit to the PTA measurements. Although the PTA likelihood is, in general, non-Gaussian and may exhibit degeneracies~\cite{NANOGrav:2023hvm}, we assume that in the vicinity of the best-fit point, the likelihood is sufficiently smooth that we approximate the distribution of the likelihood-ratio statistic using Wilks’ theorem~\cite{Wilks:1938dza}. Under this assumption, we interpret $\Delta \log {\cal L} \lsim 1.77$ as corresponding to the $1\sigma$ region, while $\Delta \log {\cal L} \lsim 4.00 \,(6.17)$ corresponds to the $2\sigma \,(3\sigma)$ regions. Across the models we have studied in this work, for fitting at the $1\sigma$ level, we found combinations of model parameters resulting in $\beta/H_* \sim 40$, $\alpha_* \sim 0.9$, and $T_*\sim 7$~MeV were preferred to achieve a fit at the $1\sigma$ level.

\begin{figure}[t!]
\centering
\includegraphics[width=0.48\textwidth]{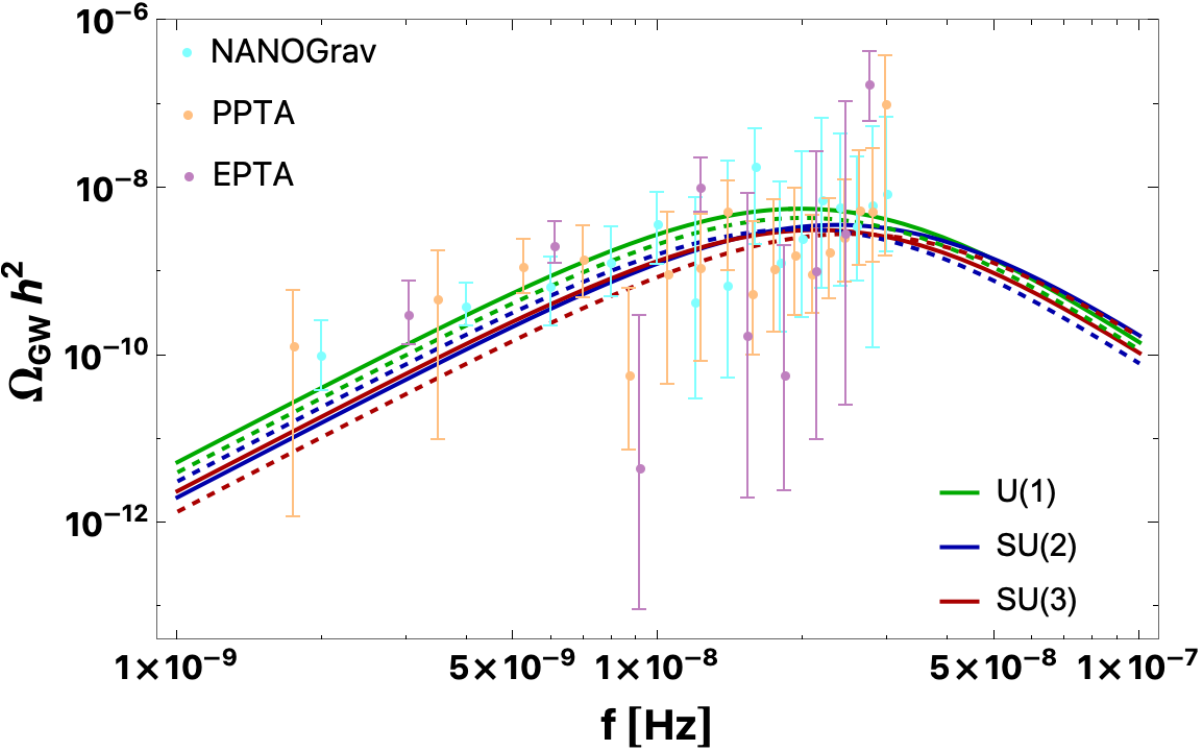}
\caption{The best-fit GW spectra to the NANOGrav 15-year dataset based on our PTArcade analysis, for the DS PTs we consider in this work. We also superimpose the representative error bars, reconstructed in Ref.~\cite{Winkler:2024olr}, for the NANOGrav, PPTA, and EPTA datasets. Colored lines indicate the gauge group, whereas dashed lines indicate that the corresponding gauge sector includes fermions (i.e., $y_D \neq 0$). We find the contribution arising from acoustic GWs, $\Omega_{\rm GW}^{\rm sw}$, to be dominant for these best-fit spectra.
}
\label{fig:best_gw}
\end{figure}

As a concrete example, we demonstrate this for the dark U(1) model without fermions in Fig.~\ref{fig:likelihood}. In the top plot, with a fixed $\mu = 4$~MeV, the PTA dataset narrows the coupling combination near the ``no nucleation'' edge to achieve a large $\alpha_*$ and small $\beta/H_*$ to minimize \eqref{eq:DeltaLogL}. The narrow region is isolated as $T_* \propto v_0 = \sqrt{\mu^2/\lambda}$, as discussed earlier (see also the top plot of Fig.~\ref{fig:PT_gauge}), and the range of $\lambda$ is isolated to obtain $T_* \sim 6-7$~MeV. Additionally, by effectively fixing the PT strength and duration as in the bottom plot of Fig.~\ref{fig:likelihood} with fixed $\lambda = 0.02$, we observe that varying $\mu$ slightly can improve the fit due to the change in $T_*$ (see middle and bottom plots of Fig.~\ref{fig:PT_gauge}), thereby shifting the peak frequency of the spectrum. Through this process, we find the best-fit model parameters for U(1) without fermions are $(\mu,\lambda,g_D) = (6.8~{\rm MeV}, 0.04, 1.2)$.

\begin{table*}
\centering
\renewcommand{\arraystretch}{1.4}
\setlength{\tabcolsep}{7.5pt}
\begin{tabular}{|c|c|c|c|}
\hline
\textbf{MODEL} & $(\mu,\,\lambda,\,g_D,\,y_D)$ & $(T_*,\,\alpha_*,\,\beta/H_*)$ & \(\Delta\log{\cal L}\) \\ \hline
\multirow{2}{*}{U$(1)$} & (6.8 MeV, 0.04, 1.2, 0)  & (6.8 MeV, 0.96, 45) &0.27 \\ \cline{2-4}
                      &(7.9 MeV, 0.06, 1.5, 0.8)  &(7.7 MeV, 0.82, 39)  & 0.36 \\ \hline
\multirow{2}{*}{SU$(2)$} & (6.8 MeV, 0.05, 1.0, 0)  &  (6.6 MeV, 0.92, 44) &0.25 \\ \cline{2-4}
                       &(7.6 MeV, 0.05, 1.2, 0.9)   & (7.7 MeV, 0.85, 40)  &0.28   \\ \hline
\multirow{2}{*}{SU$(3)$} & (7.0 MeV, 0.03, 0.7, 0)   & (6.9 MeV, 1.0, 42) &0.30 \\ \cline{2-4}
                       &(9.6 MeV, 0.06, 0.9, 0.9) &(9.4 MeV, 0.9, 34)  &0.35 \\ \hline
\end{tabular}
\caption{Approximate values of the best-fit model parameters corresponding to the GW spectra in Fig.~\ref{fig:best_gw}, with the calculated PT parameters in the third column. Given the value of $\Delta \log {\cal L} < 1.77$ for these points, these correspond to a fit at $1\sigma$ confidence level.}
\label{tab:best_gw}
\end{table*}

The PTA-preferred spectra found in this work are dominated by the sound wave contribution, with subdominant contributions from turbulence and bubble collisions. Given that the parameter region is confined to the region of very strong transitions to achieve the required amplitude, as indicated by Fig.~\ref{fig:likelihood}, we expect that uncertainties in the GW spectrum modeling would modify this region only modestly at the quantitative level and not alter our main conclusions. For example, an underestimation of the lifetime of the sound wave source would lead to an overestimation of the suppression factor $\Upsilon$ (see Eq.~\eqref{eq:upsilon} in Appendix~\ref{app:gw_spec}). Correspondingly, this would slightly expand the allowed parameter region in the top plot of Fig.~\ref{fig:likelihood}, permitting slightly weaker transitions with lower $\alpha_*$ and larger $\beta/H_*$. For example, the transitions featuring a slightly smaller $g_D$ for fixed $\lambda$ would now be allowed, cf. Fig.~\ref{fig:PT_gauge}. Similarly, overestimating the efficiency factors for the conversion of PT energy released into GW energy (see Eq.~\eqref{eq:kappasw} in Appendix~\ref{app:gw_spec}) would shrink this region, preferring now only the strongest of transitions, such as those with values of $g_D$ closest to the ``no nucleation'' edge. Finally, we comment on uncertainties in the modeling of the turbulence contribution. Given that $\Omega_{\rm GW}^{\rm turb}$ affects the spectra at higher frequencies (cf. Appendix~\ref{app:gw_spec} and Refs.~\cite{Caprini:2018mtu, Caprini:2019egz}), we expect that this would have minimal impact on our results, due to preference of fitting the sound wave peak, as indicated by Fig.~\ref{fig:best_gw}.

According to Fig.~\ref{fig:likelihood}, once the phenomenological parameters $\alpha_*$ and $\beta/H_*$ are pinned down by the PTA data, the underlying model parameters isolate distinct regions of parameter space, a common feature for all the models we considered in this work. In Fig.~\ref{fig:best_gw}, we show the GW spectra that fit best to the PTA data for various models considered in this work, with the corresponding model parameters provided in Table~\ref{tab:best_gw}. Finally, in Appendix~\ref{app:best_points}, we present an auxiliary analysis in which the PTA likelihood is approximated as Gaussian. Following Ref.~\cite{Winkler:2024olr}, we construct a $\chi^2$ statistic using the mean values and extracted error bars from the published PTA spectra of NANOGrav~\cite{NANOGrav:2023gor}, EPTA\footnote{In this auxiliary analysis, we adopt the 10.3-year EPTA dataset~\cite{EPTA:2023fyk} rather than the full 24.7-year dataset~\cite{EPTA:2023xxk}, following Ref.~\cite{Winkler:2024olr}, due to the latter having been reported to contain residual white-noise contamination.}~\cite{EPTA:2023fyk}, and PPTA~\cite{Reardon:2023gzh}. This treatment leads to the identification of a broadly similar region for the couplings as in the top plot of Fig.~\ref{fig:likelihood}, with a mild shift to lower values for $\mu$, reflecting differences in how the peak frequency is weighted.

We note that the SGWB signal observed by PTA collaborations may receive contributions from multiple sources; in particular, an astrophysical background from the inspiral of a population of supermassive black hole binaries (SMBHBs) could, in principle, coexist with the cosmological source described here, or itself fully explain the PTA signal. The canonical SGWB from this astrophysical source is expected to follow an approximately power-law spectrum, $\Omega_{\rm GW}^{\rm SMBHB}\propto f^{2/3}$, across the PTA frequency band~\cite{NANOGrav:2023hfp}. In contrast, the spectra arising from PTs exhibit a characteristic peaked structure, with the peak frequency determined primarily by the PT temperature $T_*$ and duration $(\beta/H_*)^{-1}$. While current PTA observations do not yet discriminate conclusively between these possibilities, future measurements with improved spectral sensitivity may help distinguish between a simple power-law background and a PT-induced signal. Finally, a dedicated analysis of mixed scenarios would require a joint fit of the astrophysical and cosmological components, which we defer to future studies.

\section{Implications for the Dark Sector Phenomenology}
\label{sec:DS_cosmo}

With the theoretical prior that the recently observed PTA signal originates from the DS models we consider, the PTA measurements provide precise conditions on the DS parameters for a given model, leading to specific implications for DS phenomenology. We will focus on the U$(1)$ model as an example for illustration. First, we identify the mass spectrum of the new particles: the dark Higgs $\phi$ has a physical mass of $\sqrt{2\mu^2}\sim 10$~MeV, the gauge boson $A^{'}_\mu$ acquires a mass of $\sim 20$–$35~\mathrm{MeV}$, while the mass of the fermion $\Psi$ ranges from $\sim 15$–$25~\mathrm{MeV}$ for $y_D \neq 0$. Therefore, in the absence of non-gravitational interactions with the SM, $\phi$ is the lightest stable particle in the DS.

The $\mathcal{O}(10)$~MeV scale masses and the large couplings $(\lambda, g_D, y_D)$ ensure that the $2-2$ scattering and annihilation rates between DS particles well exceed the Hubble expansion rate at $T_* \sim \text{MeV:}\, H(T_*)\sim T_*^2/M_{\rm pl}\sim 10^{-22}~\text{MeV}$ ($M_{\rm pl}= 1.22 \cdot 10^{19}$~GeV is the Planck mass). Even the number-changing process $3\phi \to 2\phi$, facilitated by the $\lambda\phi^4$ and $\lambda v_0\phi^3$ couplings after the U$(1)$ breaking, has a rate $n_{\phi}^2\sigma_{3\phi\to 2\phi}\sim 10^{-12}(\lambda/10^{-2})^4~$MeV, which is highly efficient. This ensures that the DS particles remain in both kinetic and chemical equilibrium, and most of the DS energy density after the PT is stored in the lightest particle $\phi$. 

As we have seen, in order to produce sufficient GW as observed by PTAs, the energy in the DS during PT shall be comparable to that of the SM, as indicated by $\alpha_*\sim 0.9$ in our best-fit results. After the PT, this latent heat is generally released through the production of DS particles. If such a large DS energy remains in the DS until later times, the model is subject to strong constraints from BBN~\cite{Nakai:2020oit,Bai:2021ibt} and CMB~\cite{Planck:2018vyg} observations (by modifying the Hubble expansion rate), which allow the DS energy density to be at most $\sim 10\%$ of the SM photon energy (often expressed in terms of $\Delta N_{\rm eff}$). Therefore, the DS cannot be fully decoupled from the SM, and its energy must be efficiently released to the SM through certain portal interactions, after the PT and before the BBN era. 

As noted above, at late times the dark sector energy density is stored in the lightest particle, the dark Higgs $\phi$. To release this energy into the SM to be compatible with the BBN bound, the simplest realization is through a renormalizable gauge-invariant quartic interaction of the form $\lambda_{\Phi H} |\Phi|^2\,|H|^2$, where $H$ is the SM Higgs doublet. The presence of the coupling $\lambda_{\Phi H}$ induces mixing between the DS scalar $\phi$ and the SM Higgs. Diagonalizing the resultant mass matrix, the mixing angle $\theta$ is obtained from
\begin{equation}
    \sin 2\theta = \frac{2\lambda_{\Phi H}\,v_0\,v_{\rm EW}}{m^2_h-m^2_{\phi}} \approx \frac{\lambda_{\Phi H}}{\lambda}\left(\frac{v_{\rm EW}}{v_0}\right)\left(\frac{m_{\phi}}{m_h}\right)^2
    \label{eq:mix_angle}
\end{equation}
where we have assumed $m_\phi \ll m_h$, and used the tree-level relation $m_{\phi}^2 = \lambda v_0^2$. Using the observed values $v_{\rm EW} \approx 246$~GeV and $m_h \approx 125$~GeV~\cite{ParticleDataGroup:2024cfk},  together with $m_{\phi}, v_0 \sim {\cal O}(10~{\rm MeV})$ and $\lambda \sim 0.01$ from the PTA fit, we estimate $\sin 2 \theta \sim (\lambda_{\Phi H}/10^{-3})$. This mixing between the SM Higgs and the dark Higgs enables the decay channels $\phi \to e^+e^-$ and $\phi \to \gamma\gamma$ with rates $\Gamma \propto \sin^2\theta$~\cite{Winkler:2018qyg}.

In the presence of such a coupling, energy released by the dark sector near the PT temperature is injected into the SM, producing particles via the aforementioned electromagnetic decays. We find that $\alpha_* \sim 0.9$ and the temperature $T_* \sim 7$~MeV from the PTA fit, are consistent with cosmological constraints on PTs~\cite{Bai:2021ibt}. Ensuring consistency with BBN requires that the energy density in $\phi$ is transferred sufficiently rapidly to the SM plasma. For $m_{\phi} \approx 10$~MeV (from the PTA fit), the cosmological condition that $\phi$ decays by the BBN era requires $\Gamma^{-1} \lsim 1$~s~\cite{Depta:2020zbh}, implying $\sin\theta~\gsim~10^{-5}$. Meanwhile, existing laboratory experimental data~\cite{Winkler:2018qyg,Fradette:2018hhl,Goudzovski:2022vbt,Batell:2022dpx,Bringmann:2026xcx} impose constraints $\sin\theta~\lsim~10^{-4}$. This identifies a viable window of portal couplings for which the DS can efficiently transfer its energy density prior to or during BBN.

Consequently, the mixed quartic coupling is constrained to be within the following range:  $10^{-8}~\lsim~\lambda_{\Phi H}~\lsim~10^{-7}$. We then find $\lambda_{\Phi H} \ll \lambda,\lambda_H$, where $\lambda_H$ is the SM Higgs self-coupling. This shows that the portal coupling is sufficiently small that its impact on the DS phase transition dynamics -- through field-dependent and temperature-dependent masses in the effective potential (see Appendix~\ref{app:rev_FOPT}) -- is parametrically suppressed. We further note that after electroweak symmetry breaking, a correction to the $\mu$ parameter in \eqref{eq:ds_lagrangian} is induced, which can be absorbed through a redefinition $-\mu^2 \to -\mu^2+\lambda_{\Phi H}v^2_{\rm EW}/2$, so that the leading features of the DS PT remain effectively unchanged.

The Higgs portal coupling also leads to potential complementarity with laboratory searches. First, the mixing induced by $\lambda_{\Phi H}$ allows for exotic Higgs decays in the form of $h \to \phi\phi$, which can be probed with collider experiments such as the LHC. For the range of portal couplings identified above, the decay width $\phi \to e^+e^-$ is suppressed by the small mixing angle, leading to a long lifetime $\tau \sim {\cal O}(0.01{\rm s})$. Taking into account the large Lorentz boost from Higgs decays, $\gamma \approx m_h/(2m_\phi) \approx 10^4$, the resulting decay length is ${\cal O}(10^{10}~{\rm m})$, significantly exceeding the size of detectors. As a result, $\phi$ behaves effectively as a stable particle at colliders, and the process $h \to \phi\phi$ manifests as an invisible Higgs decay~\cite{CMS:2018yfx,CMS:2023sdw}. In addition, the relevant parameter space can be probed at intensity frontier experiments such as HIKE~\cite{HIKE:2022qra,HIKE:2023ext}. For $m_\phi \sim \mathcal{O}(10\,\mathrm{MeV})$, Higgs mixing induces rare meson decays $K \to \pi \phi$, leading to missing-energy signatures. The projected sensitivity of HIKE, reaching $\sin\theta \sim 4 \times 10^{-5}$~\cite{Batell:2022dpx}, probes part of the cosmologically viable region identified here.

\section{Conclusion}
\label{sec:conclude}

In this work, we systematically investigate the possibility of addressing the recent PTA observations with an SGWB sourced by phase transitions in dark sectors. We identify minimal DS models with perturbative dynamics that can explain the signal: involving a U$(1)$ or SU$(N)$ gauge symmetry, a Higgs-like scalar, with or without Yukawa coupling to dark fermions. With the details of the underlying model identified and assuming that the PTA signal is entirely explained by a DS PT, the PTA data favor specific conditions for the DS properties. In particular, a likelihood analysis to fit the predicted GW signal to the NANOGrav 15-year dataset narrows the parameter space to the region of strong bosonic couplings, with the PT scale being of $\mathcal{O}$(MeV). For example, in the U$(1)$ model, the SM singlet dark Higgs mass lies in the approximate $2\sigma$ window of $m_\phi \in [5,30]$~MeV, the U$(1)$ gauge coupling $g_D \in [0.85,1.3]$, and the dark Higgs self-coupling $\lambda \in [0.015, 0.07]$. This information allows us to further elucidate the DS phenomenology following the PT. 

We found that to produce a sufficient GW signal to address the PTA data, the DS energy density around the PT must be comparable to that of the SM sector. On the other hand, to be compatible with BBN constraints in such scenarios, an effective coupling to the SM is necessary to transfer the entropy from the DS to the SM. We discuss the cosmology of the DS particles and demonstrate a simple way to satisfy the BBN bound-- by transferring the DS energy into the visible sector through a Higgs portal. We further identified a viable parameter region and demonstrated that, in benchmark scenarios, the model can also be probed by future experiments at both the energy and intensity frontiers.

Finally, although the study was inspired by the recent PTA observations, it is desirable to extend it by examining the DS across various energy scales, which could produce detectable GWs across a broad frequency range, a direction that we plan to pursue in future work.

\section*{Acknowledgments}
We thank Anish Ghoshal, Felix Kahlhoefer, and Kimmo Kainulainen for useful discussions. AB is supported by the Basic Science Research
Program through the National Research Foundation of Korea (NRF) funded by the Ministry of Education under grant numbers [NRF-2021R1C1C1005076, NRF-2020R1I1A3068803, RS-2026-25484206].
YC is supported by the US Department of Energy under award number DE-SC0008541. 
YDT thanks the generous support from the LANL Director's Fellowship. This research is partially supported by LANL's Laboratory Directed Research and
Development (LDRD) program. This research was partly supported by grant NSF PHY-2309135 to the Kavli Institute for Theoretical Physics (KITP). YT is supported by grant NSF PHY-2412701. YT would
like to thank the Tom and Carolyn Marquez Chair Fund
for its generous support.
This work was partially performed at the Aspen Center for Physics, supported by National Science Foundation grant No.~PHY-2210452. 

\appendix

\section{Thermodynamics of a first-order PT}
\label{app:rev_FOPT}
We briefly review the key quantities to describe a first-order PT. We present the effective potential at finite temperature, which we compute at one-loop in this work. As is customary, one considers background field fluctuations $\varphi$ about the vacuum $v_0$ in \eqref{eq:VEV}. One then obtains the following scalar potential in terms of this background field
\begin{equation}
    V_0(\varphi) = -\frac{1}{2}\mu^2\varphi^2 + \frac{1}{4}\lambda\varphi^4\,,
    \label{eq:V0}
\end{equation}
with the tree-level VEV occurring at the minimum of this potential, $v^2_0 = \mu^2/\lambda $. The field-dependent masses for the dark Higgs, Goldstone bosons, gauge bosons, and fermions are
\begin{align}
    &m^2_\phi(\varphi) = 3\lambda \varphi^2 - \mu^2\,,\quad m^2_\eta(\varphi) = \lambda \varphi^2 - \mu^2\,, \nonumber \\ 
    &m^2_{A'}(\varphi) = g^2_D\varphi^2\,,\quad m^2_{\psi}(\varphi) = \frac{1}{2}y^2_D\varphi^2\,.
    \label{eq:masses_phi}
\end{align}

The finite-temperature effective potential at one-loop is given by 
\begin{equation}
    V_{\rm{1-L}} (\varphi,T) = V_{0}(\varphi) + V_{\rm{CW}}(\varphi) + V_T(\varphi,T)\,.
    \label{eq:V1L}
\end{equation}
We work in Landau gauge and employ the on-shell renormalization scheme, which preserves the tree-level relations for $v_0$ and $m_\phi(v_0)$. However, this needs to be done carefully as the Goldstone bosons are massless at $v_0$, so one needs to use the running of the dark Higgs mass.  The result for the Coleman-Weinberg potential is then \cite{Anderson:1991zb,Espinosa:1992kf, Delaunay:2007wb}
\begin{align}
    &V_{\rm{CW}}(\varphi) = \displaystyle \sum_{i\neq \eta} \frac{\delta_i \,n_i}{64\pi^2} \bigg\{m_i^4(\varphi)\left[\ln\left(\frac{m^2_i(\varphi)}{m^2_i(v_0)}\right)-\frac{3}{2}\right] \nonumber\\
    &+2m^2_i(\varphi)m^2_i(v_0)\bigg\} + \frac{n_\eta}{64\pi^2} m_\eta^4(\varphi)\left[\ln\left(\frac{m^2_\eta(\varphi)}{m^2_\phi(v_0)}\right)-\frac{3}{2}\right]\,,
    \label{eq:VCW}
\end{align}
where for bosons (fermions), $\delta_i = +1\,\,(= -1)$, and $n_i$ refers to the degrees of freedom of the $i^{\rm{th}}$ particle. Next, the finite temperature corrections are given by \cite{Dolan:1973qd,Quiros:1999jp,Laine:2016hma}
\begin{align}
    &V_T(\varphi,T) = \nonumber \\
    &\frac{T^4}{2\pi^2}\displaystyle \sum_i \delta_i\,n_i \int_0^{\infty} dx\, x^2\,\ln\left[1 - \delta_i\exp\left(-\sqrt{x^2+ \frac{m^2_i(\varphi)}{T^2}}\right)\right] \,.
    \label{eq:VT}
\end{align}
%
\begin{figure}[t!]
\centering
\includegraphics[width=0.482\textwidth]{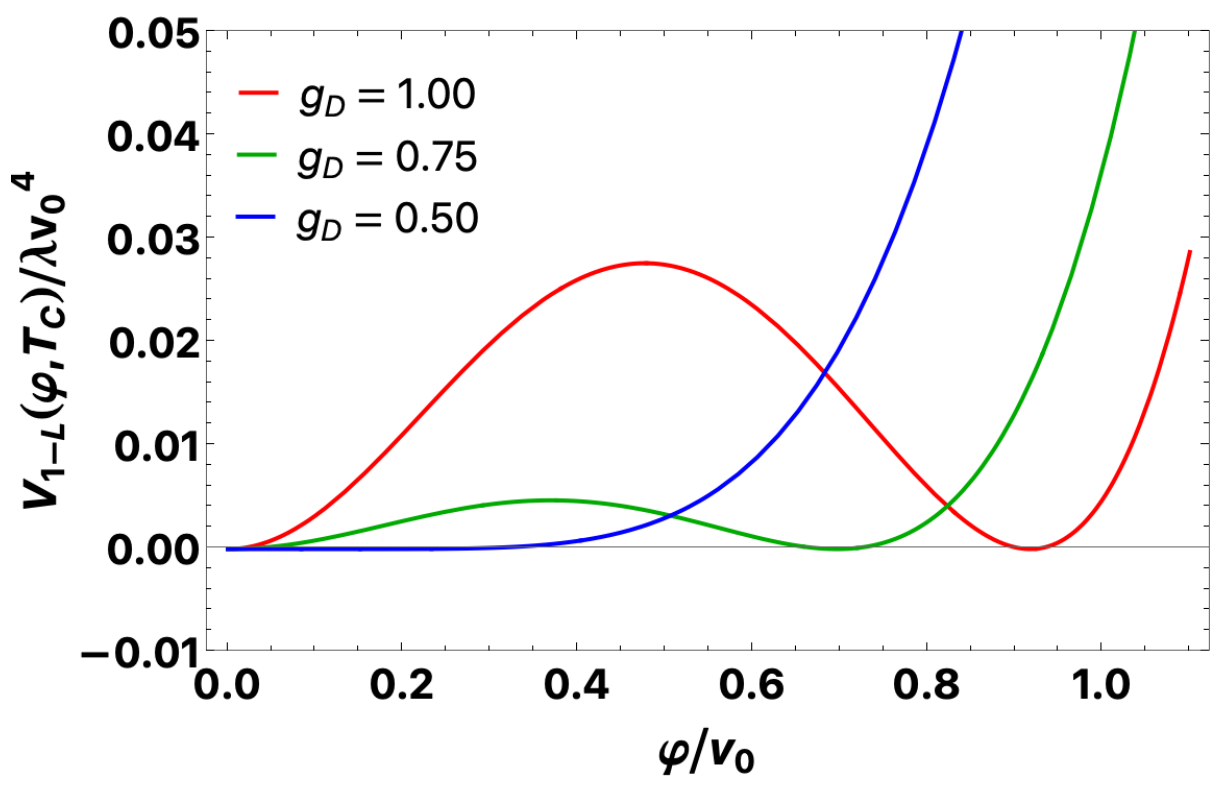}
\caption{Effective potentials for fixed $(\mu,\lambda) = (2\,{\rm MeV}, 0.02)$ and various gauge couplings $g_D$, at their respective critical temperatures. Increasing the gauge coupling increases the height of the potential barrier and the separation between the degenerate minima, which directly correlate to the strength of the first-order PT.}
\label{fig:Vcrit}
\end{figure}
%
These have the following expansions in the high-temperature limit:
\begin{align}
    V_T(\varphi,T) \approx &\displaystyle \sum_{i\,=\,\rm{bosons}} T^4\left[-\frac{\pi^2}{90} +\frac{m^2_i(\varphi)}{24\,T^2}-\frac{m^3_i(\varphi)}{12\pi\,T^3} + ...\right] \nonumber \\
    &+ \displaystyle \sum_{i\,=\,\rm{fermions}} T^4\left[\frac{7}{8}\frac{\pi^2}{90} -\frac{m^2_i(\varphi)}{48\,T^2} + ...\right], 
    \label{eq:VT_highT}
\end{align}
where we note that the cubic term from the bosons (absent for fermions) is responsible for generating a thermally-induced barrier, thus rendering the PT first-order. To quantify this, we require the minima at $\varphi = 0$ and $\varphi = v_c$ to be degenerate at the critical temperature $T_c$. We can realize these with the following conditions
\begin{align}
    V_{\rm {1-L}}(v_c,T_c) = V_{\rm {1-L}}(0,T_c)\,,\quad
    \frac{\partial V_{\rm {1-L}}(\varphi,T)}{\partial \varphi}\bigg|_{v_c,T_c} = 0\,,
\end{align}
and for stable minima $\partial^2_{\varphi}V_{\rm {1-L}}(v_c,T_c) >0$. We use the full expression \eqref{eq:VT} to track the thermal evolution accurately. To account for infrared divergences arising at small field values, one needs to re-sum the bosonic masses in \eqref{eq:masses_phi} by accounting for the thermal contribution from the zero Matsubara modes \cite{Laine:2016hma}. This is done customarily either through the re-summation of daisy diagrams \cite{Carrington:1991hz}, termed as the Arnold-Espinosa method, or using the leading order thermal masses for the bosons in \eqref{eq:VCW} and \eqref{eq:VT}, referred to as the Parwani approximation \cite{Parwani:1991gq}. However, the Arnold-Espinosa method relies on the assumption of the validity of the high-temperature limit $(m_i^2/T^2 < 1)$, and its extrapolation to low temperatures $(m_i^2/T^2 > 1)$ is inconsistent, whereas the Parwani approximation allows for continuation between both regimes, see, for e.g. Refs.~\cite{Cline:2011mm,Laine:2017hdk,Kainulainen:2019kyp, Kainulainen:2021eki}. Therefore, to consistently track the thermal evolution, in this work, we employ the latter which amounts to performing the following replacements for the bosonic masses $m^2_i(\varphi) \to m^2_i(\varphi,T)$ where
\begin{align}
    &m^2_{\phi,\,\eta} (\varphi,T) = m^2_{\phi,\,\eta}(\varphi) \nonumber\\
    &\qquad \qquad + \frac{((3 + n_\eta)\lambda + 3n_{A'}\,g_D^2 + n_\psi y^2_D)\,T^2}{12}\,,  \\
    &m^2_{A'}(\varphi,T) = m^2_{A'}(\varphi) + \frac{(n_{A'} + n_{\psi})}{3}g^2_D\, T^2\,,
\end{align}
and, if the model does not include fermions, $n_{\psi} = 0$. Note that for fermions, no such resummation is required (at the lowest order), due to the lack of a zero Matsubara mode \cite{Laine:2016hma}. 

As an example, we consider the simplest case of a spontaneously broken $U(1)_D$, which has the fewest number of free parameters: $\mu$, $\lambda$, and $g_D$, to provide initial insight. With the parameters $\mu$ and $\lambda$ fixed in the scalar potential \eqref{eq:V0}, the height of the thermally-induced barrier is determined by the gauge coupling $g_D$, as depicted in Fig.~\ref{fig:Vcrit}. In our numerical scheme, we therefore quantify a PT as ``not first-order'' if the barrier height at the corresponding critical temperature $\Delta V_c/(\lambda v_0^4) < 10^{-6}$. A proper determination of whether the transition in these scenarios is second-order or cross-over would require alternative methods, such as lattice calculations~\cite{Aarts:2023vsf}, and is beyond the scope of this work, where we have focused on (strong) first-order PTs.

\section{PT Parameters for Dark Sectors}
\label{app:PT_params}

\begin{figure}[t!]
\centering
\includegraphics[width=0.482\textwidth]{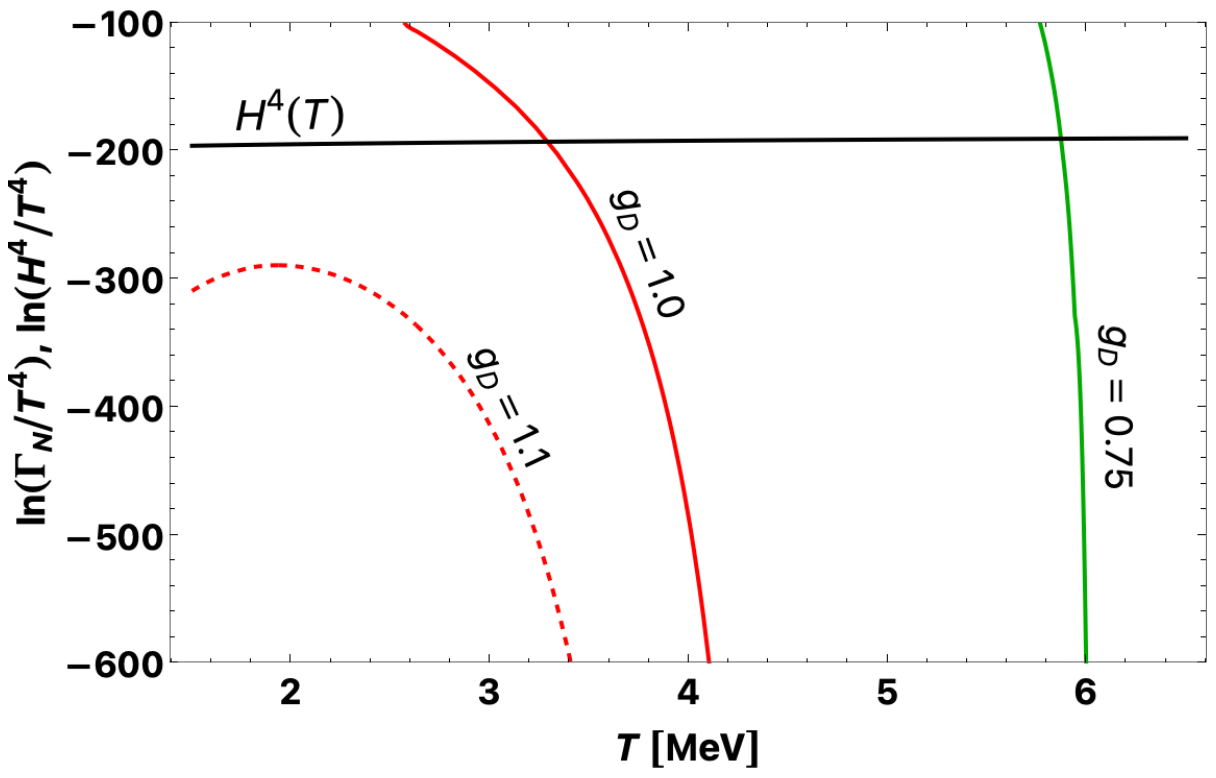}
\caption{The Hubble rate per Hubble volume $H^4(T)$, along with nucleation rates for the U(1) model with $\mu = 2$~MeV and $\lambda = 0.02$ with different values of $g_D$. The intersection point implies the nucleation criterion \eqref{eq:nucl_cond} is satisfied for $g_D = 0.75,\,1.0$, whereas this fails for $g_D = 1.1$.}
\label{fig:nucl_criterion}
\end{figure}

Below the critical temperature, the first-order phase transition proceeds through the nucleation of bubbles of the true vacuum. This occurs when the Hubble parameter becomes comparable to the nucleation rate. In Fig.~\ref{fig:nucl_criterion}, we demonstrate through examples of the U(1) case, with fixed $\mu = 2$~MeV and $\lambda = 0.02$, two scenarios where this nucleation criterion is met $(g_D = 1.0)$ and is not met $(g_D = 1.1)$. Note that in the latter case, a barrier is induced at low temperatures and the transition may occur through quantum tunneling from the false to the true vacuum~\cite{Coleman:1977py}. We have verified within the scanned parameter space that the thermal transition rate~\eqref{eq:Gamma} is always dominant. Finally, if $g_D$ is pushed to an even higher value for fixed $\lambda$, a scenario is eventually reached where loop-level effects at zero-temperature shift the minimum to $\varphi = 0$. We have classified these cases as ``no nucleation" to aid in the computational speed of our parameter scans.

\begin{figure*}
\centering
\includegraphics[width=0.35\textwidth]{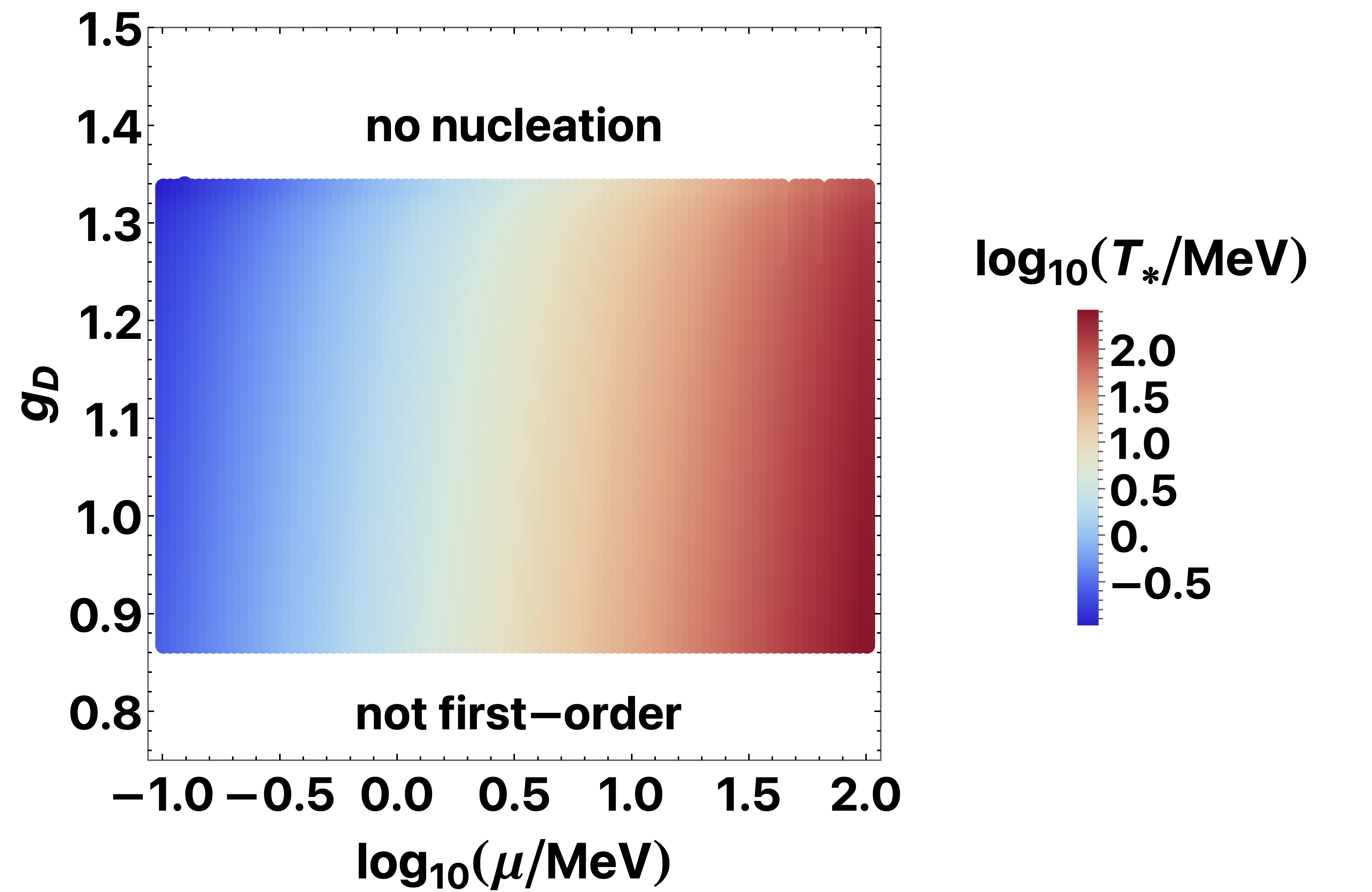}\,
\includegraphics[width=0.3\textwidth]{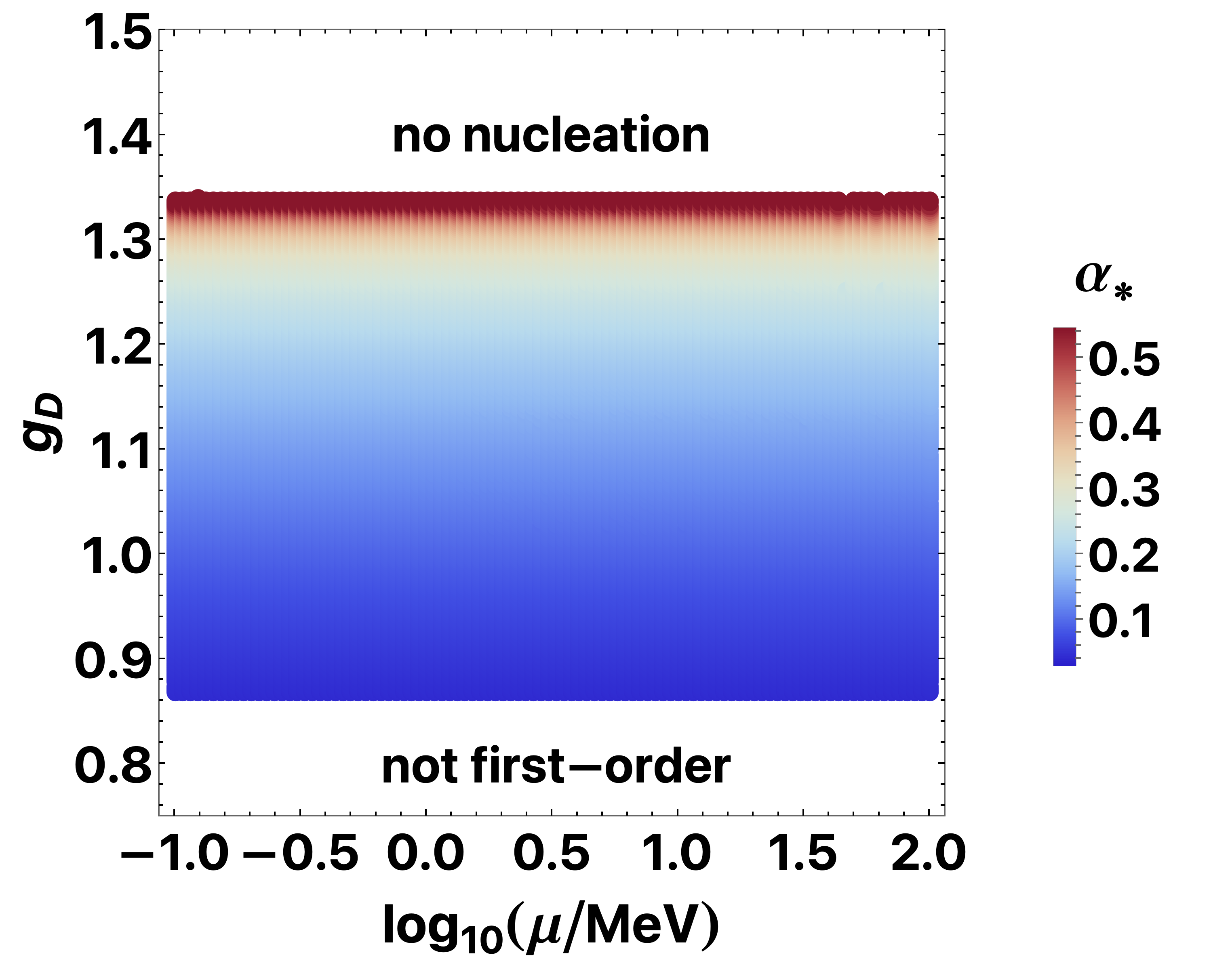}\,
\includegraphics[width=0.3\textwidth]{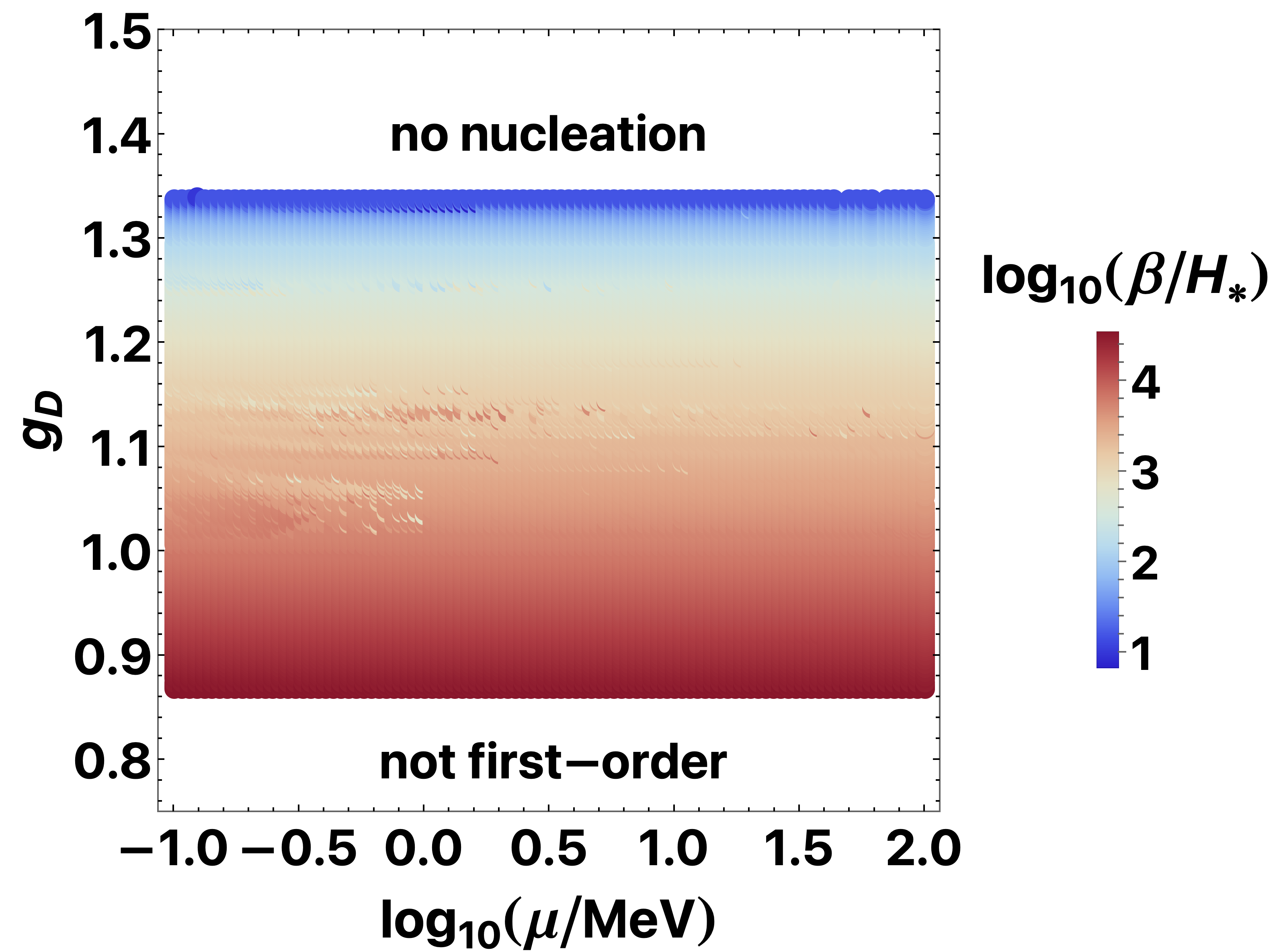}\\[8mm]
\includegraphics[width=0.35\textwidth]{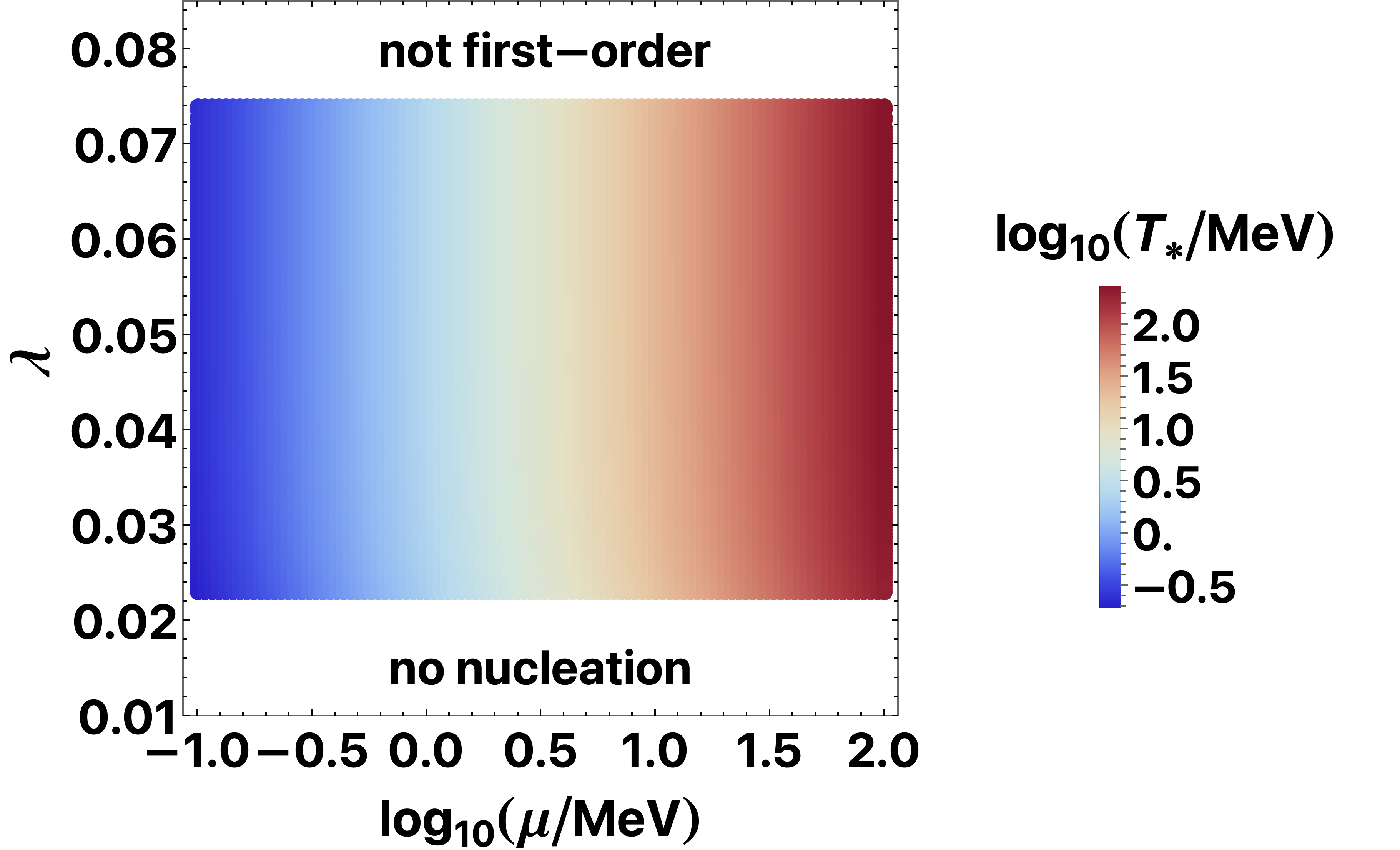}\,
\includegraphics[width=0.3\textwidth]{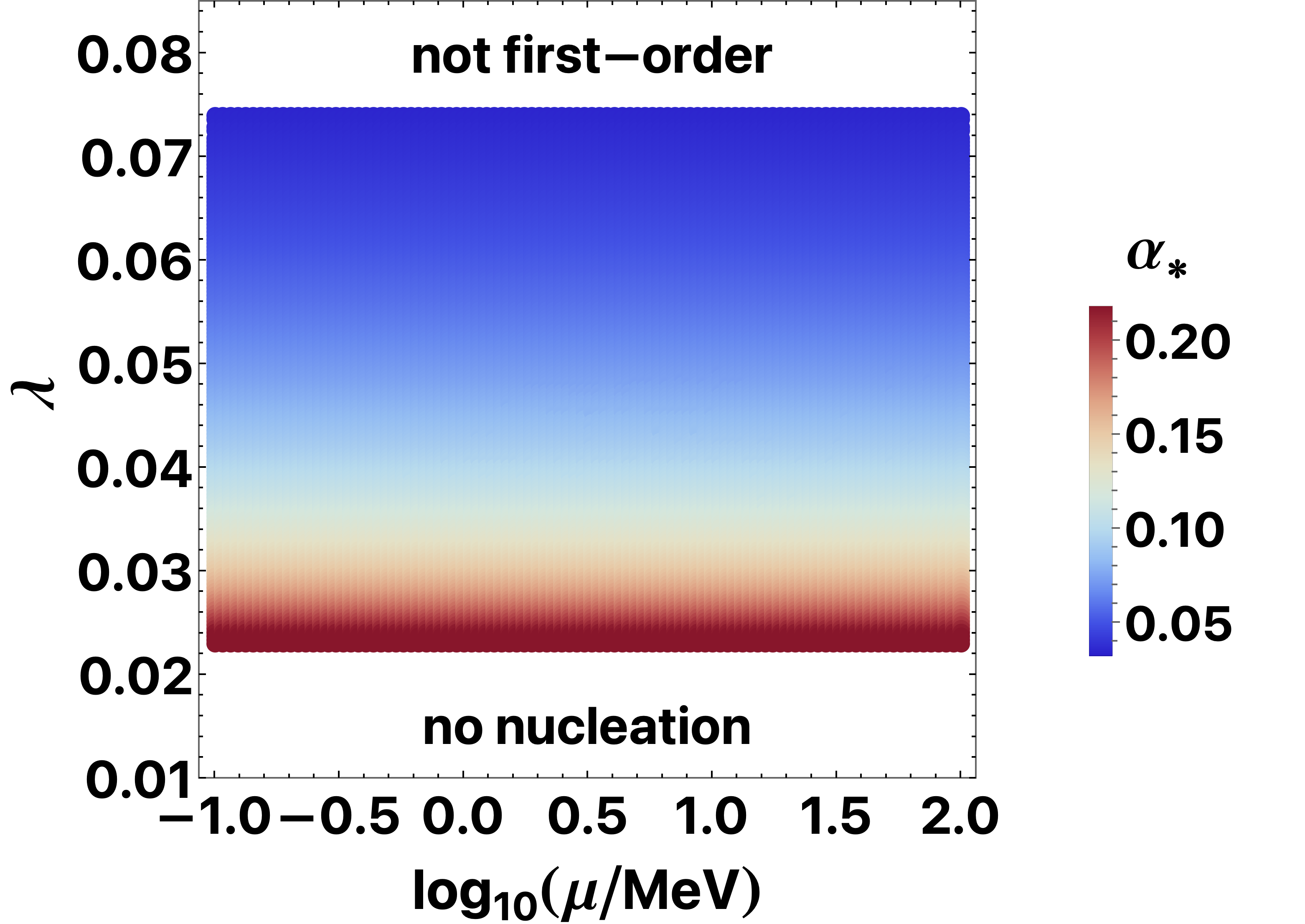}\,
\includegraphics[width=0.3\textwidth]{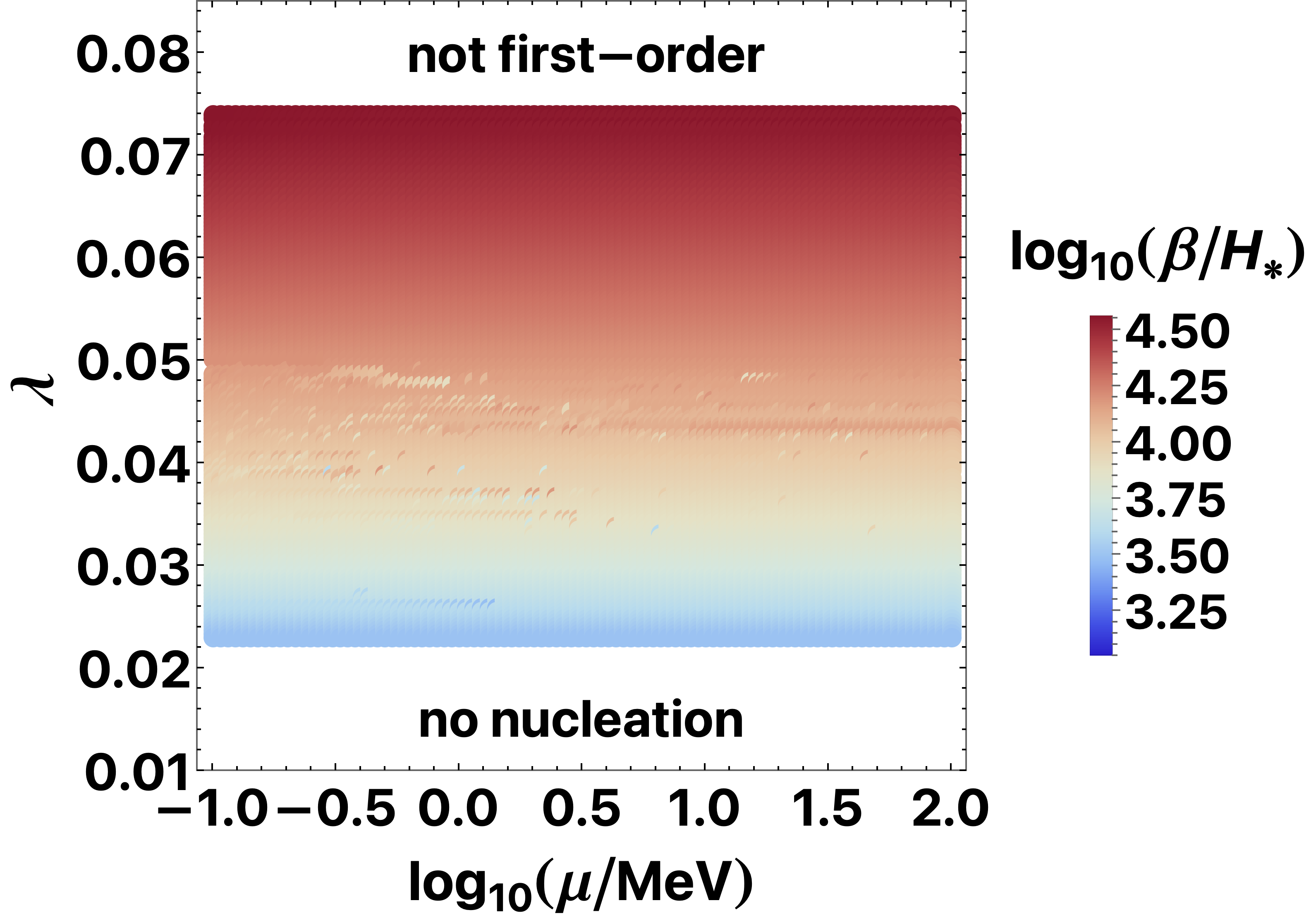}
\caption{Dependence of the various PT parameters on the model parameters for the U(1) model without fermions $(y_D = 0)$. \textit{Top row:} The PT parameters for fixed $\lambda = 0.05$ in the $(\mu, g_D)$ plane. \textit{Bottom row:} The PT parameters for fixed $g_D = 1.0$, and giving the dependence in the $(\mu, \lambda)$ plane.}
\label{fig:model_PTparam1}
\end{figure*}

\begin{figure*}
\centering
\includegraphics[width=0.35\textwidth]{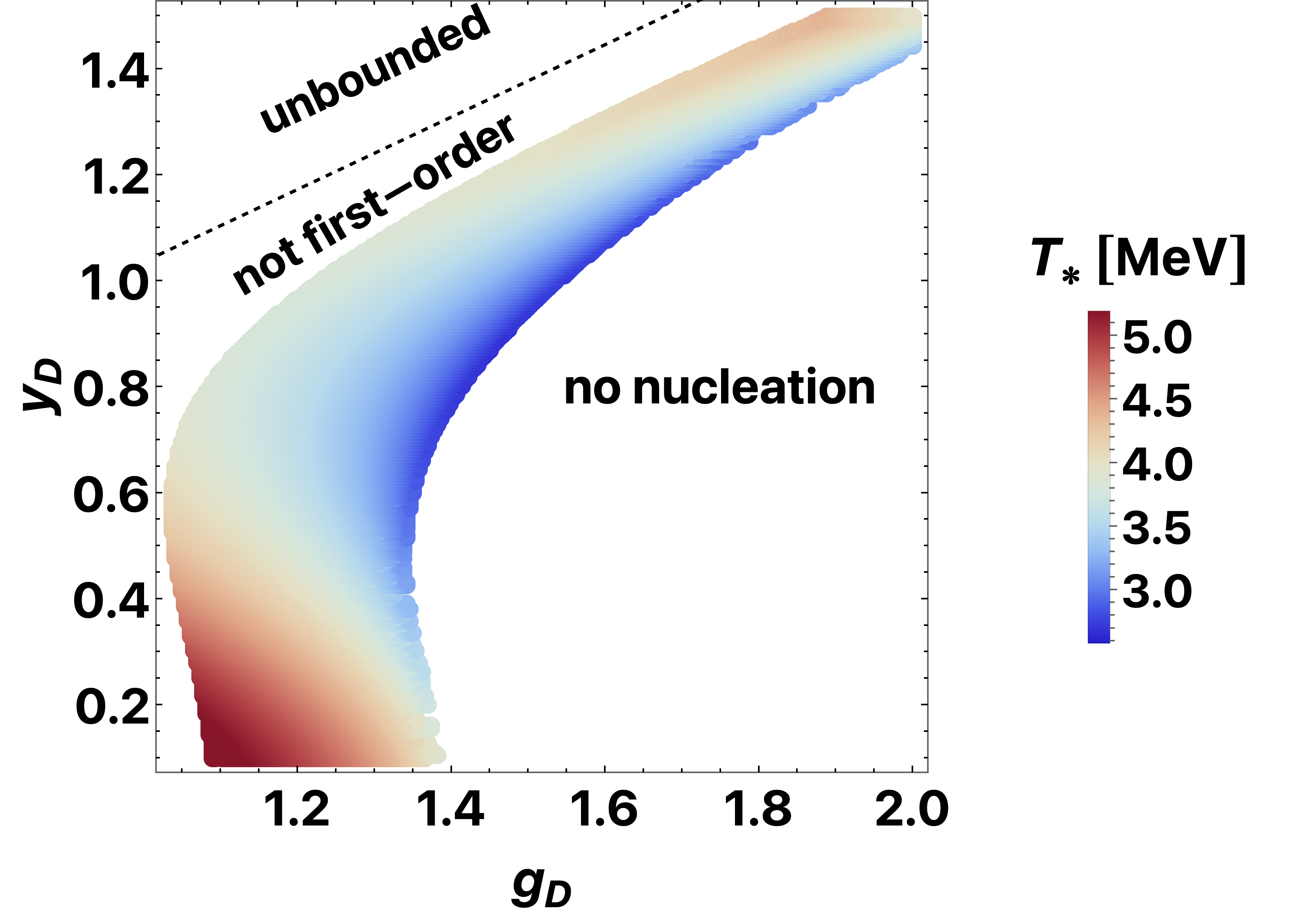}\,
\includegraphics[width=0.35\textwidth]{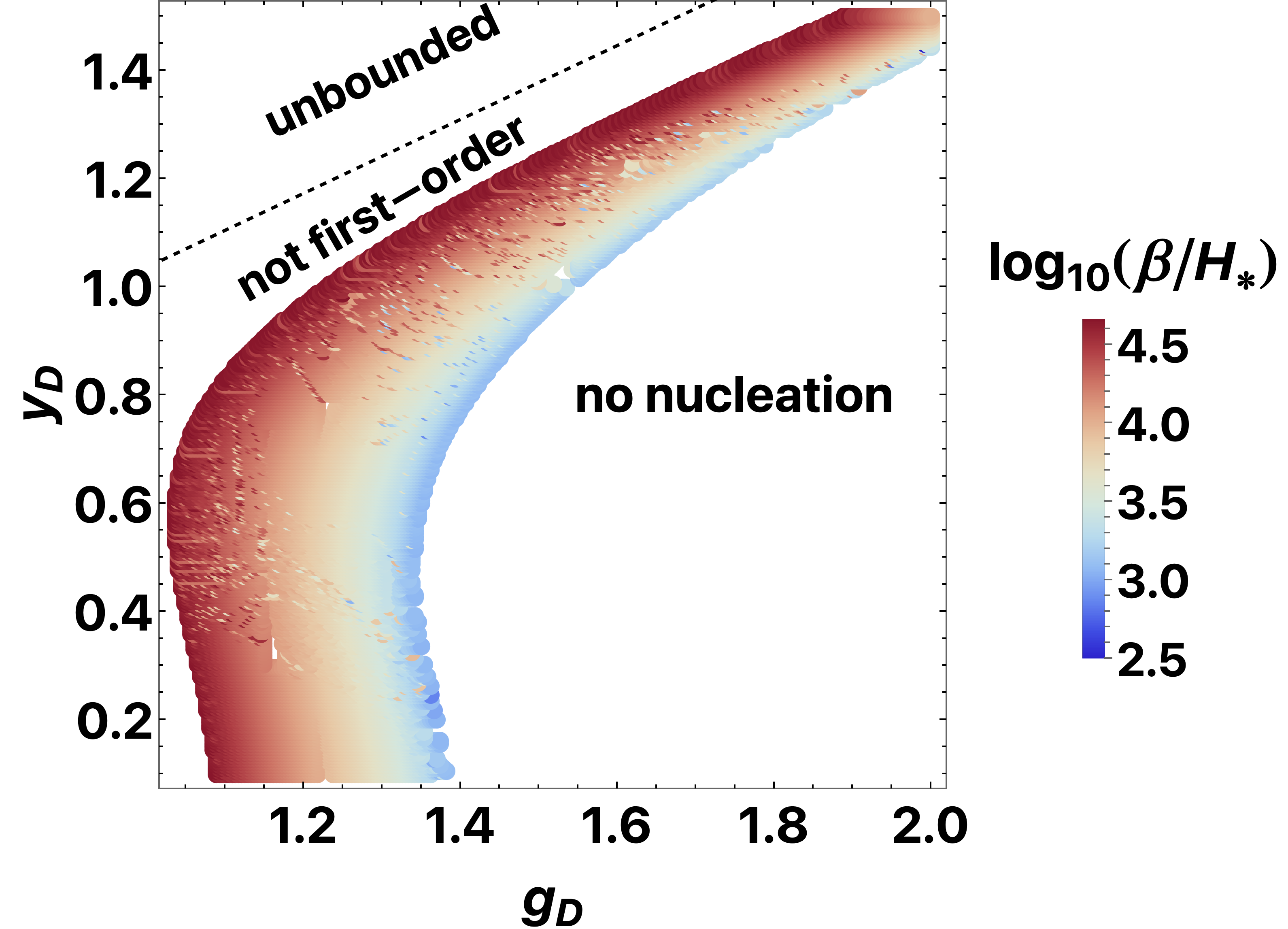}\\[8mm]
\includegraphics[width=0.35\textwidth]{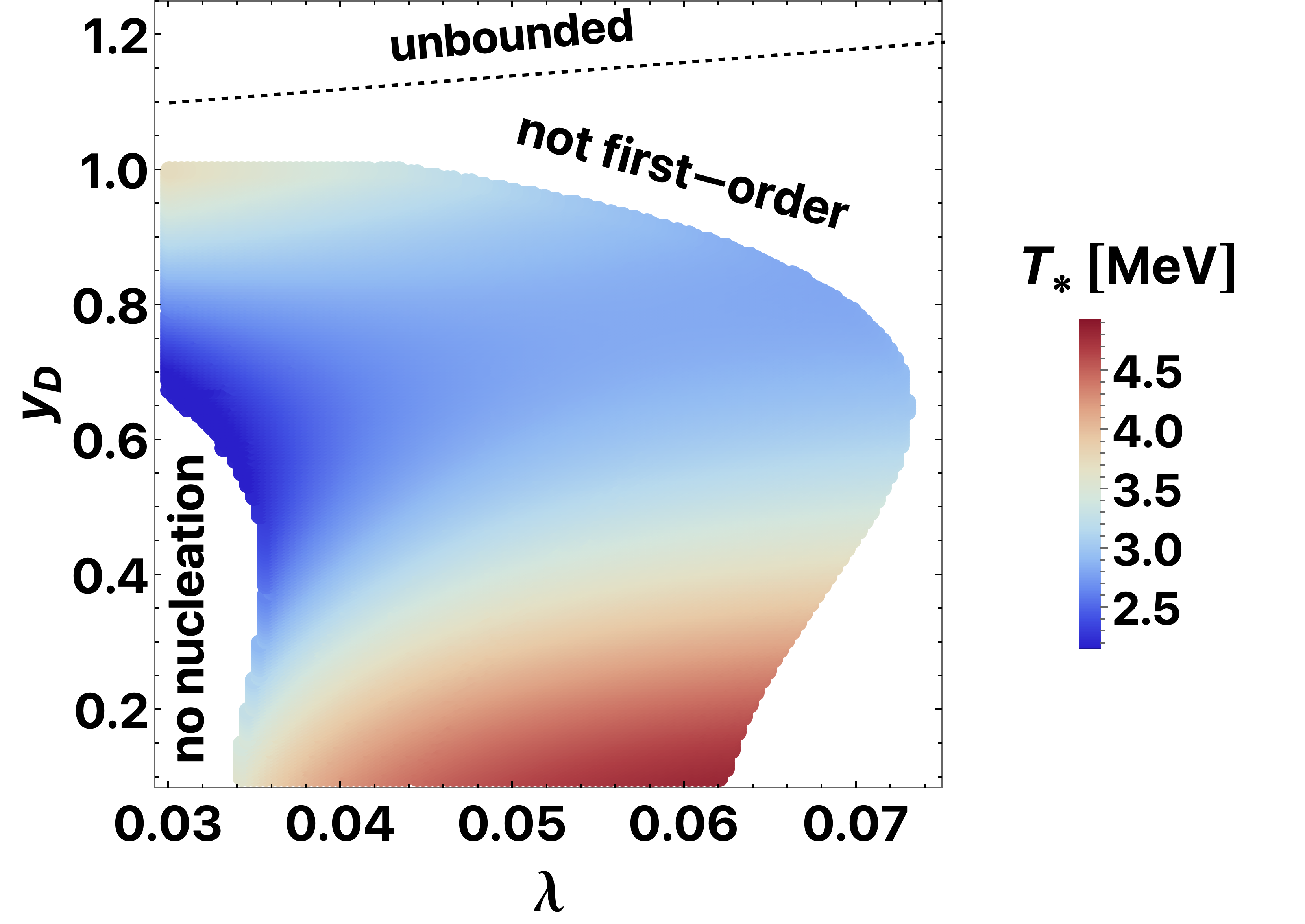}\,
\includegraphics[width=0.35\textwidth]{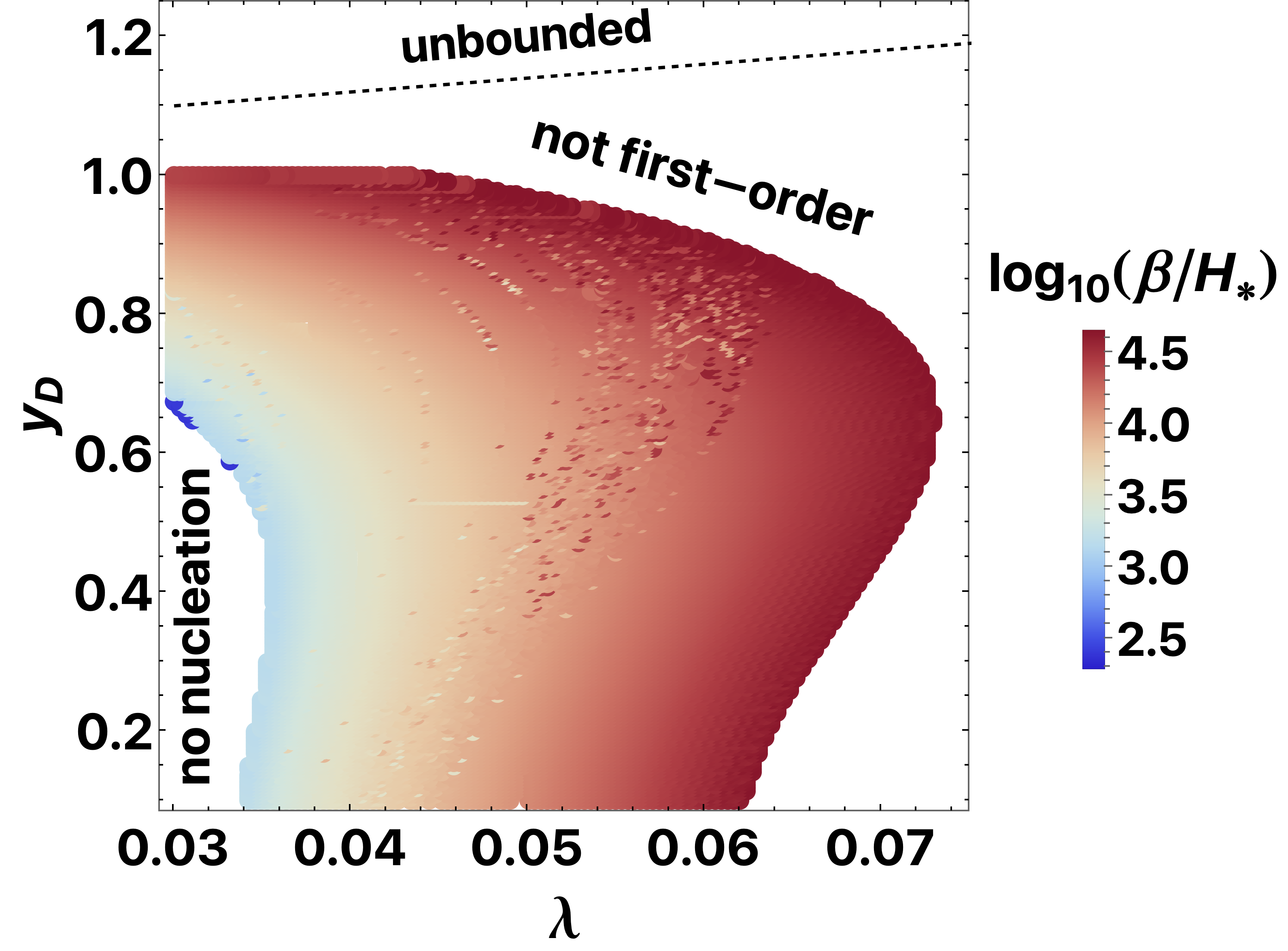}
\caption{Dependence of the various PT parameters on the model parameters for the U(1) model with fermions $(y_D \neq 0)$. \textit{Top row:} The dependence of $T_*$ and $\beta/H_*$ in the $(g_D,y_D)$ plane, with fixed $\lambda = 0.05$ and $\mu = 2$~MeV. \textit{Bottom row:} The dependence in the $(\lambda,y_D)$ plane for fixed $g_D = 1.2$ and $\mu = 2$~MeV. The dashed black lines demarcate the regions where the effective potential becomes unbounded, due to the large Yukawa coupling.}
\label{fig:model_PTparam2}
\end{figure*}

In the calculation of the PT parameters, it should be noted that since the inverse duration, $\beta/H_*$, involves a derivative of the action, it can be subject to numeric instabilities. Accordingly, we carefully smoothed the interpolation of $S_3/T$, prior to differentiation, to mitigate these effects. However, we still observe numeric artifacts for isolated points in our contours of $\beta/H_*$, arising from our dense parameter scans. Typically, these are apparent in regions where the PT is marginally first-order. We have ensured that these instabilities do not affect our conclusions concerning points that best fit the PTA dataset. Furthermore, we reject points in our parameter scans where $\beta/H_* \geq 10^6$ (indicating very short PTs). These typically correspond to very weak transitions and are therefore categorized as not first-order PTs. Finally, we also exclude cases where $\beta/H_* \lesssim 3$ (i.e., very long PTs) following Refs.~\cite{Freese:2022qrl, Winkler:2024olr}, as such values may lead to a phase of eternal inflation~\cite{Guth:1982pn}.

In Figs.~\ref{fig:model_PTparam1} and ~\ref{fig:model_PTparam2}, we present the dependence of the PT parameters on the model parameters of the U(1) group, categorized into cases without fermions $(y_D = 0)$ and with fermions $(y_D \neq 0)$. Focusing first on the case without fermions in Fig.~\ref{fig:model_PTparam1}, we observe that for fixed values of the other couplings, varying the $\mu$ parameter has little impact on the strength $(\alpha_*)$ and inverse duration $(\beta/H_*)$ of the PT. Its primary role is to set the scale of the transition, $T_*$, as depicted in the first column of Fig.~\ref{fig:model_PTparam1}. For the case including fermions, as previously discussed in the main text, the negative contribution from the fermions to the effective potential necessitates compensation by typically increasing the bosonic couplings. For completeness, we have included the dependence of the PT parameters $T_*$ and $\beta/H_*$ in the $(g_D,y_D)$ and $(\lambda,y_D)$ planes here, with the dependence of $\alpha_*$ displayed in the main text, in Fig.~\ref{fig:model_PTparam2}. Finally, regarding the increase in rank $N$ of the gauge group: this dependence is straightforward, as noted in the main text, with the effect being a simple shift of the parameter space due to the additional contribution from the increased number of gauge bosons in the effective potential.

\section{GW spectrum from a first-order PT}
\label{app:gw_spec}
Here, we review the various contributions to SGWB arising from first-order PTs. Starting from the analytic form of GW spectrum for sound waves, this is typically obtained by fitting to numerical simulations, and can be described by \cite{Hindmarsh:2017gnf, Caprini:2018mtu,Caprini:2019egz} 
\begin{align}
    &\Omega^{\rm{sw}}_{\rm{GW}}h^2(f) = \Omega^{\rm{sw, peak}}_{\rm{GW}}h^2\left(\frac{f}{f^{\rm{sw}}_{\rm{peak}}}\right)^3\left(\frac{7}{4+3\left(\frac{f}{f^{\rm{sw}}_{\rm{peak}}}\right)^2}\right)^{\frac{7}{2}}\,,
    \label{eq:omega_sw}
\end{align}
where the peak density and peak frequency after appropriate red-shifting to today, assuming a radiation-dominated era, are given by 
\begin{align}
    \Omega^{\rm{sw, peak}}_{\rm{GW}}h^2 =& 5.71\times10^{-8}\,v_w\,\left(\frac{10}{g_{*,\rm{tot}}}\right)^{\frac{1}{3}}\nonumber\\
    &\times\left(\frac{\beta/H_*}{100}\right)^{-1} \left(\frac{\kappa_{\rm{sw}}\,\alpha_{\ast}}{1+\alpha_{\ast}}\right)^{2}\,,
    \label{eq:omega_swpeak}
\end{align}
\begin{align}
    &f^{\rm{sw}}_{\rm{peak}} = \frac{5.96\times10^{-9}\,\text{Hz}}{v_w}\left(\frac{\beta/H_*}{100}\right)\left(\frac{g_{*,\rm{tot}}}{10}\right)^{\frac{1}{6}}\left(\frac{T_*}{1\,\rm{MeV}}\right)\,.
    \label{eq:fswpeak}
\end{align}
Here, $g_{*,\rm{tot}}$ are the total relativistic degrees of freedom at $T_*$. We assume that the first-order PT occurs in the ``detonation regime", where the bubble wall velocity exceeds the speed of sound, $v_w > 1/\sqrt{3}$. The efficiency factor for the conversion of the latent heat to the plasma motion in this limit is given by \cite{Steinhardt:1981ct,Espinosa:2010hh}
\begin{equation}
    \kappa_{\rm{sw}} \approx \frac{\alpha_{\rm{DS},\ast}}{0.73 + 0.083\sqrt{\alpha_{\rm{DS},\ast}}+\alpha_{\rm{DS},\ast}}\,,
    \label{eq:kappasw}
\end{equation}
and the wall velocity can be estimated using the Jouguet velocity,
\begin{equation}
    v_w \approx \frac{1/\sqrt{3}+\sqrt{\alpha_{\rm{DS},\ast}+2\alpha_{\rm{DS},\ast}/3}}{1+\alpha_{\rm{DS},\ast}}\,.
    \label{eq:wall_vel}
\end{equation}
Note that these parameters depend on the DS dynamics and thus have been calculated using the latent heat normalized to the DS radiation density, $\alpha_{{\rm DS},*}$~\cite{Fairbairn:2019xog, Bringmann:2023opz}. Typically, we find that the large $\alpha_{\rm{DS},\ast}$, thus implying $v_w \sim 1$, is consistent with strong phase transitions \cite{Bodeker:2017cim}. Finally, we multiply \eqref{eq:omega_sw} with a suppression factor,
\begin{equation}
    \Upsilon = \min\left[1,\frac{(8\pi)^{1/3}}{(\beta/H_*)}\sqrt{\frac{4(1+\alpha_*)}{3\kappa_{\rm sw}\alpha_*}}\right]\,.
    \label{eq:upsilon}
\end{equation}
to account for the finite lifetime of the source \cite{Ellis:2020awk, Guo:2020grp}.

The contribution from turbulence can be estimated as \cite{Caprini:2009yp, Caprini:2018mtu}
\begin{align}
    &\Omega^{\rm{turb}}_{\rm{GW}}h^2(f) = \Omega^{\rm{turb, peak}}_{\rm{GW}}h^2\,\frac{(f/f_{\rm{turb}})^3}{(1+8\pi f/h_*)\left(1+\frac{f}{f^{\rm{turb}}_{\rm{peak}}}\right)^{11/3}}\,,
    \label{eq:omega_turb}
\end{align}
where 
\begin{align}
    &h_* = 1.13\times10^{-10}\,\text{Hz} \left(\frac{g_{*,\,\text{tot}}}{10}\right)^{\frac{1}{6}}\left(\frac{T_*}{1\,\rm{MeV}}\right)
\end{align}
is the Hubble rate at the time of the PT, red-shifted to today. This factor arises from the fact that the turbulence as a source for GWs can last several Hubble times. While the spectral shape is well-modeled, the amplitude is subject to uncertainties \cite{Kosowsky:2001xp, Gogoberidze:2007an, Niksa:2018ofa}, 
\begin{align}
    \Omega^{\rm{turb, peak}}_{\rm{GW}}h^2 =& 7.22\times10^{-6}\,v_w\,\left(\frac{10}{g_{*,\rm{tot}}}\right)^{\frac{1}{3}}\nonumber\\
    &\times \left(\frac{\beta/H_*}{100}\right)^{-1} \left(\frac{\kappa_{\rm{turb}}\,\alpha_{\ast}}{1+\alpha_{\ast}}\right)^{\frac{3}{2}}\,.
    \label{eq:omega_turbpeak}
\end{align}
The efficiency factor of the conversion of the kinetic energy to the vortical motion of the plasma is given by $\kappa_{\rm{turb}} \equiv \epsilon\,\kappa_{\rm sw}$ where we have the vorticity $\epsilon \approx 0.05$. The peak frequency, slightly higher than \eqref{eq:fswpeak}, is given by 
\begin{align}
    &f^{\rm{turb}}_{\rm{peak}} = \frac{1.84\times10^{-8}\,\text{Hz}}{v_w}\left(\frac{\beta/H_*}{100}\right)\left(\frac{g_{*,\rm{tot}}}{10}\right)^{\frac{1}{6}}\left(\frac{T_*}{1\,\rm{MeV}}\right)\,.
    \label{eq:fturbpeak}
\end{align}

\begin{figure}[t!]
\centering
\includegraphics[width=0.482\textwidth]{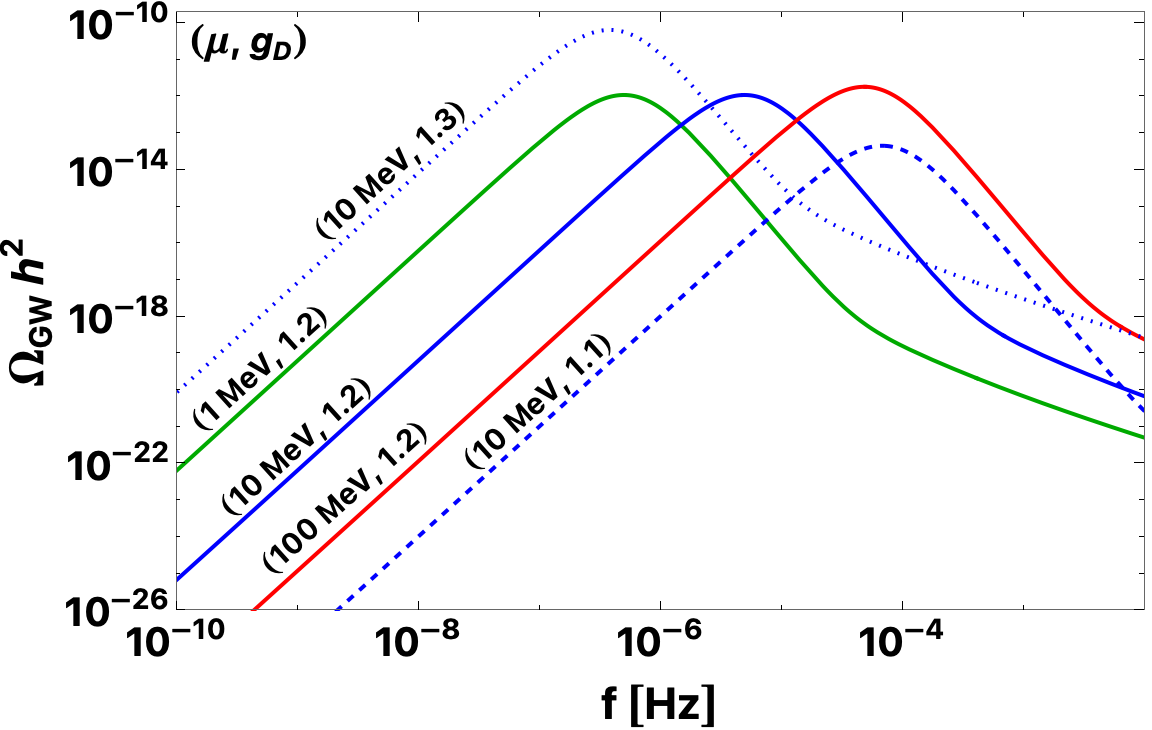}
\caption{Examples of the GW spectrum for various parameters in the U$(1)$ case. Here, we have fixed $\lambda = 0.05$ and $y_D = 0$ for all curves and $(\mu, g_D)$ is indicated in parentheses. The solid-colored curves have the same $g_D = 1.2$, but different $\mu$. The dotted and dashed curves have the same $\mu$, but different $g_D$.}
\label{fig:example_gw}
\end{figure}

For completeness, we also include the contribution from bubble collisions, which is typically 4-5 orders of magnitude smaller than the sound wave contribution in our analysis, due to the thermal phase transitions considered and the lack of a high degree of supercooling in the scanned parameter space~\cite{Caprini:2019egz}. The spectral shape obtained from employing the ``envelope'' approximation \cite{Kosowsky:1992rz, Kosowsky:1992vn, Huber:2008hg, Weir:2016tov} is given by \cite{Caprini:2018mtu, Caprini:2019egz}  
\begin{align}
    &\Omega^{\rm{bc}}_{\rm{GW}}h^2(f) =\Omega^{\rm{bc, peak}}_{\rm{GW}}h^2\, \left[\frac{3.8 \, (f/f^{\rm{bc}}_{\rm{peak}})^{2.8}}{1 + 2.8 \,(f / f^{\rm{bc}}_{\rm{peak}})^{3.8}}\right]\,,
    \label{eq:omega_bc}
\end{align}
with the peak amplitude 
\begin{align}
    \Omega^{\text{bc,peak}}_{\rm{GW}}h^2 =& 3.6 \times 10^{-9} \left( \frac{0.11 \, v_w^ 3}{0.42 + v_w^2} \right)\left( \frac{10}{g_{*,\rm{tot}}} \right)^{\frac{1}{3}}\nonumber\\
    &\times \left( \frac{\beta/H_*}{100} \right)^{-2}
    \left( \frac{\kappa_{\rm{bc}}\, \alpha}{1+\alpha} \right)^2
\,,
\label{eq:omega_bcpeak}
\end{align}
suppressed by an additional factor of $\beta/H_*$ in comparison to \eqref{eq:omega_swpeak} and \eqref{eq:omega_turbpeak}. The peak frequency is given  by 
\begin{align}
    f^{\rm{bc}}_{\rm{peak}} = 1.12 \times 10^{-8} \, \text{Hz} 
    \left( \frac{k\,\beta/H_*}{100} \right) \left( \frac{T_*}{1 \, \text{MeV}} \right) \left( \frac{g_{*,\rm{tot}}}{10} \right)^{\frac{1}{6}}\,
    \label{eq:fbcpeak},
\end{align}
with $k = 0.62/(1.8 - 0.1 v_w + v_w^2)$. This implies that the peak frequency is almost an order of magnitude lower than \eqref{eq:fswpeak} for $v_w \sim 1$. Finally, given the relativistic wall velocities considered~\cite{Ellis:2019oqb}, we set $\kappa_{\rm bc} = 1$ for the spectra, assuming highly efficient energy transfer to the GWs from the bubble collisions. 

In Fig.~\ref{fig:example_gw}, we illustrate the dependence of the GW spectrum on the Lagrangian parameters for the simplest case of the U$(1)$ model with $y_D = 0$ and $\lambda = 0.05$. Varying $\mu$ affects $T_*$ and thus changes the location of the peak frequency. A decrease (increase) in $\beta/H_*$ causes the spectrum to shift diagonally to the right (left), while an increase (decrease) in $\alpha_*$ directly increases (decreases) the amplitude. Therefore, the gauge coupling $g_D$ (for a fixed $\lambda$) essentially determines $\alpha_*$ and $\beta/H_*$ (see also Fig.~2 in the main text). Finally, the turbulence effect typically leads to a component in the spectrum with a higher peak frequency, modifying the tails of the spectra.

\section{Gaussian Approximation of the PTA Likelihood}
\label{app:best_points}

\begin{figure}[t!]
\centering
\includegraphics[width=0.4\textwidth]{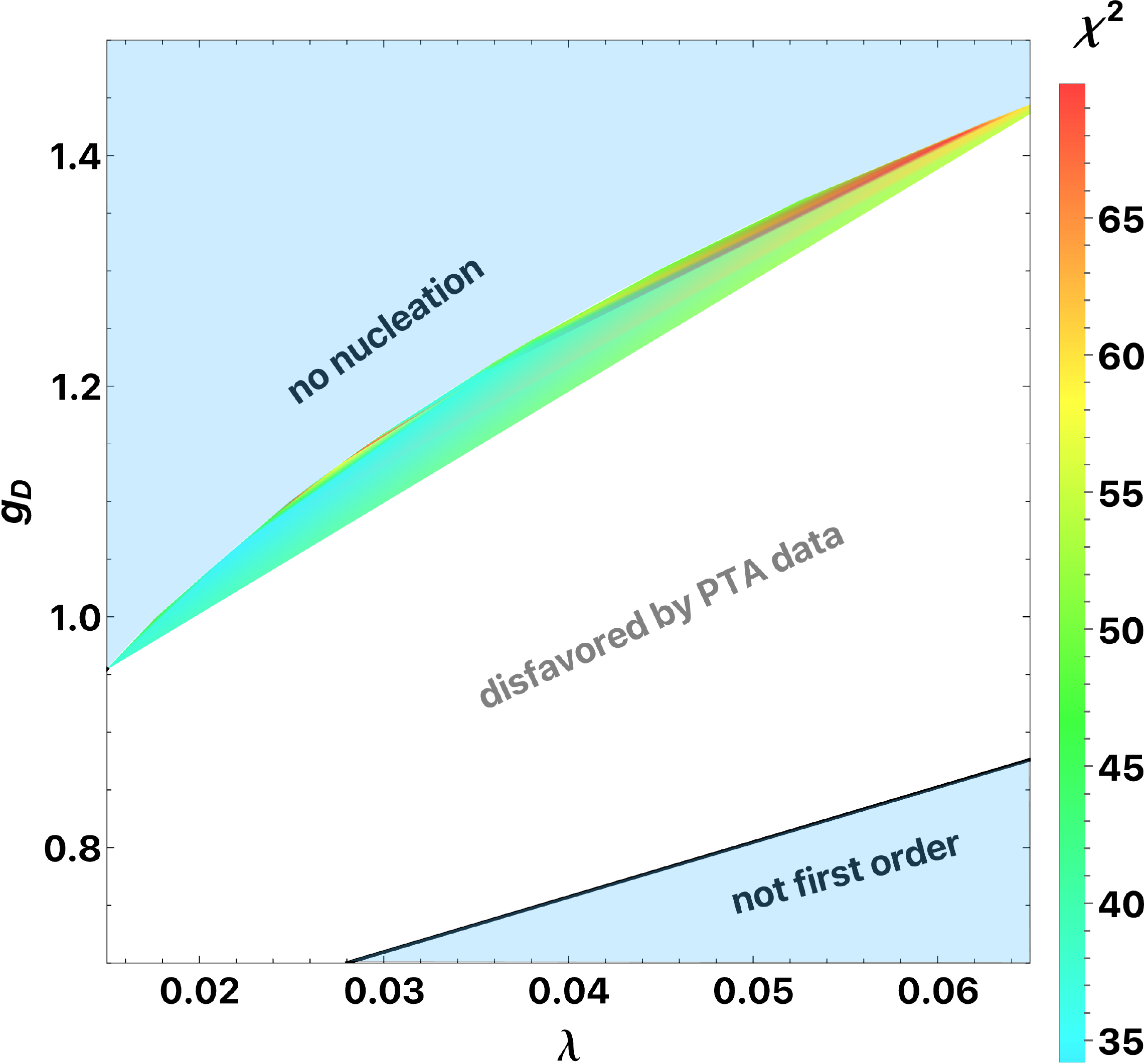}
\caption{
The coupling combination $(\lambda, g_D)$ preferred by the PTA dataset, with fixed $\mu = 4$~MeV as an example, for the U$(1)$ model with $y_D = 0$. We show the $\chi^2$ values for up to a $3\sigma$ fit to PTA data. The white region is disfavored by the PTA fit.}
\label{fig:chisq_U1}
\end{figure}

\begin{table*}
\centering
\renewcommand{\arraystretch}{1.4}
\setlength{\tabcolsep}{7.5pt}
\begin{tabular}{|c|c|c|c|}
\hline
\textbf{MODEL} & $(\mu,\,\lambda,\,g_D,\,y_D)$ & $(T_*,\,\alpha_*,\,\beta/H_*)$ & \(\chi^2\) \\ \hline
\multirow{2}{*}{U$(1)$} & (4.9 MeV, 0.04, 1.2, 0)  & (5.0 MeV, 0.96, 45) &37 \\ \cline{2-4}
                      &(6.0 MeV, 0.06, 1.5, 0.8)  &(5.8 MeV, 0.82, 39)  & 36 \\ \hline
\multirow{2}{*}{SU$(2)$} & (5.0 MeV, 0.05, 1.0, 0)  &  (4.8 MeV, 0.92, 44) &38 \\ \cline{2-4}
                       &(5.3 MeV, 0.05, 1.2, 0.9)   & (5.1 MeV, 0.85, 40)  &40   \\ \hline
\multirow{2}{*}{SU$(3)$} & (4.5 MeV, 0.03, 0.7, 0)   & (4.5 MeV, 1.0, 42) &41 \\ \cline{2-4}
                       &(5.6 MeV, 0.06, 0.9, 0.9) &(5.3 MeV, 0.9, 34)  &43 \\ \hline
\end{tabular}
\caption{Approximate values of the best-fit model parameters, with the calculated PT parameters in the third column. The d.o.f. is on average 40, implying all these points correspond to a fit at $1\sigma$ confidence level.}
\label{tab:best_gw_chisq}
\end{table*}

In this alternative analysis, we consider the mean values and $1\sigma$ error bars extracted from the PTA violin plots, as computed in Ref.~\cite{Winkler:2024olr}, which approximates the PTA likelihood as a Gaussian distribution. To quantify the compatibility of the model with the PTA data, we employ a goodness-to-fit metric $\chi^2$, defined as
\begin{align}
    \chi^2 = \sum_i \left[\frac{\log_{10}\Omega_{\text{GW}}h^2(f_i)-\log_{10}\Omega^{\rm{data}}_{\text{GW}}h^2(f_i)}{\Delta \log_{10}\Omega^{\rm{data}}_{\text{GW}}h^2(f_i)}\right]^2\,,
    \label{eq:chisq}
\end{align}
where $\Omega^{\rm{data}}_{\text{GW}}(f)$ refers to the mean value of the PTA data. $f_i$ refers to the frequency of the $i^{\rm{th}}$ bin and we sum over the full PTA data set, consisting of 41 bins. As the error bars $\Delta \log_{10}\Omega^{\rm{data}}_{\text{GW}}(f_i)$ are asymmetric, we use the upper or lower error value depending on whether the model prediction lies above or below the mean value.

To evaluate the significance level of the fit, we first identify the number of degrees of freedom (d.o.f.). Assuming no prior constraints on the model parameters, the d.o.f. is given by $n = 41 - n_{\rm{params}}$.  For a systematic study, we approach the model parameter space as follows: we fix the gauge group (i.e. $N$) and then consider the case with and without $y_D = 0$. This is demonstrated in Fig. \ref{fig:chisq_U1} for the U$(1)$ case with $y_D = 0$. Since the PT strength and duration are less sensitive to $\mu$ as discussed earlier, we first make use of the results of Fig. \ref{fig:PT_gauge} with fixed $\mu = 2$~MeV, to identify the $(\lambda,\,,g_D)$ that leads to an optimal combination of $\beta/H_*$ and $\alpha_*$ to fix the amplitude of the signal. With these couplings determined, we then vary $\mu$ to find its best-fit value, which determines the $T_*$ and therefore anchors the peak frequency. For our likelihood analysis, we note that a fit at the significance level of $1\sigma$ corresponds to $\chi^2 \sim 45$, while $\chi^2 \sim 60 \, (70)$ corresponds to a fit at the level of $2\sigma \,(3\sigma)$ with the d.o.f. $n = 40$. We adopt a similar approach when analyzing the fitting in the cases of SU$(2)$ and SU$(3)$ models. For fitting at the $1\sigma$ level, we found the preferred parameter values as: $\beta/H_* \sim 40$, $\alpha_* \sim 0.9$ and $T_*\sim5$~MeV. The main distinction arises in the value of $T_*$, which is slightly smaller than the preferred value $T_* \sim 7$~MeV, obtained in the main text based on the actual PTA likelihood distribution. Finally, we provide the obtained best-fit parameters through this alternate analysis for the models considered in this work in Table~\ref{tab:best_gw_chisq}.


\bibliographystyle{elsarticle-num}
\bibliography{lit}

@article{NANOGrav:2020bcs,
    author = "Arzoumanian, Zaven and others",
    collaboration = "NANOGrav",
    title = "{The NANOGrav 12.5 yr Data Set: Search for an Isotropic Stochastic Gravitational-wave Background}",
    eprint = "2009.04496",
    archivePrefix = "arXiv",
    primaryClass = "astro-ph.HE",
    doi = "10.3847/2041-8213/abd401",
    journal = "Astrophys. J. Lett.",
    volume = "905",
    number = "2",
    pages = "L34",
    year = "2020"
}

@article{NANOGrav:2023gor,
    author = "Agazie, Gabriella and others",
    collaboration = "NANOGrav",
    title = "{The NANOGrav 15 yr Data Set: Evidence for a Gravitational-wave Background}",
    eprint = "2306.16213",
    archivePrefix = "arXiv",
    primaryClass = "astro-ph.HE",
    doi = "10.3847/2041-8213/acdac6",
    journal = "Astrophys. J. Lett.",
    volume = "951",
    number = "1",
    pages = "L8",
    year = "2023"
}

@article{EPTA:2021crs,
    author = "Chen, S. and others",
    collaboration = "EPTA",
    title = "{Common-red-signal analysis with 24-yr high-precision timing of the European Pulsar Timing Array: inferences in the stochastic gravitational-wave background search}",
    eprint = "2110.13184",
    archivePrefix = "arXiv",
    primaryClass = "astro-ph.HE",
    doi = "10.1093/mnras/stab2833",
    journal = "Mon. Not. Roy. Astron. Soc.",
    volume = "508",
    number = "4",
    pages = "4970--4993",
    year = "2021"
}

@article{EPTA:2023fyk,
    author = "Antoniadis, J. and others",
    collaboration = "EPTA, InPTA:",
    title = "{The second data release from the European Pulsar Timing Array - III. Search for gravitational wave signals}",
    eprint = "2306.16214",
    archivePrefix = "arXiv",
    primaryClass = "astro-ph.HE",
    doi = "10.1051/0004-6361/202346844",
    journal = "Astron. Astrophys.",
    volume = "678",
    pages = "A50",
    year = "2023"
}

@article{EPTA:2023xxk,
    author = "Antoniadis, J. and others",
    collaboration = "EPTA, InPTA",
    title = "{The second data release from the European Pulsar Timing Array - IV. Implications for massive black holes, dark matter, and the early Universe}",
    eprint = "2306.16227",
    archivePrefix = "arXiv",
    primaryClass = "astro-ph.CO",
    doi = "10.1051/0004-6361/202347433",
    journal = "Astron. Astrophys.",
    volume = "685",
    pages = "A94",
    year = "2024"
}

@article{Goncharov:2021oub,
    author = "Goncharov, Boris and others",
    title = "{On the Evidence for a Common-spectrum Process in the Search for the Nanohertz Gravitational-wave Background with the Parkes Pulsar Timing Array}",
    eprint = "2107.12112",
    archivePrefix = "arXiv",
    primaryClass = "astro-ph.HE",
    doi = "10.3847/2041-8213/ac17f4",
    journal = "Astrophys. J. Lett.",
    volume = "917",
    number = "2",
    pages = "L19",
    year = "2021"
}

@article{Reardon:2023gzh,
    author = "Reardon, Daniel J. and others",
    title = "{Search for an Isotropic Gravitational-wave Background with the Parkes Pulsar Timing Array}",
    eprint = "2306.16215",
    archivePrefix = "arXiv",
    primaryClass = "astro-ph.HE",
    doi = "10.3847/2041-8213/acdd02",
    journal = "Astrophys. J. Lett.",
    volume = "951",
    number = "1",
    pages = "L6",
    year = "2023"
}

@article{Xu:2023wog,
    author = "Xu, Heng and others",
    title = "{Searching for the Nano-Hertz Stochastic Gravitational Wave Background with the Chinese Pulsar Timing Array Data Release I}",
    eprint = "2306.16216",
    archivePrefix = "arXiv",
    primaryClass = "astro-ph.HE",
    doi = "10.1088/1674-4527/acdfa5",
    journal = "Res. Astron. Astrophys.",
    volume = "23",
    number = "7",
    pages = "075024",
    year = "2023"
}

@article{Miles:2024seg,
    author = "Miles, Matthew T. and others",
    title = "{The MeerKAT Pulsar Timing Array: the first search for gravitational waves with the MeerKAT radio telescope}",
    eprint = "2412.01153",
    archivePrefix = "arXiv",
    primaryClass = "astro-ph.HE",
    doi = "10.1093/mnras/stae2571",
    journal = "Mon. Not. Roy. Astron. Soc.",
    volume = "536",
    number = "2",
    pages = "1489--1500",
    year = "2024"
}

@article{Hellings:1983fr,
    author = "Hellings, R. w. and Downs, G. s.",
    title = "{UPPER LIMITS ON THE ISOTROPIC GRAVITATIONAL RADIATION BACKGROUND FROM PULSAR TIMING ANALYSIS}",
    doi = "10.1086/183954",
    journal = "Astrophys. J. Lett.",
    volume = "265",
    pages = "L39--L42",
    year = "1983"
}

@article{Haehnelt:1994wt,
    author = "Haehnelt, Martin G.",
    title = "{Low frequency gravitational waves from supermassive black holes}",
    eprint = "astro-ph/9405032",
    archivePrefix = "arXiv",
    doi = "10.1093/mnras/269.1.199",
    journal = "Mon. Not. Roy. Astron. Soc.",
    volume = "269",
    pages = "199",
    year = "1994"
}

@article{Rajagopal:1994zj,
    author = "Rajagopal, Mohan and Romani, Roger W.",
    title = "{Ultralow frequency gravitational radiation from massive black hole binaries}",
    eprint = "astro-ph/9412038",
    archivePrefix = "arXiv",
    doi = "10.1086/175813",
    journal = "Astrophys. J.",
    volume = "446",
    pages = "543--549",
    year = "1995"
}

@article{Jaffe:2002rt,
    author = "Jaffe, Andrew H. and Backer, Donald C.",
    title = "{Gravitational waves probe the coalescence rate of massive black hole binaries}",
    eprint = "astro-ph/0210148",
    archivePrefix = "arXiv",
    doi = "10.1086/345443",
    journal = "Astrophys. J.",
    volume = "583",
    pages = "616--631",
    year = "2003"
}

@article{Wyithe:2002ep,
    author = "Wyithe, J. Stuart B. and Loeb, Abraham",
    title = "{Low - frequency gravitational waves from massive black hole binaries: Predictions for LISA and pulsar timing arrays}",
    eprint = "astro-ph/0211556",
    archivePrefix = "arXiv",
    doi = "10.1086/375187",
    journal = "Astrophys. J.",
    volume = "590",
    pages = "691--706",
    year = "2003"
}

@article{Sesana:2004sp,
    author = "Sesana, Alberto and Haardt, Francesco and Madau, Piero and Volonteri, Marta",
    title = "{Low - frequency gravitational radiation from coalescing massive black hole binaries in hierarchical cosmologies}",
    eprint = "astro-ph/0401543",
    archivePrefix = "arXiv",
    doi = "10.1086/422185",
    journal = "Astrophys. J.",
    volume = "611",
    pages = "623--632",
    year = "2004"
}

@article{Sesana:2008mz,
    author = "Sesana, Alberto and Vecchio, Alberto and Colacino, Carlo Nicola",
    title = "{The stochastic gravitational-wave background from massive black hole binary systems: implications for observations with Pulsar Timing Arrays}",
    eprint = "0804.4476",
    archivePrefix = "arXiv",
    primaryClass = "astro-ph",
    doi = "10.1111/j.1365-2966.2008.13682.x",
    journal = "Mon. Not. Roy. Astron. Soc.",
    volume = "390",
    pages = "192",
    year = "2008"
}

@article{Burke-Spolaor:2018bvk,
    author = "Burke-Spolaor, Sarah and others",
    title = "{The Astrophysics of Nanohertz Gravitational Waves}",
    eprint = "1811.08826",
    archivePrefix = "arXiv",
    primaryClass = "astro-ph.HE",
    doi = "10.1007/s00159-019-0115-7",
    journal = "Astron. Astrophys. Rev.",
    volume = "27",
    number = "1",
    pages = "5",
    year = "2019"
}

@article{Middleton:2020asl,
    author = "Middleton, H. and Sesana, A. and Chen, S. and Vecchio, A. and Del Pozzo, W. and Rosado, P. A.",
    title = "{Retracted: Correction to: Massive black hole binary systems and the NANOGrav 12.5 yr results}",
    eprint = "2011.01246",
    archivePrefix = "arXiv",
    primaryClass = "astro-ph.HE",
    doi = "10.1093/mnras/stad1487",
    journal = "Mon. Not. Roy. Astron. Soc.",
    volume = "502",
    number = "1",
    pages = "L99--L103",
    year = "2021",
    note = "[Erratum: Mon.Not.Roy.Astron.Soc. 526, L34 (2023)]"
}

@article{NANOGrav:2023hfp,
    author = "Agazie, Gabriella and others",
    collaboration = "NANOGrav",
    title = "{The NANOGrav 15 yr Data Set: Constraints on Supermassive Black Hole Binaries from the Gravitational-wave Background}",
    eprint = "2306.16220",
    archivePrefix = "arXiv",
    primaryClass = "astro-ph.HE",
    doi = "10.3847/2041-8213/ace18b",
    journal = "Astrophys. J. Lett.",
    volume = "952",
    number = "2",
    pages = "L37",
    year = "2023"
}

@ARTICLE{1980Natur.287..307B,
       author = {{Begelman}, M.~C. and {Blandford}, R.~D. and {Rees}, M.~J.},
        title = "{Massive black hole binaries in active galactic nuclei}",
        journal = "Nature",
         year = 1980,
        month = sep,
       volume = {287},
       number = {5780},
        pages = {307-309},
          doi = {10.1038/287307a0},
       adsurl = {https://ui.adsabs.harvard.edu/abs/1980Natur.287..307B}
}

@article{Milosavljevic:2002ht,
    author = "Milosavljevic, Milos and Merritt, David",
    editor = "Centrella, J. M.",
    title = "{The Final parsec problem}",
    eprint = "astro-ph/0212270",
    archivePrefix = "arXiv",
    doi = "10.1063/1.1629432",
    journal = "AIP Conf. Proc.",
    volume = "686",
    number = "1",
    pages = "201--210",
    year = "2003"
}

@article{NANOGrav:2023hvm,
    author = "Afzal, Adeela and others",
    collaboration = "NANOGrav",
    title = "{The NANOGrav 15 yr Data Set: Search for Signals from New Physics}",
    eprint = "2306.16219",
    archivePrefix = "arXiv",
    primaryClass = "astro-ph.HE",
    reportNumber = "FERMILAB-PUB-23-589-T",
    doi = "10.3847/2041-8213/acdc91",
    journal = "Astrophys. J. Lett.",
    volume = "951",
    number = "1",
    pages = "L11",
    year = "2023",
    note = "[Erratum: Astrophys.J.Lett. 971, L27 (2024), Erratum: Astrophys.J. 971, L27 (2024)]"
}

@article{Bringmann:2023opz,
    author = "Bringmann, Torsten and Depta, Paul Frederik and Konstandin, Thomas and Schmidt-Hoberg, Kai and Tasillo, Carlo",
    title = "{Does NANOGrav observe a dark sector phase transition?}",
    eprint = "2306.09411",
    archivePrefix = "arXiv",
    primaryClass = "astro-ph.CO",
    reportNumber = "DESY-23-077",
    doi = "10.1088/1475-7516/2023/11/053",
    journal = "JCAP",
    volume = "11",
    pages = "053",
    year = "2023"
}

@article{Winkler:2024olr,
    author = "Winkler, Martin Wolfgang and Freese, Katherine",
    title = "{Origin of the stochastic gravitational wave background: First-order phase transition versus black hole mergers}",
    eprint = "2401.13729",
    archivePrefix = "arXiv",
    primaryClass = "astro-ph.CO",
    reportNumber = "NORDITA-2024-002, UT-WI-02-2024",
    doi = "10.1103/PhysRevD.111.083509",
    journal = "Phys. Rev. D",
    volume = "111",
    number = "8",
    pages = "083509",
    year = "2025"
}

@article{Ellis:2020ena,
    author = "Ellis, John and Lewicki, Marek",
    title = "{Cosmic String Interpretation of NANOGrav Pulsar Timing Data}",
    eprint = "2009.06555",
    archivePrefix = "arXiv",
    primaryClass = "astro-ph.CO",
    reportNumber = "KCL-PH-TH/2020-53, CERN-TH-2020-150",
    doi = "10.1103/PhysRevLett.126.041304",
    journal = "Phys. Rev. Lett.",
    volume = "126",
    number = "4",
    pages = "041304",
    year = "2021"
}

@article{Ellis:2023tsl,
    author = "Ellis, John and Lewicki, Marek and Lin, Chunshan and Vaskonen, Ville",
    title = "{Cosmic superstrings revisited in light of NANOGrav 15-year data}",
    eprint = "2306.17147",
    archivePrefix = "arXiv",
    primaryClass = "astro-ph.CO",
    reportNumber = "KCL-PH-TH/2023-38, CERN-TH-2023-126, AION-REPORT/2023-07",
    doi = "10.1103/PhysRevD.108.103511",
    journal = "Phys. Rev. D",
    volume = "108",
    number = "10",
    pages = "103511",
    year = "2023"
}

@article{Kume:2024adn,
    author = "Kume, Jun'ya and Hindmarsh, Mark",
    title = "{Revised bounds on local cosmic strings from NANOGrav observations}",
    eprint = "2404.02705",
    archivePrefix = "arXiv",
    primaryClass = "astro-ph.CO",
    reportNumber = "RESCEU-4/24, HIP-2024-11/TH",
    month = "4",
    year = "2024"
}

@article{Ferreira:2022zzo,
    author = "Ferreira, Ricardo Z. and Notari, Alessio and Pujolas, Oriol and Rompineve, Fabrizio",
    title = "{Gravitational waves from domain walls in Pulsar Timing Array datasets}",
    eprint = "2204.04228",
    archivePrefix = "arXiv",
    primaryClass = "astro-ph.CO",
    reportNumber = "CERN-TH-2022-214",
    doi = "10.1088/1475-7516/2023/02/001",
    journal = "JCAP",
    volume = "02",
    pages = "001",
    year = "2023"
}

@article{Kitajima:2023cek,
    author = "Kitajima, Naoya and Lee, Junseok and Murai, Kai and Takahashi, Fuminobu and Yin, Wen",
    title = "{Gravitational waves from domain wall collapse, and application to nanohertz signals with QCD-coupled axions}",
    eprint = "2306.17146",
    archivePrefix = "arXiv",
    primaryClass = "hep-ph",
    reportNumber = "TU-1198",
    doi = "10.1016/j.physletb.2024.138586",
    journal = "Phys. Lett. B",
    volume = "851",
    pages = "138586",
    year = "2024"
}

@article{Bai:2023cqj,
    author = "Bai, Yang and Chen, Ting-Kuo and Korwar, Mrunal",
    title = "{QCD-collapsed domain walls: QCD phase transition and gravitational wave spectroscopy}",
    eprint = "2306.17160",
    archivePrefix = "arXiv",
    primaryClass = "hep-ph",
    doi = "10.1007/JHEP12(2023)194",
    journal = "JHEP",
    volume = "12",
    pages = "194",
    year = "2023"
}

@article{Benetti:2021uea,
    author = "Benetti, Micol and Graef, Leila Lobato and Vagnozzi, Sunny",
    title = "{Primordial gravitational waves from NANOGrav: A broken power-law approach}",
    eprint = "2111.04758",
    archivePrefix = "arXiv",
    primaryClass = "astro-ph.CO",
    doi = "10.1103/PhysRevD.105.043520",
    journal = "Phys. Rev. D",
    volume = "105",
    number = "4",
    pages = "043520",
    year = "2022"
}

@article{Schwaller:2015tja,
    author = "Schwaller, Pedro",
    title = "{Gravitational Waves from a Dark Phase Transition}",
    eprint = "1504.07263",
    archivePrefix = "arXiv",
    primaryClass = "hep-ph",
    reportNumber = "CERN-PH-TH-2015-093",
    doi = "10.1103/PhysRevLett.115.181101",
    journal = "Phys. Rev. Lett.",
    volume = "115",
    number = "18",
    pages = "181101",
    year = "2015"
}

@article{Jaeckel:2016jlh,
    author = "Jaeckel, Joerg and Khoze, Valentin V. and Spannowsky, Michael",
    title = "{Hearing the signal of dark sectors with gravitational wave detectors}",
    eprint = "1602.03901",
    archivePrefix = "arXiv",
    primaryClass = "hep-ph",
    reportNumber = "IPPP-16-12, DCPT-16-24",
    doi = "10.1103/PhysRevD.94.103519",
    journal = "Phys. Rev. D",
    volume = "94",
    number = "10",
    pages = "103519",
    year = "2016"
}

@article{Breitbach:2018ddu,
    author = "Breitbach, Moritz and Kopp, Joachim and Madge, Eric and Opferkuch, Toby and Schwaller, Pedro",
    title = "{Dark, Cold, and Noisy: Constraining Secluded Hidden Sectors with Gravitational Waves}",
    eprint = "1811.11175",
    archivePrefix = "arXiv",
    primaryClass = "hep-ph",
    reportNumber = "CERN-TH-2018-255, MITP/18-115",
    doi = "10.1088/1475-7516/2019/07/007",
    journal = "JCAP",
    volume = "07",
    pages = "007",
    year = "2019"
}

@article{Dent:2022bcd,
    author = "Dent, James B. and Dutta, Bhaskar and Ghosh, Sumit and Kumar, Jason and Runburg, Jack",
    title = "{Sensitivity to dark sector scales from gravitational wave signatures}",
    eprint = "2203.11736",
    archivePrefix = "arXiv",
    primaryClass = "hep-ph",
    reportNumber = "MI-HET-774, KIAS-P22016",
    doi = "10.1007/JHEP08(2022)300",
    journal = "JHEP",
    volume = "08",
    pages = "300",
    year = "2022"
}

@article{Morgante:2022zvc,
    author = "Morgante, Enrico and Ramberg, Nicklas and Schwaller, Pedro",
    title = "{Gravitational waves from dark SU(3) Yang-Mills theory}",
    eprint = "2210.11821",
    archivePrefix = "arXiv",
    primaryClass = "hep-ph",
    doi = "10.1103/PhysRevD.107.036010",
    journal = "Phys. Rev. D",
    volume = "107",
    number = "3",
    pages = "036010",
    year = "2023"
}

@article{Pasechnik:2023hwv,
    author = "Pasechnik, Roman and Reichert, Manuel and Sannino, Francesco and Wang, Zhi-Wei",
    title = "{Gravitational waves from composite dark sectors}",
    eprint = "2309.16755",
    archivePrefix = "arXiv",
    primaryClass = "hep-ph",
    doi = "10.1007/JHEP02(2024)159",
    journal = "JHEP",
    volume = "02",
    pages = "159",
    year = "2024"
}

@article{Addazi:2023jvg,
    author = "Addazi, Andrea and Cai, Yi-Fu and Marciano, Antonino and Visinelli, Luca",
    title = "{Have pulsar timing array methods detected a cosmological phase transition?}",
    eprint = "2306.17205",
    archivePrefix = "arXiv",
    primaryClass = "astro-ph.CO",
    reportNumber = "CA21106; CA21136",
    doi = "10.1103/PhysRevD.109.015028",
    journal = "Phys. Rev. D",
    volume = "109",
    number = "1",
    pages = "015028",
    year = "2024"
}

@article{Koutroulis:2023wit,
    author = "Koutroulis, Fotis and McCullough, Matthew and Merchand, Marco and Pokorski, Stefan and Sakurai, Kazuki",
    title = "{Phases of Pseudo-Nambu-Goldstone bosons}",
    eprint = "2309.15749",
    archivePrefix = "arXiv",
    primaryClass = "hep-ph",
    doi = "10.1007/JHEP05(2024)095",
    journal = "JHEP",
    volume = "05",
    pages = "095",
    year = "2024"
}

@article{DiBari:2023upq,
    author = "Di Bari, Pasquale and Rahat, Moinul Hossain",
    title = "{Split Majoron model confronts the NANOGrav signal and cosmological tensions}",
    eprint = "2307.03184",
    archivePrefix = "arXiv",
    primaryClass = "hep-ph",
    doi = "10.1103/PhysRevD.110.055019",
    journal = "Phys. Rev. D",
    volume = "110",
    number = "5",
    pages = "055019",
    year = "2024"
}

@article{Feng:2024pab,
    author = "Feng, Wan-Zhe and Li, Jinzheng and Nath, Pran",
    title = "{Cosmologically consistent analysis of gravitational waves from hidden sectors}",
    eprint = "2403.09558",
    archivePrefix = "arXiv",
    primaryClass = "hep-ph",
    doi = "10.1103/PhysRevD.110.015020",
    journal = "Phys. Rev. D",
    volume = "110",
    number = "1",
    pages = "015020",
    year = "2024"
}

@article{Han:2023olf,
    author = "Han, Chengcheng and Xie, Ke-Pan and Yang, Jin Min and Zhang, Mengchao",
    title = "{Self-interacting dark matter implied by nano-Hertz gravitational waves}",
    eprint = "2306.16966",
    archivePrefix = "arXiv",
    primaryClass = "hep-ph",
    doi = "10.1103/PhysRevD.109.115025",
    journal = "Phys. Rev. D",
    volume = "109",
    number = "11",
    pages = "115025",
    year = "2024"
}

@article{Li:2023bxy,
    author = "Li, Shao-Ping and Xie, Ke-Pan",
    title = "{Collider test of nano-Hertz gravitational waves from pulsar timing arrays}",
    eprint = "2307.01086",
    archivePrefix = "arXiv",
    primaryClass = "hep-ph",
    doi = "10.1103/PhysRevD.108.055018",
    journal = "Phys. Rev. D",
    volume = "108",
    number = "5",
    pages = "055018",
    year = "2023"
}

@article{Nakai:2020oit,
    author = "Nakai, Yuichiro and Suzuki, Motoo and Takahashi, Fuminobu and Yamada, Masaki",
    title = "{Gravitational Waves and Dark Radiation from Dark Phase Transition: Connecting NANOGrav Pulsar Timing Data and Hubble Tension}",
    eprint = "2009.09754",
    archivePrefix = "arXiv",
    primaryClass = "astro-ph.CO",
    reportNumber = "TU-1109; IPMU20-0100",
    doi = "10.1016/j.physletb.2021.136238",
    journal = "Phys. Lett. B",
    volume = "816",
    pages = "136238",
    year = "2021"
}

@article{Bai:2021ibt,
    author = "Bai, Yang and Korwar, Mrunal",
    title = "{Cosmological constraints on first-order phase transitions}",
    eprint = "2109.14765",
    archivePrefix = "arXiv",
    primaryClass = "hep-ph",
    doi = "10.1103/PhysRevD.105.095015",
    journal = "Phys. Rev. D",
    volume = "105",
    number = "9",
    pages = "095015",
    year = "2022"
}

@article{Kamionkowski:1993fg,
    author = "Kamionkowski, Marc and Kosowsky, Arthur and Turner, Michael S.",
    title = "{Gravitational radiation from first order phase transitions}",
    eprint = "astro-ph/9310044",
    archivePrefix = "arXiv",
    reportNumber = "IASSNS-HEP-93-44, FERMILAB-PUB-93-235-A",
    doi = "10.1103/PhysRevD.49.2837",
    journal = "Phys. Rev. D",
    volume = "49",
    pages = "2837--2851",
    year = "1994"
}

@article{Apreda:2001us,
    author = "Apreda, Riccardo and Maggiore, Michele and Nicolis, Alberto and Riotto, Antonio",
    title = "{Gravitational waves from electroweak phase transitions}",
    eprint = "gr-qc/0107033",
    archivePrefix = "arXiv",
    reportNumber = "UGVA-DPT-07-1096",
    doi = "10.1016/S0550-3213(02)00264-X",
    journal = "Nucl. Phys. B",
    volume = "631",
    pages = "342--368",
    year = "2002"
}

@article{Grojean:2006bp,
    author = "Grojean, Christophe and Servant, Geraldine",
    title = "{Gravitational Waves from Phase Transitions at the Electroweak Scale and Beyond}",
    eprint = "hep-ph/0607107",
    archivePrefix = "arXiv",
    reportNumber = "CERN-PH-TH-2006-125",
    doi = "10.1103/PhysRevD.75.043507",
    journal = "Phys. Rev. D",
    volume = "75",
    pages = "043507",
    year = "2007"
}

@article{Ashoorioon:2009nf,
    author = "Ashoorioon, A. and Konstandin, T.",
    title = "{Strong electroweak phase transitions without collider traces}",
    eprint = "0904.0353",
    archivePrefix = "arXiv",
    primaryClass = "hep-ph",
    reportNumber = "UAB-FT-666, MCTP-09-10",
    doi = "10.1088/1126-6708/2009/07/086",
    journal = "JHEP",
    volume = "07",
    pages = "086",
    year = "2009"
}

@article{Kakizaki:2015wua,
    author = "Kakizaki, Mitsuru and Kanemura, Shinya and Matsui, Toshinori",
    title = "{Gravitational waves as a probe of extended scalar sectors with the first order electroweak phase transition}",
    eprint = "1509.08394",
    archivePrefix = "arXiv",
    primaryClass = "hep-ph",
    reportNumber = "UT-HET-106",
    doi = "10.1103/PhysRevD.92.115007",
    journal = "Phys. Rev. D",
    volume = "92",
    number = "11",
    pages = "115007",
    year = "2015"
}

@article{Vaskonen:2016yiu,
    author = "Vaskonen, Ville",
    title = "{Electroweak baryogenesis and gravitational waves from a real scalar singlet}",
    eprint = "1611.02073",
    archivePrefix = "arXiv",
    primaryClass = "hep-ph",
    doi = "10.1103/PhysRevD.95.123515",
    journal = "Phys. Rev. D",
    volume = "95",
    number = "12",
    pages = "123515",
    year = "2017"
}

@article{Dorsch:2016nrg,
    author = "Dorsch, G. C. and Huber, S. J. and Konstandin, T. and No, J. M.",
    title = "{A Second Higgs Doublet in the Early Universe: Baryogenesis and Gravitational Waves}",
    eprint = "1611.05874",
    archivePrefix = "arXiv",
    primaryClass = "hep-ph",
    reportNumber = "DESY-16-213",
    doi = "10.1088/1475-7516/2017/05/052",
    journal = "JCAP",
    volume = "05",
    pages = "052",
    year = "2017"
}

@article{Beniwal:2017eik,
    author = "Beniwal, Ankit and Lewicki, Marek and Wells, James D. and White, Martin and Williams, Anthony G.",
    title = "{Gravitational wave, collider and dark matter signals from a scalar singlet electroweak baryogenesis}",
    eprint = "1702.06124",
    archivePrefix = "arXiv",
    primaryClass = "hep-ph",
    reportNumber = "ADP-17-08-T1014, ADP--17--08-T1014",
    doi = "10.1007/JHEP08(2017)108",
    journal = "JHEP",
    volume = "08",
    pages = "108",
    year = "2017"
}

@Article{LISA:2017pwj,
  author        = {Amaro-Seoane, Pau and others},
  title         = {Laser Interferometer Space Antenna},
  year          = {2017},
  month         = {2},
  archiveprefix = {arXiv},
  collaboration = {LISA},
  eprint        = {1702.00786},
  primaryclass  = {astro-ph.IM},
}

@article{LISA:2022kgy,
    author = "Arun, K. G. and others",
    collaboration = "LISA",
    title = "{New horizons for fundamental physics with LISA}",
    eprint = "2205.01597",
    archivePrefix = "arXiv",
    primaryClass = "gr-qc",
    doi = "10.1007/s41114-022-00036-9",
    journal = "Living Rev. Rel.",
    volume = "25",
    number = "1",
    pages = "4",
    year = "2022"
}

@article{Caprini:2015zlo,
    author = "Caprini, Chiara and others",
    title = "{Science with the space-based interferometer eLISA. II: Gravitational waves from cosmological phase transitions}",
    eprint = "1512.06239",
    archivePrefix = "arXiv",
    primaryClass = "astro-ph.CO",
    reportNumber = "DESY-15-246",
    doi = "10.1088/1475-7516/2016/04/001",
    journal = "JCAP",
    volume = "04",
    pages = "001",
    year = "2016"
}

@article{Caprini:2018mtu,
    author = "Caprini, Chiara and Figueroa, Daniel G.",
    title = "{Cosmological Backgrounds of Gravitational Waves}",
    eprint = "1801.04268",
    archivePrefix = "arXiv",
    primaryClass = "astro-ph.CO",
    doi = "10.1088/1361-6382/aac608",
    journal = "Class. Quant. Grav.",
    volume = "35",
    number = "16",
    pages = "163001",
    year = "2018"
}

@article{Caprini:2019egz,
    author = "Caprini, Chiara and others",
    title = "{Detecting gravitational waves from cosmological phase transitions with LISA: an update}",
    eprint = "1910.13125",
    archivePrefix = "arXiv",
    primaryClass = "astro-ph.CO",
    reportNumber = "DESY-19-159, IPPP/19/27, HIP-2019-14/TH, MITP/19-066, IFT-UAM/CSIC-19-139",
    doi = "10.1088/1475-7516/2020/03/024",
    journal = "JCAP",
    volume = "03",
    pages = "024",
    year = "2020"
}

@article{Kosowsky:1992rz,
    author = "Kosowsky, Arthur and Turner, Michael S. and Watkins, Richard",
    title = "{Gravitational waves from first order cosmological phase transitions}",
    reportNumber = "FERMILAB-PUB-91-333-A-REV, FERMILAB-PUB-91-333-A",
    doi = "10.1103/PhysRevLett.69.2026",
    journal = "Phys. Rev. Lett.",
    volume = "69",
    pages = "2026--2029",
    year = "1992"
}

@article{Kosowsky:1992vn,
    author = "Kosowsky, Arthur and Turner, Michael S.",
    title = "{Gravitational radiation from colliding vacuum bubbles: envelope approximation to many bubble collisions}",
    eprint = "astro-ph/9211004",
    archivePrefix = "arXiv",
    reportNumber = "FERMILAB-PUB-92-295-A",
    doi = "10.1103/PhysRevD.47.4372",
    journal = "Phys. Rev. D",
    volume = "47",
    pages = "4372--4391",
    year = "1993"
}

@article{Huber:2008hg,
    author = "Huber, Stephan J. and Konstandin, Thomas",
    title = "{Gravitational Wave Production by Collisions: More Bubbles}",
    eprint = "0806.1828",
    archivePrefix = "arXiv",
    primaryClass = "hep-ph",
    doi = "10.1088/1475-7516/2008/09/022",
    journal = "JCAP",
    volume = "09",
    pages = "022",
    year = "2008"
}

@article{Weir:2016tov,
    author = "Weir, David J.",
    title = "{Revisiting the envelope approximation: gravitational waves from bubble collisions}",
    eprint = "1604.08429",
    archivePrefix = "arXiv",
    primaryClass = "astro-ph.CO",
    doi = "10.1103/PhysRevD.93.124037",
    journal = "Phys. Rev. D",
    volume = "93",
    number = "12",
    pages = "124037",
    year = "2016"
}

@article{Hindmarsh:2013xza,
    author = "Hindmarsh, Mark and Huber, Stephan J. and Rummukainen, Kari and Weir, David J.",
    title = "{Gravitational waves from the sound of a first order phase transition}",
    eprint = "1304.2433",
    archivePrefix = "arXiv",
    primaryClass = "hep-ph",
    reportNumber = "HIP-2013-07-TH",
    doi = "10.1103/PhysRevLett.112.041301",
    journal = "Phys. Rev. Lett.",
    volume = "112",
    pages = "041301",
    year = "2014"
}

@article{Hindmarsh:2015qta,
    author = "Hindmarsh, Mark and Huber, Stephan J. and Rummukainen, Kari and Weir, David J.",
    title = "{Numerical simulations of acoustically generated gravitational waves at a first order phase transition}",
    eprint = "1504.03291",
    archivePrefix = "arXiv",
    primaryClass = "astro-ph.CO",
    reportNumber = "HIP-2015-13-TH",
    doi = "10.1103/PhysRevD.92.123009",
    journal = "Phys. Rev. D",
    volume = "92",
    number = "12",
    pages = "123009",
    year = "2015"
}

@article{Hindmarsh:2017gnf,
    author = "Hindmarsh, Mark and Huber, Stephan J. and Rummukainen, Kari and Weir, David J.",
    title = "{Shape of the acoustic gravitational wave power spectrum from a first order phase transition}",
    eprint = "1704.05871",
    archivePrefix = "arXiv",
    primaryClass = "astro-ph.CO",
    reportNumber = "HIP-2017-02-TH, HIP-2017-02/TH",
    doi = "10.1103/PhysRevD.96.103520",
    journal = "Phys. Rev. D",
    volume = "96",
    number = "10",
    pages = "103520",
    year = "2017",
    note = "[Erratum: Phys.Rev.D 101, 089902 (2020)]"
}

@article{Steinhardt:1981ct,
    author = "Steinhardt, Paul Joseph",
    title = "{Relativistic Detonation Waves and Bubble Growth in False Vacuum Decay}",
    reportNumber = "UPR-0181T",
    doi = "10.1103/PhysRevD.25.2074",
    journal = "Phys. Rev. D",
    volume = "25",
    pages = "2074",
    year = "1982"
}

@article{Espinosa:2010hh,
    author = "Espinosa, Jose R. and Konstandin, Thomas and No, Jose M. and Servant, Geraldine",
    title = "{Energy Budget of Cosmological First-order Phase Transitions}",
    eprint = "1004.4187",
    archivePrefix = "arXiv",
    primaryClass = "hep-ph",
    reportNumber = "CERN-PH-TH-2010-027",
    doi = "10.1088/1475-7516/2010/06/028",
    journal = "JCAP",
    volume = "06",
    pages = "028",
    year = "2010"
}

@article{Bodeker:2017cim,
    author = "Bodeker, Dietrich and Moore, Guy D.",
    title = "{Electroweak Bubble Wall Speed Limit}",
    eprint = "1703.08215",
    archivePrefix = "arXiv",
    primaryClass = "hep-ph",
    doi = "10.1088/1475-7516/2017/05/025",
    journal = "JCAP",
    volume = "05",
    pages = "025",
    year = "2017"
}

@article{Fairbairn:2019xog,
    author = "Fairbairn, Malcolm and Hardy, Edward and Wickens, Alastair",
    title = "{Hearing without seeing: gravitational waves from hot and cold hidden sectors}",
    eprint = "1901.11038",
    archivePrefix = "arXiv",
    primaryClass = "hep-ph",
    reportNumber = "KCL-PH-TH/2019-12",
    doi = "10.1007/JHEP07(2019)044",
    journal = "JHEP",
    volume = "07",
    pages = "044",
    year = "2019"
}

@article{Ellis:2020awk,
    author = "Ellis, John and Lewicki, Marek and No, Jos\'e Miguel",
    title = "{Gravitational waves from first-order cosmological phase transitions: lifetime of the sound wave source}",
    eprint = "2003.07360",
    archivePrefix = "arXiv",
    primaryClass = "hep-ph",
    reportNumber = "KCL-PH-TH/2020-04, CERN-TH-2020-016, IFT-UAM/CSIC-20-35",
    doi = "10.1088/1475-7516/2020/07/050",
    journal = "JCAP",
    volume = "07",
    pages = "050",
    year = "2020"
}

@article{Guo:2020grp,
    author = "Guo, Huai-Ke and Sinha, Kuver and Vagie, Daniel and White, Graham",
    title = "{Phase Transitions in an Expanding Universe: Stochastic Gravitational Waves in Standard and Non-Standard Histories}",
    eprint = "2007.08537",
    archivePrefix = "arXiv",
    primaryClass = "hep-ph",
    doi = "10.1088/1475-7516/2021/01/001",
    journal = "JCAP",
    volume = "01",
    pages = "001",
    year = "2021"
}

@article{Kosowsky:2001xp,
    author = "Kosowsky, Arthur and Mack, Andrew and Kahniashvili, Tinatin",
    title = "{Gravitational radiation from cosmological turbulence}",
    eprint = "astro-ph/0111483",
    archivePrefix = "arXiv",
    reportNumber = "RAP-334",
    doi = "10.1103/PhysRevD.66.024030",
    journal = "Phys. Rev. D",
    volume = "66",
    pages = "024030",
    year = "2002"
}

@article{Dolgov:2002ra,
    author = "Dolgov, Alexander D. and Grasso, Dario and Nicolis, Alberto",
    title = "{Relic backgrounds of gravitational waves from cosmic turbulence}",
    eprint = "astro-ph/0206461",
    archivePrefix = "arXiv",
    doi = "10.1103/PhysRevD.66.103505",
    journal = "Phys. Rev. D",
    volume = "66",
    pages = "103505",
    year = "2002"
}

@article{Gogoberidze:2007an,
    author = "Gogoberidze, Grigol and Kahniashvili, Tina and Kosowsky, Arthur",
    title = "{The Spectrum of Gravitational Radiation from Primordial Turbulence}",
    eprint = "0705.1733",
    archivePrefix = "arXiv",
    primaryClass = "astro-ph",
    doi = "10.1103/PhysRevD.76.083002",
    journal = "Phys. Rev. D",
    volume = "76",
    pages = "083002",
    year = "2007"
}

@article{Caprini:2009yp,
    author = "Caprini, Chiara and Durrer, Ruth and Servant, Geraldine",
    title = "{The stochastic gravitational wave background from turbulence and magnetic fields generated by a first-order phase transition}",
    eprint = "0909.0622",
    archivePrefix = "arXiv",
    primaryClass = "astro-ph.CO",
    doi = "10.1088/1475-7516/2009/12/024",
    journal = "JCAP",
    volume = "12",
    pages = "024",
    year = "2009"
}

@article{Niksa:2018ofa,
    author = {Niksa, Peter and Schlederer, Martin and Sigl, G\"unter},
    title = "{Gravitational Waves produced by Compressible MHD Turbulence from Cosmological Phase Transitions}",
    eprint = "1803.02271",
    archivePrefix = "arXiv",
    primaryClass = "astro-ph.CO",
    doi = "10.1088/1361-6382/aac89c",
    journal = "Class. Quant. Grav.",
    volume = "35",
    number = "14",
    pages = "144001",
    year = "2018"
}

@article{Ghosh:2023aum,
    author = "Ghosh, Tathagata and Ghoshal, Anish and Guo, Huai-Ke and Hajkarim, Fazlollah and King, Stephen F. and Sinha, Kuver and Wang, Xin and White, Graham",
    title = "{Did we hear the sound of the Universe boiling? Analysis using the full fluid velocity profiles and NANOGrav 15-year data}",
    eprint = "2307.02259",
    archivePrefix = "arXiv",
    primaryClass = "astro-ph.HE",
    doi = "10.1088/1475-7516/2024/05/100",
    journal = "JCAP",
    volume = "05",
    pages = "100",
    year = "2024"
}

@article{Anderson:1991zb,
    author = "Anderson, Greg W. and Hall, Lawrence J.",
    title = "{The Electroweak phase transition and baryogenesis}",
    reportNumber = "LBL-31169, UCB-PTH-91-41",
    doi = "10.1103/PhysRevD.45.2685",
    journal = "Phys. Rev. D",
    volume = "45",
    pages = "2685--2698",
    year = "1992"
}

@article{Espinosa:1992kf,
    author = "Espinosa, J. R. and Quiros, M. and Zwirner, F.",
    title = "{On the nature of the electroweak phase transition}",
    eprint = "hep-ph/9212248",
    archivePrefix = "arXiv",
    reportNumber = "CERN-TH-6577-92, IEM-FT-58-92",
    doi = "10.1016/0370-2693(93)90450-V",
    journal = "Phys. Lett. B",
    volume = "314",
    pages = "206--216",
    year = "1993"
}

@article{Delaunay:2007wb,
    author = "Delaunay, Cedric and Grojean, Christophe and Wells, James D.",
    title = "{Dynamics of Non-renormalizable Electroweak Symmetry Breaking}",
    eprint = "0711.2511",
    archivePrefix = "arXiv",
    primaryClass = "hep-ph",
    reportNumber = "CERN-PH-TH-2007-219, MCTP-07-31, SACLAY-T07-141",
    doi = "10.1088/1126-6708/2008/04/029",
    journal = "JHEP",
    volume = "04",
    pages = "029",
    year = "2008"
}

@article{Dolan:1973qd,
    author = "Dolan, L. and Jackiw, R.",
    title = "{Symmetry Behavior at Finite Temperature}",
    reportNumber = "MIT-CTP-406",
    doi = "10.1103/PhysRevD.9.3320",
    journal = "Phys. Rev. D",
    volume = "9",
    pages = "3320--3341",
    year = "1974"
}

@inproceedings{Quiros:1999jp,
    author = "Quiros, Mariano",
    title = "{Finite temperature field theory and phase transitions}",
    booktitle = "{ICTP Summer School in High-Energy Physics and Cosmology}",
    eprint = "hep-ph/9901312",
    archivePrefix = "arXiv",
    reportNumber = "IEM-FT-187-99",
    pages = "187--259",
    month = "1",
    year = "1999"
}

@book{Laine:2016hma,
    author = "Laine, Mikko and Vuorinen, Aleksi",
    title = "{Basics of Thermal Field Theory}",
    eprint = "1701.01554",
    archivePrefix = "arXiv",
    primaryClass = "hep-ph",
    doi = "10.1007/978-3-319-31933-9",
    publisher = "Springer",
    volume = "925",
    year = "2016"
}

@article{Parwani:1991gq,
    author = "Parwani, Rajesh R.",
    title = "{Resummation in a hot scalar field theory}",
    eprint = "hep-ph/9204216",
    archivePrefix = "arXiv",
    reportNumber = "ITP-SB-91-64",
    doi = "10.1103/PhysRevD.45.4695",
    journal = "Phys. Rev. D",
    volume = "45",
    pages = "4695",
    year = "1992",
    note = "[Erratum: Phys.Rev.D 48, 5965 (1993)]"
}

@article{Carrington:1991hz,
    author = "Carrington, M. E.",
    title = "{The Effective potential at finite temperature in the Standard Model}",
    reportNumber = "TPI-MINN-91-48-T-REV, TPI-MINN-91-48-T",
    doi = "10.1103/PhysRevD.45.2933",
    journal = "Phys. Rev. D",
    volume = "45",
    pages = "2933--2944",
    year = "1992"
}

@article{Kainulainen:2021eki,
    author = "Kainulainen, Kimmo and Koskivaara, Olli",
    title = "{Non-equilibrium dynamics of a scalar field with quantum backreaction}",
    eprint = "2105.09598",
    archivePrefix = "arXiv",
    primaryClass = "hep-ph",
    doi = "10.1007/JHEP12(2021)190",
    journal = "JHEP",
    volume = "12",
    pages = "190",
    year = "2021"
}

@article{Kainulainen:2019kyp,
    author = "Kainulainen, Kimmo and Keus, Venus and Niemi, Lauri and Rummukainen, Kari and Tenkanen, Tuomas V. I. and Vaskonen, Ville",
    title = "{On the validity of perturbative studies of the electroweak phase transition in the Two Higgs Doublet model}",
    eprint = "1904.01329",
    archivePrefix = "arXiv",
    primaryClass = "hep-ph",
    doi = "10.1007/JHEP06(2019)075",
    journal = "JHEP",
    volume = "06",
    pages = "075",
    year = "2019"
}

@article{Cline:2011mm,
    author = "Cline, James M. and Kainulainen, Kimmo and Trott, Michael",
    title = "{Electroweak Baryogenesis in Two Higgs Doublet Models and B meson anomalies}",
    eprint = "1107.3559",
    archivePrefix = "arXiv",
    primaryClass = "hep-ph",
    doi = "10.1007/JHEP11(2011)089",
    journal = "JHEP",
    volume = "11",
    pages = "089",
    year = "2011"
}

@article{Laine:2017hdk,
    author = "Laine, M. and Meyer, M. and Nardini, G.",
    title = "{Thermal phase transition with full 2-loop effective potential}",
    eprint = "1702.07479",
    archivePrefix = "arXiv",
    primaryClass = "hep-ph",
    doi = "10.1016/j.nuclphysb.2017.04.023",
    journal = "Nucl. Phys. B",
    volume = "920",
    pages = "565--600",
    year = "2017"
}

@article{Coleman:1977py,
    author = "Coleman, Sidney R.",
    title = "{The Fate of the False Vacuum. 1. Semiclassical Theory}",
    reportNumber = "HUTP-77-A004",
    doi = "10.1103/PhysRevD.16.1248",
    journal = "Phys. Rev. D",
    volume = "15",
    pages = "2929--2936",
    year = "1977",
    note = "[Erratum: Phys.Rev.D 16, 1248 (1977)]"
}

@article{Callan:1977pt,
    author = "Callan, Jr., Curtis G. and Coleman, Sidney R.",
    title = "{The Fate of the False Vacuum. 2. First Quantum Corrections}",
    reportNumber = "HUTP-77-A032",
    doi = "10.1103/PhysRevD.16.1762",
    journal = "Phys. Rev. D",
    volume = "16",
    pages = "1762--1768",
    year = "1977"
}

@article{Linde:1980tt,
    author = "Linde, Andrei D.",
    title = "{Fate of the False Vacuum at Finite Temperature: Theory and Applications}",
    reportNumber = "LEBEDEV-80-92",
    doi = "10.1016/0370-2693(81)90281-1",
    journal = "Phys. Lett. B",
    volume = "100",
    pages = "37--40",
    year = "1981"
}

@article{Linde:1981zj,
    author = "Linde, Andrei D.",
    title = "{Decay of the False Vacuum at Finite Temperature}",
    reportNumber = "LEBEDEV-81-265",
    doi = "10.1016/0550-3213(83)90072-X",
    journal = "Nucl. Phys. B",
    volume = "216",
    pages = "421",
    year = "1983",
    note = "[Erratum: Nucl.Phys.B 223, 544 (1983)]"
}

@article{Guada:2020xnz,
    author = "Guada, Victor and Nemev\v{s}ek, Miha and Pintar, Matev\v{z}",
    title = "{FindBounce: Package for multi-field bounce actions}",
    eprint = "2002.00881",
    archivePrefix = "arXiv",
    primaryClass = "hep-ph",
    doi = "10.1016/j.cpc.2020.107480",
    journal = "Comput. Phys. Commun.",
    volume = "256",
    pages = "107480",
    year = "2020"
}

@article{Ellis:2018mja,
    author = "Ellis, John and Lewicki, Marek and No, Jos\'e Miguel",
    title = "{On the Maximal Strength of a First-Order Electroweak Phase Transition and its Gravitational Wave Signal}",
    eprint = "1809.08242",
    archivePrefix = "arXiv",
    primaryClass = "hep-ph",
    reportNumber = "KCL-PH-TH/2018-46, CERN-TH/2018-197, IFT-UAM/CSIC-18-94, CERN-TH-2018-197",
    doi = "10.1088/1475-7516/2019/04/003",
    journal = "JCAP",
    volume = "04",
    pages = "003",
    year = "2019"
}

@article{CMS:2018yfx,
    author = "Sirunyan, Albert M and others",
    collaboration = "CMS",
    title = "{Search for invisible decays of a Higgs boson produced through vector boson fusion in proton-proton collisions at $\sqrt{s} =$ 13 TeV}",
    eprint = "1809.05937",
    archivePrefix = "arXiv",
    primaryClass = "hep-ex",
    reportNumber = "CMS-HIG-17-023, CERN-EP-2018-139",
    doi = "10.1016/j.physletb.2019.04.025",
    journal = "Phys. Lett. B",
    volume = "793",
    pages = "520--551",
    year = "2019"
}

@article{Depta:2020zbh,
    author = "Depta, Paul Frederik and Hufnagel, Marco and Schmidt-Hoberg, Kai",
    title = "{Updated BBN constraints on electromagnetic decays of MeV-scale particles}",
    eprint = "2011.06519",
    archivePrefix = "arXiv",
    primaryClass = "hep-ph",
    reportNumber = "DESY-20-160, DESY 20-160, ULB-TH/20-15",
    doi = "10.1088/1475-7516/2021/04/011",
    journal = "JCAP",
    volume = "04",
    pages = "011",
    year = "2021"
}

@article{Planck:2018vyg,
    author = "Aghanim, N. and others",
    collaboration = "Planck",
    title = "{Planck 2018 results. VI. Cosmological parameters}",
    eprint = "1807.06209",
    archivePrefix = "arXiv",
    primaryClass = "astro-ph.CO",
    doi = "10.1051/0004-6361/201833910",
    journal = "Astron. Astrophys.",
    volume = "641",
    pages = "A6",
    year = "2020",
    note = "[Erratum: Astron.Astrophys. 652, C4 (2021)]"
}

@article{Kawasaki:2004qu,
    author = "Kawasaki, Masahiro and Kohri, Kazunori and Moroi, Takeo",
    title = "{Big-Bang nucleosynthesis and hadronic decay of long-lived massive particles}",
    eprint = "astro-ph/0408426",
    archivePrefix = "arXiv",
    reportNumber = "ICRR-REPORT-508-2004-6, OU-TAP-234, TU-727",
    doi = "10.1103/PhysRevD.71.083502",
    journal = "Phys. Rev. D",
    volume = "71",
    pages = "083502",
    year = "2005"
}

@article{Hannestad:2004px,
    author = "Hannestad, Steen",
    title = "{What is the lowest possible reheating temperature?}",
    eprint = "astro-ph/0403291",
    archivePrefix = "arXiv",
    doi = "10.1103/PhysRevD.70.043506",
    journal = "Phys. Rev. D",
    volume = "70",
    pages = "043506",
    year = "2004"
}

@article{Hasegawa:2019jsa,
    author = "Hasegawa, Takuya and Hiroshima, Nagisa and Kohri, Kazunori and Hansen, Rasmus S. L. and Tram, Thomas and Hannestad, Steen",
    title = "{MeV-scale reheating temperature and thermalization of oscillating neutrinos by radiative and hadronic decays of massive particles}",
    eprint = "1908.10189",
    archivePrefix = "arXiv",
    primaryClass = "hep-ph",
    reportNumber = "KEK-TH-2149, KEK-Cosmo-242, RIKEN-iTHEMS-Report-19, IPMU19-0120",
    doi = "10.1088/1475-7516/2019/12/012",
    journal = "JCAP",
    volume = "12",
    pages = "012",
    year = "2019"
}

@article{Ashoorioon:2022raz,
    author = "Ashoorioon, Amjad and Rezazadeh, Kazem and Rostami, Abasalt",
    title = "{NANOGrav signal from the end of inflation and the LIGO mass and heavier primordial black holes}",
    eprint = "2202.01131",
    archivePrefix = "arXiv",
    primaryClass = "astro-ph.CO",
    reportNumber = "IPM/P-2022/06",
    doi = "10.1016/j.physletb.2022.137542",
    journal = "Phys. Lett. B",
    volume = "835",
    pages = "137542",
    year = "2022"
}

@article{Chowdhury:2023xvy,
    author = "Chowdhury, Debtosh and Hait, Arpan and Mohanty, Subhendra and Prakash, Suraj",
    title = "{Ultralight dark matter explanation of NANOGrav observations}",
    eprint = "2311.10148",
    archivePrefix = "arXiv",
    primaryClass = "hep-ph",
    doi = "10.1103/PhysRevD.110.083023",
    journal = "Phys. Rev. D",
    volume = "110",
    number = "8",
    pages = "083023",
    year = "2024"
}

@article{Fujikura:2023lkn,
    author = "Fujikura, Kohei and Girmohanta, Sudhakantha and Nakai, Yuichiro and Suzuki, Motoo",
    title = "{NANOGrav signal from a dark conformal phase transition}",
    eprint = "2306.17086",
    archivePrefix = "arXiv",
    primaryClass = "hep-ph",
    reportNumber = "UT-Komaba/23-6",
    doi = "10.1016/j.physletb.2023.138203",
    journal = "Phys. Lett. B",
    volume = "846",
    pages = "138203",
    year = "2023"
}

@article{Chatterjee:2022pxf,
    author = "Chatterjee, Arindam and Datta, AseshKrishna and Roy, Subhojit",
    title = "{Electroweak phase transition in the Z$_{3}$-invariant NMSSM: Implications of LHC and Dark matter searches and prospects of detecting the gravitational waves}",
    eprint = "2202.12476",
    archivePrefix = "arXiv",
    primaryClass = "hep-ph",
    reportNumber = "HRI-RECAPP-2022-001",
    doi = "10.1007/JHEP06(2022)108",
    journal = "JHEP",
    volume = "06",
    pages = "108",
    year = "2022"
}

@article{Athron:2023xlk,
    author = "Athron, Peter and Bal\'azs, Csaba and Fowlie, Andrew and Morris, Lachlan and Wu, Lei",
    title = "{Cosmological phase transitions: From perturbative particle physics to gravitational waves}",
    eprint = "2305.02357",
    archivePrefix = "arXiv",
    primaryClass = "hep-ph",
    doi = "10.1016/j.ppnp.2023.104094",
    journal = "Prog. Part. Nucl. Phys.",
    volume = "135",
    pages = "104094",
    year = "2024"
}

@article{Salvio:2023ynn,
    author = "Salvio, Alberto",
    title = "{Supercooling in radiative symmetry breaking: theory extensions, gravitational wave detection and primordial black holes}",
    eprint = "2307.04694",
    archivePrefix = "arXiv",
    primaryClass = "hep-ph",
    doi = "10.1088/1475-7516/2023/12/046",
    journal = "JCAP",
    volume = "12",
    pages = "046",
    year = "2023"
}

@article{Salvio:2023blb,
    author = "Salvio, Alberto",
    title = "{Pulsar timing arrays and primordial black holes from a supercooled phase transition}",
    eprint = "2312.04628",
    archivePrefix = "arXiv",
    primaryClass = "hep-ph",
    doi = "10.1016/j.physletb.2024.138639",
    journal = "Phys. Lett. B",
    volume = "852",
    pages = "138639",
    year = "2024"
}

@article{Aarts:2023vsf,
    author = "Aarts, Gert and others",
    title = "{Phase Transitions in Particle Physics}: {Results and Perspectives from Lattice Quantum Chromo-Dynamics}",
    eprint = "2301.04382",
    archivePrefix = "arXiv",
    primaryClass = "hep-lat",
    doi = "10.1016/j.ppnp.2023.104070",
    journal = "Prog. Part. Nucl. Phys.",
    volume = "133",
    pages = "104070",
    year = "2023"
}

@article{Freese:2022qrl,
    author = "Freese, Katherine and Winkler, Martin Wolfgang",
    title = "{Have pulsar timing arrays detected the hot big bang: Gravitational waves from strong first order phase transitions in the early Universe}",
    eprint = "2208.03330",
    archivePrefix = "arXiv",
    primaryClass = "astro-ph.CO",
    reportNumber = "UTTG 11-2022, NORDITA 2022-056",
    doi = "10.1103/PhysRevD.106.103523",
    journal = "Phys. Rev. D",
    volume = "106",
    number = "10",
    pages = "103523",
    year = "2022"
}

@article{Guth:1982pn,
    author = "Guth, Alan H. and Weinberg, Erick J.",
    title = "{Could the Universe Have Recovered from a Slow First Order Phase Transition?}",
    reportNumber = "MIT-CTP-950",
    doi = "10.1016/0550-3213(83)90307-3",
    journal = "Nucl. Phys. B",
    volume = "212",
    pages = "321--364",
    year = "1983"
}

@article{Baldes:2018emh,
    author = "Baldes, Iason and Garcia-Cely, Camilo",
    title = "{Strong gravitational radiation from a simple dark matter model}",
    eprint = "1809.01198",
    archivePrefix = "arXiv",
    primaryClass = "hep-ph",
    reportNumber = "DESY 18-155, DESY-18-155",
    doi = "10.1007/JHEP05(2019)190",
    journal = "JHEP",
    volume = "05",
    pages = "190",
    year = "2019"
}

@article{Madge:2018gfl,
    author = "Madge, Eric and Schwaller, Pedro",
    title = "{Leptophilic dark matter from gauged lepton number: Phenomenology and gravitational wave signatures}",
    eprint = "1809.09110",
    archivePrefix = "arXiv",
    primaryClass = "hep-ph",
    reportNumber = "MITP/18-088",
    doi = "10.1007/JHEP02(2019)048",
    journal = "JHEP",
    volume = "02",
    pages = "048",
    year = "2019"
}

@article{Ellis:2019oqb,
    author = "Ellis, John and Lewicki, Marek and No, Jos{\'e} Miguel and Vaskonen, Ville",
    title = "{Gravitational wave energy budget in strongly supercooled phase transitions}",
    eprint = "1903.09642",
    archivePrefix = "arXiv",
    primaryClass = "hep-ph",
    reportNumber = "KCL-PH-TH/2019-32, CERN-TH-2019-032, IFT-UAM/CSIC-19-32",
    doi = "10.1088/1475-7516/2019/06/024",
    journal = "JCAP",
    volume = "06",
    pages = "024",
    year = "2019"
}

@article{Croon:2020cgk,
    author = "Croon, Djuna and Gould, Oliver and Schicho, Philipp and Tenkanen, Tuomas V. I. and White, Graham",
    title = "{Theoretical uncertainties for cosmological first-order phase transitions}",
    eprint = "2009.10080",
    archivePrefix = "arXiv",
    primaryClass = "hep-ph",
    reportNumber = "HIP-2020-26/TH",
    doi = "10.1007/JHEP04(2021)055",
    journal = "JHEP",
    volume = "04",
    pages = "055",
    year = "2021"
}

@article{Gould:2021ccf,
    author = "Gould, Oliver and Hirvonen, Joonas",
    title = "{Effective field theory approach to thermal bubble nucleation}",
    eprint = "2108.04377",
    archivePrefix = "arXiv",
    primaryClass = "hep-ph",
    reportNumber = "HIP-2020-19/TH",
    doi = "10.1103/PhysRevD.104.096015",
    journal = "Phys. Rev. D",
    volume = "104",
    number = "9",
    pages = "096015",
    year = "2021"
}

@article{Gould:2021dzl,
    author = "Gould, Oliver",
    title = "{Real scalar phase transitions: a nonperturbative analysis}",
    eprint = "2101.05528",
    archivePrefix = "arXiv",
    primaryClass = "hep-ph",
    reportNumber = "HIP-2021-2/TH",
    doi = "10.1007/JHEP04(2021)057",
    journal = "JHEP",
    volume = "04",
    pages = "057",
    year = "2021"
}

@article{Mitridate:2023oar,
    author = {Mitridate, Andrea and Wright, David and von Eckardstein, Richard and Schr{\"o}der, Tobias and Nay, Jonathan and Olum, Ken and Schmitz, Kai and Trickle, Tanner},
    title = "{PTArcade}",
    eprint = "2306.16377",
    archivePrefix = "arXiv",
    primaryClass = "hep-ph",
    reportNumber = "FERMILAB-PUB-23-588-T",
    month = "6",
    year = "2023"
}

@article{Wilks:1938dza,
    author = "Wilks, S. S.",
    title = "{The Large-Sample Distribution of the Likelihood Ratio for Testing Composite Hypotheses}",
    doi = "10.1214/aoms/1177732360",
    journal = "Annals Math. Statist.",
    volume = "9",
    number = "1",
    pages = "60--62",
    year = "1938"
}

@article{ParticleDataGroup:2024cfk,
    author = "{[Particle Data Group]} and Navas, S. and others",
    collaboration = "Particle Data Group",
    title = "{Review of particle physics}",
    doi = "10.1103/PhysRevD.110.030001",
    journal = "Phys. Rev. D",
    volume = "110",
    number = "3",
    pages = "030001",
    year = "2024"
}

@article{Winkler:2018qyg,
    author = "Winkler, Martin Wolfgang",
    title = "{Decay and detection of a light scalar boson mixing with the Higgs boson}",
    eprint = "1809.01876",
    archivePrefix = "arXiv",
    primaryClass = "hep-ph",
    reportNumber = "NORDITA-2018-087",
    doi = "10.1103/PhysRevD.99.015018",
    journal = "Phys. Rev. D",
    volume = "99",
    number = "1",
    pages = "015018",
    year = "2019"
}

@article{Fradette:2018hhl,
    author = "Fradette, Anthony and Pospelov, Maxim and Pradler, Josef and Ritz, Adam",
    title = "{Cosmological beam dump: constraints on dark scalars mixed with the Higgs boson}",
    eprint = "1812.07585",
    archivePrefix = "arXiv",
    primaryClass = "hep-ph",
    doi = "10.1103/PhysRevD.99.075004",
    journal = "Phys. Rev. D",
    volume = "99",
    number = "7",
    pages = "075004",
    year = "2019"
}

@article{Goudzovski:2022vbt,
    author = "Goudzovski, Evgueni and others",
    title = "{New physics searches at kaon and hyperon factories}",
    eprint = "2201.07805",
    archivePrefix = "arXiv",
    primaryClass = "hep-ph",
    reportNumber = "FERMILAB-PUB-22-057-T",
    doi = "10.1088/1361-6633/ac9cee",
    journal = "Rept. Prog. Phys.",
    volume = "86",
    number = "1",
    pages = "016201",
    year = "2023"
}

@inproceedings{Batell:2022dpx,
    author = "Batell, Brian and Blinov, Nikita and Hearty, Christopher and McGehee, Robert",
    title = "{Exploring Dark Sector Portals with High Intensity Experiments}",
    booktitle = "{Snowmass 2021}",
    eprint = "2207.06905",
    archivePrefix = "arXiv",
    primaryClass = "hep-ph",
    month = "7",
    year = "2022"
}

@article{CMS:2023sdw,
    author = "Tumasyan, Armen and others",
    collaboration = "CMS",
    title = "{A search for decays of the Higgs boson to invisible particles in events with a top-antitop quark pair or a vector boson in proton-proton collisions at $\sqrt{s} = 13\,\text {Te}\hspace{-.08em}\text {V} $}",
    eprint = "2303.01214",
    archivePrefix = "arXiv",
    primaryClass = "hep-ex",
    reportNumber = "CMS-HIG-21-007, CERN-EP-2023-004",
    doi = "10.1140/epjc/s10052-023-11952-7",
    journal = "Eur. Phys. J. C",
    volume = "83",
    number = "10",
    pages = "933",
    year = "2023"
}

@article{HIKE:2022qra,
    author = "Cortina Gil, E. and others",
    collaboration = "HIKE",
    title = "{HIKE, High Intensity Kaon Experiments at the CERN SPS}: {Letter of Intent}",
    eprint = "2211.16586",
    archivePrefix = "arXiv",
    primaryClass = "hep-ex",
    reportNumber = "CERN-SPSC-2022-031, SPSC-I-257",
    month = "11",
    year = "2022"
}

@article{HIKE:2023ext,
    author = "Ashraf, M. U. and others",
    collaboration = "HIKE",
    title = "{High Intensity Kaon Experiments (HIKE) at the CERN SPS Proposal for Phases 1 and 2}",
    eprint = "2311.08231",
    archivePrefix = "arXiv",
    primaryClass = "hep-ex",
    reportNumber = "CERN-SPSC-2023-031",
    month = "11",
    year = "2023"
}

@article{Bringmann:2026xcx,
    author = "Bringmann, Torsten and Konstandin, Thomas and Matuszak, Jonas and Schmidt-Hoberg, Kai and Tasillo, Carlo",
    title = "{Tuning the violins: dark sector phase transition models for the PTA signal}",
    eprint = "2602.09092",
    archivePrefix = "arXiv",
    primaryClass = "hep-ph",
    reportNumber = "DESY-26-014, P3H-26-009, TTP26-002",
    month = "2",
    year = "2026"
}

\end{document}